\documentclass[acmsmall,screen]{acmart}
\AtBeginDocument{%
  \providecommand\BibTeX{{%
    \normalfont B\kern-0.5em{\scshape i\kern-0.25em b}\kern-0.8em\TeX}}}

\usepackage{multirow} 
\usepackage{makecell} 
\usepackage{xcolor}

\setcopyright{acmlicensed}
\acmJournal{PACMHCI}
\acmYear{2021} \acmVolume{5} \acmNumber{CSCW1} \acmArticle{9} \acmMonth{4} \acmPrice{15.00}\acmDOI{10.1145/3449083}
\newcommand\revision[1]{{\color{black} #1}}

\begin{document}

\title{CASS: Towards Building a Social-Support Chatbot for Online Health Community}
% \title{CASS: a Generalize Chatbot Architecture for supporting member well-being and community engagement}

% \title{Learn from the Community, For the Community: Building a Neural-Network-Based Generic Chatbot Architecture to Support Online Health Community}

% \title{By the People, For the People: Explore, Build, and Deploy a Neural-Network-Based Chatbot with an Online Health Community Data and to Provide Social Support for the Community}

\author{Liuping Wang}
\authornote{Both authors contributed equally to this research.}
\email{liuping2019@iscas.ac.cn}
\affiliation{
\institution{Institute of Software, Chinese Academy of Sciences}
}
\affiliation{
    \institution{University of Chinese Academy of Sciences}
    \country{China}
}

\author{Dakuo Wang}
\authornotemark[1]
\email{dakuo.wang@ibm.com}
\affiliation{
    \institution{IBM Research}
    \country{USA}
}

\author{Feng Tian}
\authornote{Corresponding author}
\email{tianfeng@iscas.ac.cn}
\affiliation{%
    \institution{Institute of Software, Chinese Academy of Sciences}
    \country{China}
}

\author{Zhenhui Peng}
\affiliation{
    \institution{The Hong Kong University of Science and Technology}
    \country{Hong Kong}
}

\author{Xiangmin Fan}
\affiliation{
    \institution{Institute of Software, Chinese Academy of Sciences}
    \country{China}
}
 
\author{Zhan Zhang}
\affiliation{
    \institution{Pace University}
    \country{USA}
}

\author{Shuai Ma}
\affiliation{
    \institution{The Hong Kong University of Science and Technology}
    \country{Hong Kong}
}

\author{Mo Yu}
\affiliation{
    \institution{IBM Research}
    \country{USA}
}

\author{Xiaojuan Ma}
\affiliation{
    \institution{The Hong Kong University of Science and Technology}
    \country{Hong Kong}
}

\author{Hongan Wang}
\affiliation{
    \institution{Institute of Software, Chinese Academy of Sciences}
    \country{China}
}

\renewcommand{\shortauthors}{Wang and Wang, et al.}

\begin{abstract}
\revision{Chatbots systems, despite their popularity in today's HCI and CSCW research, fall short for one of the two reasons: 1) many of the systems use a rule-based dialog flow, thus they can only respond to a limited number of pre-defined inputs with pre-scripted responses; or 2) they are designed with a focus on single-user scenarios, thus it is unclear how these systems may affect other users or the community. In this paper, we develop a generalizable chatbot architecture (CASS) to provide social support for community members in an online health community. The CASS architecture is based on advanced neural network algorithms, thus it can handle new inputs from users and generate a variety of responses to them. CASS is also generalizable as it can be easily migrate to other online communities. 
With a follow-up field experiment, CASS is proven useful in supporting individual members who seek emotional support. Our work also contributes to fill the research gap on how a chatbot may influence the whole community's engagement.
}
% with three studies: 1) Exploring and analyzing an online health community for pregnant women to gather contextual knowledge and data as the foundation of system building research; 2) Developing a chatbot system with an advanced neural network algorithm (CNN and LSTM in particular) trained on the community data, in return designing the \textbf{C}h\textbf{A}tbot with a goal to provide \textbf{S}ocial \textbf{S}upport for individual community members (\textbf{CASS}), as well as to foster community contribution and commitment; 3) Lastly, conducting a follow-up field experiment with the chatbot deployed back into the community to evaluate its functional effectiveness, benefit to individual members, and supports for community well-being. Each study used different research methods (e.g., content analysis, descriptive statistical analysis, and system deployment), but these studies as a whole exemplify an Action Research (AR) that the researcher's learning and doing, gaining research insights and contributing back to the community can be achieved at the same time.

\end{abstract}




\maketitle

\section{Introduction}
Chatbot systems\footnote{\revision{In this paper, we consider functional bots (e.g., Twitter bots~\cite{TwitterBotGilani:2017:BIT:3041021.3054255}, and Wikipedia bots~\cite{ Wikipedia:Johnson:2016:HRP:2858036.2858123}) out of our research scope, as these bots' primary function focuses on completing tasks (e.g., broadcasting an information or editing an article), as opposed to communicating with users through a flow of conversations, which is the primary feature of chabots.}} have been increasingly adopted in many fields (e.g., healthcare~\cite{doctorbot}, human resources (HR)~\cite{liao2018all}, and customer service~\cite{xu2017new}), since the first chatbot system---ELIZA---emerged in 1964 to provide consulting sessions as a computer therapist~\cite{ELIZA:Weizenbaum:1966:ECP:365153.365168}. \revision{In recent years, an increasing number of chatbot systems are being developed in various research labs and companies with a premise that these systems can have more powerful capabilities and support} more user scenarios~\cite{hu2018touch,TaskManagement:Toxtli:2018:UCT:3173574.3173632,cranshaw2017calendar}. For example, Hu et al. ~\cite{hu2018touch} built an experimental chatbot system that can understand the tones in a text input (e.g., sad or polite) and generate responses with an appropriate tone. 

\revision{Following these system development efforts, many recent Human-Computer Interaction (HCI) and Computer-Supported Cooperative Work (CSCW) studies have examined various aspects of chatbots from the end users' perspective, such as human-in-the-loop chatbot design~\cite{cranshaw2017calendar}, user perception of chatbots ~\cite{Typefaces10.1145/3025453.3025919}, playful usage of chatbots ~\cite{liao2018all}, and human trust in chatbots~\cite{Jakesch:2019:ACP:3290605.3300469}. However, most of these studies have inherent limitations: 1) many chatbot systems (e.g.,~\cite{xu161same,liao2018all}) use a rule-based architecture, which makes the chatbot capable of understanding only a limited number of user inputs and responding with prescripted sentences, hindering its generalization; and 2) most chatbots are deployed and tested only in single-user scenarios, but how these systems interacting with and impacting a group of users (or a community) is understudied}.

The first limitation of current chatbots -- only returning a pre-defined list of responses to a user -- is partially caused by the use of traditional heuristic rule-based algorithms or information retrieval techniques~\cite{Recommend_comment_Morris2018TowardsAA}. Even with some advanced chatbot-development toolkit's help (e.g., Microsoft Cognitive Service~\cite{MicrosoftLUIS} and IBM Watson~\cite{IBMWatson}), a chatbot can only use neural-network based approaches (NN) to understand the text, but its responding function is still limited to a rule-based selection process. 
% Besides, there is a good reason for not using NN in these commercial platforms---without a careful design and \textbf{human-in-the-loop} monitoring (e.g.,~\cite{cranshaw2017calendar}), a chatbot designed using such NN capability can easily turn into a disaster (e.g., Miscrosoft Tay~\cite{wolf2017we}). 
\revision{In this work, we propose a NN-based chatbot architecture based on which a chatbot system can accurately handle unseen questions and generate various forms of responses with the same meaning. 
This architecture is designed to have a high scalability and generalizability, so that other researchers and developers can take our code\footnote{We have made the code repository open source: https://github.com/liupingw/CASS-Framework}, provide it with a cleaned and labeled training dataset, retrain it, and deploy it for different online communities. Inspired by previous literature~\cite{wolf2017we}, we also build a human-in-the-loop module so a human operator can monitor and intervene the fully automated architecture, if needed.}

The second research gap is that most of today's HCI and CSCW research primarily focus on an individual user's interaction with and perception of a chatbot system~\cite{xu2017new, customer_engagement_tone,liao2018all,woebot,DBT:Schroeder:2018:PSC:3173574.3173972,Care_for_bot_Lee:2019:CVC:3290605.3300932}, and it is unclear how a chatbot system may affect a group of users or a community. 
\revision{In this paper, we aim to address this research gap by building and deploying a social-support chatbot system, \textbf{CASS} (\textbf{C}h\textbf{A}tbot for \textbf{S}ocial \textbf{S}upport), and evaluating its impact on the individuals who need social-support as well as on the other members in the community. Motivated by existing literature that online communities often suffer low engagement from the members due to that many posts can not get a timely response, our chatbot's primary function is designed to engage in conversations with those un-replied social-support-needing posts. CASS automates the entire end-to-end process: retrieving new posts from the community, classifying these posts as with (or without) social-support needs, and generating appropriate responses to the posts with social-support needs. }

This study is exploratory in nature. We choose an online pregnancy healthcare community as our research site (\textit{YouBaoBao}\footnote{https://youbaobao.meiyou.com/}), because this type of online community is extensively studied in recent research~\cite{Pregnant_community_Gui:2017:ISS:3171581.3134685,hci_and_intimate10.1145/2858036.2858187}, allowing us to leverage these existing knowledge to inform the system design and implementation. The whole research project consists of three parts, and we will organize this paper following the order of these three studies: 

Study 1 is an empirical exploration study, where we conduct both qualitative content analysis and descriptive statistical analysis  to understand the context of the research site. A qualitative open coding of the posts suggests that there are three primary categories of posts in YouBaoBao:\textit{ Emotional-Support Seeking, Informational-Support Seeking, and Sharing Daily Life.} The quantitative statistical analysis also shows that half of the posts in this community can not get a timely response, which may cause the already-stressful community members (i.e., pregnant women) to be at a higher risk~\cite{Pregnant_community_Gui:2017:ISS:3171581.3134685}.

% It may also lead to lower community participation or community commitment, which are prevalent problems adversely affecting the online community healthiness. 
\revision{To provide social supports to this community, in Study 2, we build the CASS chatbot system, using the proposed generalizable NN-based architecture. The CASS design is tailored based on the findings from Study 1, so that it can understand what posts need social support, and what posts count as an un-replied case that needs immediate intervention. In Study 2, we also adopt the standard evaluation practices (using both automated metrics and human evaluators) to show that the core NN-based algorithms' performance is satisfactory.}
%  detect and respond to those emotional support seeking

\revision{At last, in Study 3, CASS is deployed back to the YouBaoBao community, with a human-in-the-loop module to filter inappropriate AI-generated responses. Through a 7-day field experiment, the result shows that the CASS system can provide the desired emotional support to the individual community members who need help. In addition, we find evidence to support that the deployment of CASS indeed have positive impacts on other members and the entire community, as other community members are more likely to participate in the conversations intervened by the chatbot system. This finding suggests that such social-support chatbot can not only support individual member's well-being, but also improve the community engagement level, which is also an important dimension according to McGrath's TIP theory of groups~\cite{mcgrath1991time}.}

\revision{In summary, this paper makes the following contributions: 
\begin{itemize}
  \item An empirical understanding of the challenges and user needs in an online pregnancy healthcare community, and how these findings can be used to tailor a chatbot system design;
  \item A scalable and generalizable chatbot development architecture, with which researchers and developers can easily build a fully automated chatbot system with NN-based models to be deployed in another online community; 
 \item Insights and recommendations for designing and deploying future chatbots systems to interact or collaborate with humans in the context of online communities.
\end{itemize}
}

\section{Related Work}
\revision{The literature review is divided into three subsections: we first review selected HCI work on social support scenarios in online health communities. Then, we focused on the group of literature about human and chatbot interaction. Lastly, we switch to the literature that specifically addresses challenges and issues of chatbot systems deployment in real world.}

\subsection{Online Health Community and Emotional Social Support}
Online community has been a longstanding research topic for HCI and CSCW researchers (e.g.,~\cite{zhu2013effects,social_support_article,To_stay_or_leave_Wang:2012:SLR:2145204.2145329,linguisticSharma:2018:MHS:3173574.3174215,Channel:Yang:2019:CMS:3290605.3300261,Social_support_definition:Bambina:2007:OSS:1534606,Emotional_support_UdenKraan2012BreastC,Pregnant_community_Gui:2017:ISS:3171581.3134685,seekers:Yang:2019:SPW:3290605.3300574,Norms_matter:Chancellor:2018:NMC:3173574.3174240,discovering10.1145/2389176.2389216,donath2002identity,Actor_Network10.1145/3290605.3300914}). Existing studies have looked at a variety of topics including community structure~\cite{resnick2012starting,kraut2011encouraging,shaping_pro_10.1145/2998181.2998277}, community activities~\cite{kraut2011encouraging,Pregnant_community_Gui:2017:ISS:3171581.3134685,collaborative_10.1145/2145204.2145331,Logie2011AskedAA,collective_10.1145/2702123.2702566,wen2012understanding,Pseudonymous_10.1145/3173574.3174063}, members' commitment and contribution~\cite{ren2012encouraging,Commitment10.1145/3025453.3026008}, engaging newcomers~\cite{kraut2010dealing,kraut2012building,ren2012building}, rewarding mechanism design~\cite{resnick2012starting,kraut2011encouraging,predictors_of_answer10.1145/1357054.1357191}, and the cold start problem~\cite{resnick2012starting}. Recently, a number of studies has focused on a special type of online community -- the online health community for pregnant women ~\cite{predicting_postpartum10.1145/2470654.2466447,characterizing_predicting10.1145/2531602.2531675,Pregnant_community_Gui:2017:ISS:3171581.3134685,social_capital_doi:10.1111/j.1467-9566.2005.00464.x,connected_motherhood_doi:10.1089/tmj.2014.0118}. This is a special group of users. In addition to the significant changes on their bodies, their mental state also changes a lot over the trajectory of their pregnancy. They often have a much higher stress level than before getting pregnant, thus their mental health is at high stake ~\cite{Ishibe2006Incidence,depression_pregnancy:Barry:2017:MMM:3025453.3025918, pregnancy_depressionarticle,xiang2020sedentary}. Banti et al. ~\cite{Banti2011From} reported that 12.4\% of pregnant women had presented some depression symptoms during the pregnancy, and 9.6\% of them encountered depression in the postpartum period.

It is known that pregnant women often go to online health communities to seek social support from peers~\cite{social_sharing_10.1145/2675133.2675297,social_support_EVANS2012405,mobile_phone_10.1145/2702123.2702258}. Previous literature reveals that members in such health communities actively seek social support from others, and they are also willing to volunteer their time to provide social support to other help seekers~\cite{de2014mental,yang2017self}. Prior research roughly divided the social supports into two categories: \textit{informational social support and emotional social support}~\cite{To_stay_or_leave_Wang:2012:SLR:2145204.2145329,cutrona1994social,biyani2014identifying}. Informational support refers to posters seeking information or knowledge about the course of their disease, treatments, side effects, communication with physicians, and other burdens (e.g., financial problems) ~\cite{To_stay_or_leave_Wang:2012:SLR:2145204.2145329}. Emotional support refers to posters seeking encouragement and empathy when experienced an emotional disturbance ~\cite{Pregnant_community_Gui:2017:ISS:3171581.3134685}. In this project, as an illustration, our chatbot system focuses on providing non-informational social support to community members.

It is often difficult to motivate community members to actively reply to others members' posts in a timely manner~\cite{ackerman1996answer,nam2009questions, lee2013analyzing,mamykina2011design,wang2016answerer}. Seminal research has explored various ways to solve this problem~\cite{engaging_10.1145/2702123.2702124}. In more traditional online communities (e.g., Wikipedia), researchers have attempted to stimulate member's intrinsic and extrinsic motivations~\cite{ackerman1996answer} with monetary reward~\cite{lee2013analyzing} or virtual badges and reputation rewards~\cite{mamykina2011design}. In the online health communities, when a user posts a support seeking post, he/she is recommended to use simpler language and express the needs more explicitly ~\cite{talk_to_me:Arguello:2006:TMF:1124772.1124916}. It is also suggested that posts with more detailed user profile information and a photo are more likely to get replies~\cite{face_attract_Bakhshi:2014:FEU:2556288.2557403}. Even so, there are many posts may never get a reply. For example, Wang et al. ~\cite{no_reply_community:article} reported that at least 10\% posts in an online community never received a response. 

When a pregnant woman posts a support seeking post and never gets a response, it may have more severe harms to the user and to the community. Because pregnancy women are already stressful, overlooking their support seeking may make things worse ~\cite{Depressed_women_Leigh2008RiskFF,Design_Opportunities_for_Mental_Health_Peer_Support_TechnologiesO'Leary:2017:DOM:2998181.2998349}. Furthermore, when community members constantly fail to get the needed social support, they are less likely to contribute to the community, so the community engagement level decreases accordingly, and even worse, the members may leave the community over time~\cite{zhu2014impact,kraut2012building,To_stay_or_leave_Wang:2012:SLR:2145204.2145329}.

In this paper, we will illustrate how to  leverage on the latest AI technology to build a chatbot system that can automatically detect non-informational support-seeking posts, and respond to it with appropriate sentences. The system architecture is scalable and generalizable so it can be easily migrated to other online communities.

\subsection{Human Interaction with Chatbots}

Chatbot is an increasingly popular research topic in recent years. Most of today's chatbot systems are built to interact with a single user~\cite{xu2017new, customer_engagement_tone,liao2018all,woebot,DBT:Schroeder:2018:PSC:3173574.3173972,Care_for_bot_Lee:2019:CVC:3290605.3300932}. For example, the famous ELIZA and its successors can provide individual cognitive therapy sessions to users with a purpose of relieving their stress and anxiety, as well as helping them gain self-compassion~\cite{woebot,DBT:Schroeder:2018:PSC:3173574.3173972,Care_for_bot_Lee:2019:CVC:3290605.3300932}. Many commercial chatbots in customer service domain are designed to answer customers' frequently asked questions or perform a simple function (e.g., check bank account balance) ~\cite{xu2017new, customer_engagement_tone}. There are also human resources (HR) chatbots serving as a process guider to lead new employees to go through their onboarding process~\cite{liao2018all}. There are also some chatbot applications in the healthcare domain which can help patients to better understand their symptoms~\cite{doctorbot}. 

Only till recently, a number of researchers have started to explore and build chatbots that can interact with a group of people~\cite{dissecting10.1145/2675133.2675208,Could_you_10.1145/3025453.3025830,TaskManagement:Toxtli:2018:UCT:3173574.3173632,MakeSenseGroupChat:Zhang:2018:MSG:3290265.3274465,Face_value:Shamekhi:2018:FVE:3173574.3173965,cranshaw2017calendar,seering_village10.1145/3313831.3376708}. For example, Toxtli et al. ~\cite{TaskManagement:Toxtli:2018:UCT:3173574.3173632} built a chatbot as a group chat facilitator to assign tasks to group members. Zhang et al. ~\cite{MakeSenseGroupChat:Zhang:2018:MSG:3290265.3274465} developed a chatbot for an online communication application to automatically summarize group chat messages. Cranshaw et al. ~\cite{cranshaw2017calendar} developed a chatbot system that can serve as an assistant to coordinate meetings for multiple people via email. All these studies expand the chatbot user scenario from supporting single user to multiple users. Notably, Seering et al. ~\cite{seering_village10.1145/3313831.3376708} recently built a chatbot in an online gaming community on Twitch. They designed four different versions of the chatbot -- ``baby'', ``toddler'', ``adolescent'', and  ``tennager'' -- to simulate a chatbot's grownup process in a 3-week deployment. However, the user interaction with the chatbot is quite primitive that users need to input command-line text (e.g., ``@Babybot'' or ``!feed''), thus it is more like a digital pet (tamagotchi) ~\cite{pettman2009love} than a chatbot. 
% as they found the introduction of the chatbot sparked both human-chatbot interactions and human-human interactions, suggesting that a chatbot can have positive impact on a community.

Along this research line, our work builds a chatbot system that can provide emotional support to pregenant women members in an online community; also, we design a field experiment to reveal findings on how such deployment of the chatbot can impact the whole community. Different from Seering's work~\cite{seering_village10.1145/3313831.3376708}, where they deployed the ``digital pet'' into a Twitch community and users can have chitchat with the system, we aim to build a chatbot to meet community members' existing needs --- social-support seeking. We hope our chatbot can provide functional benefits to the users through a conversational communication. Another difference is that Seering's system~\cite{seering_village10.1145/3313831.3376708} used an information retrieval (IR) approach, and in turn, it could only return limited responses that were pre-defined by the researchers. The rule-based approach (e.g., IR) constraints the potential of chatbot in providing social support for a community~\cite{li2016deep}, and users may perceive the responses from the chatbot not as useful as from people ~\cite{Recommend_comment_Morris2018TowardsAA}. In contrast, we build an architecture that leverages on the state-of-the-art NN-based models for the development of more powerful chabot systems. 
% This approach has been demonstrated feasible in the study conducted by Hu et al. ~\cite{hu2018touch} where they exemplified the use of a sequence-to-sequence model to build a language generation module for their tone-aware customer service chatbot.

\revision{\subsection{Impacts of Chatbot System Deployment}
While reviewing recent work on building and deploying chatbots~\cite{Race:Schlesinger:2018:LTR:3173574.3173889, DBLP:journals/corr/abs-1810-09590, halfaker2019ores}, we found that despite many of these chatbots were designed with a good will, the deployment of certain systems may negatively impact the stakeholders or the intended users. 
% For example, during the development of customer service chatbots, it needs a significant amount of data, for which human customer service experts need to label  (e.g.,~\cite{hu2018touch,xu2017new}), but the deployment of such chatbots may cause companies to hire less and less customer service workers in the future. 

One example is the chatbot system developed by ~\cite{seering_village10.1145/3313831.3376708} and deployed on Twitch. It designs a novel interaction approach that users can ``raise'' the chatbot as a pet through a number of commands, such as ``!feeding''. However, the deployment of such a chatbot may distract users' attention from their original goal of using the platform, which is to watch videos and socialize with the host and the other community members. Thus, with the chabot, users may engage with the platform but not with each other. Such close bonding with a chatbot may even hurt the individual user's benefits in the long term\footnote{A report says there are users interacting with Microsoft's XiaoIce for more than six hours restless, and treat XiaoIce as his girlfriend~\cite{zhou2020design}} and also negatively impact the community's engagement level~\cite{zhou2020design}.

The unexpected consequence of a chatbot's deployment is not uncommon when putting AI and machine learning systems to practical use. Often, today's NN-based algorithm research requires a large amount of data. Such data may come from an online community (e.g., Reddit), and they need to be tagged with ground truth labels by human annotators. However, the benefit of the original human annotators may be neglected during the training and deployment of the algorithm, as the algorithm's performance and optimization is developer's most important objective. For example, developing a functional customer service chatbots needs a significant amount of training data labeled by human customer service experts or obtained from their practices (e.g.,~\cite{hu2018touch,xu2017new}). But the deployment of such chatbots may cause companies to hire fewer customer service workers in the future, which may also in return reduce the sources of training data.

Fortunately, some HCI researchers have noticed this challenge and they jointly work on an emerging research topic -- \textbf{Human-Centered AI} -- that aims to take an algorithm's impact on human users into the consideration of the algorithm design~\cite{robert2020designing, woodruff2018qualitative, zhu2018value, bratteteig2018does, lee2017human,halfaker2019ores, halfaker2014snuggle}. For example, Woodruff et al. interview 44 participants from several marginalized populations in the United States. Participants indicate that algorithmic fairness (or lack thereof) could substantially affect their trust in a company or product, if such application is deployed~\cite{woodruff2018qualitative}. Thus, such fairness considerations should be taken into account during the algorithm design~\cite{lee2017human, robert2020designing}. In addition to the fairness, various other design considerations may also influence the eventual consequence of an AI system in the real world deployment, such as stakeholders's tacit knowledge~\cite{zhu2018value} and community involvement~\cite{lee2019webuildai}. It is also important to build an in-depth understanding of user needs or even involve them in the design process, as exemplified by a few recent work adopting participatory design research method~\cite{halfaker2019ores,bratteteig2018does, lee2019webuildai}. 

In addition to incorporating the various considerations into an AI system design, there are also some good practices that one can follow in the deployment of the AI system~\cite{halfaker2014snuggle,DBLP:journals/corr/abs-1810-09590,halfaker2019ores}. For example, a group of researchers designed and developed ORES -- an algorithmic scoring service that supports real-time scoring of wiki edits using multiple independent classifiers trained on different datasets. This paper proposes an example of deploying AI algorithms in an online community, but their algorithm is less explicit to the users, compared to the chatbot systems that we aim to develop and deploy in this paper.

% In addition, there are also many practices that learnt from community and for the community on wikipedia,such as ORES~\cite{halfaker2019ores} which is an algorithmic scoring service that supports real-time scoring of wiki edits), Wikipedia's bots~\cite{DBLP:journals/corr/abs-1810-09590} which operate in complex social and technical environment, Snuggle~\cite{halfaker2014snuggle} which is a new interface that promote the development of wikipedians.} 
% This philosophy is well-documented in \textbf{Action Research} studies~\cite{hayes2011relationship}, suggesting that researchers should not only learn novel research insights from the studied community, but also contribute benefits back to the same community.

In this paper, we design and develop a chatbot system that can provide social support for an online health community. Besides the objective of ensuring a good functional performance, we are also interested in evaluating its potential impacts on the online community after its deployment. This work joins the recent \textbf{Human-AI Collaboration} research effort~\cite{wang2019humanai,xu161same,doctorbot,autods} that aims to develop and deploy AI systems that can work together with people, instead of replacing people. It differs from the Human-AI Interaction discussion~\cite{amershi2019guidelines}, as it goes beyond the usability and interactive design of AI systems, but focuses more on the cooperative nature of AI systems with human partners and their context (e.g.,~\cite{horvitz1999principles,grudin2017tool}). }

\section{Study 1: Empirical Understanding of the research site}

\subsection {Research Site}
In this subsection, we first provide an overview for the research context---YouBaoBao---one of the largest and most popular online health communities for pregnancy and parenting discussions in China. Users are often pregnant women or couples that are expecting a baby. They can discuss a variety of topics on this platform, such as pregnancy, childbirth, childcare, and early education. This community has 164 sub-forums (similar to ``sub-reddits'' in Reddit), such as specific sub-forums for different stages of the pregnancy: ``First Trimester'', ``Second Trimester'', and ``Third Trimester''. YouBaoBao is a self-organized health community, where community members can publish a post to seek advice and information, and they can reply to a post to provide support and advice. Community members have an option to disclose their current \textit{pregnancy status}, which is always displayed next to their user name. There are three pre-defined categories of pregnancy status from which a user can choose: preparation stage (e.g., ``planning to have a baby''), three-trimester stage (e.g., ``due in 8 months''), and postpartum stage (e.g., ``having a 4-month old baby''). It is worth noting that members can post, browse, and reply in any of the sub-forums regardless of their labeled pregnancy status. For example, we saw instances where new moms posting and replying in the ``first-trimester'' sub-forum, despite they have already delivered. 

\begin{figure}
\centering
 \includegraphics[width=\columnwidth]{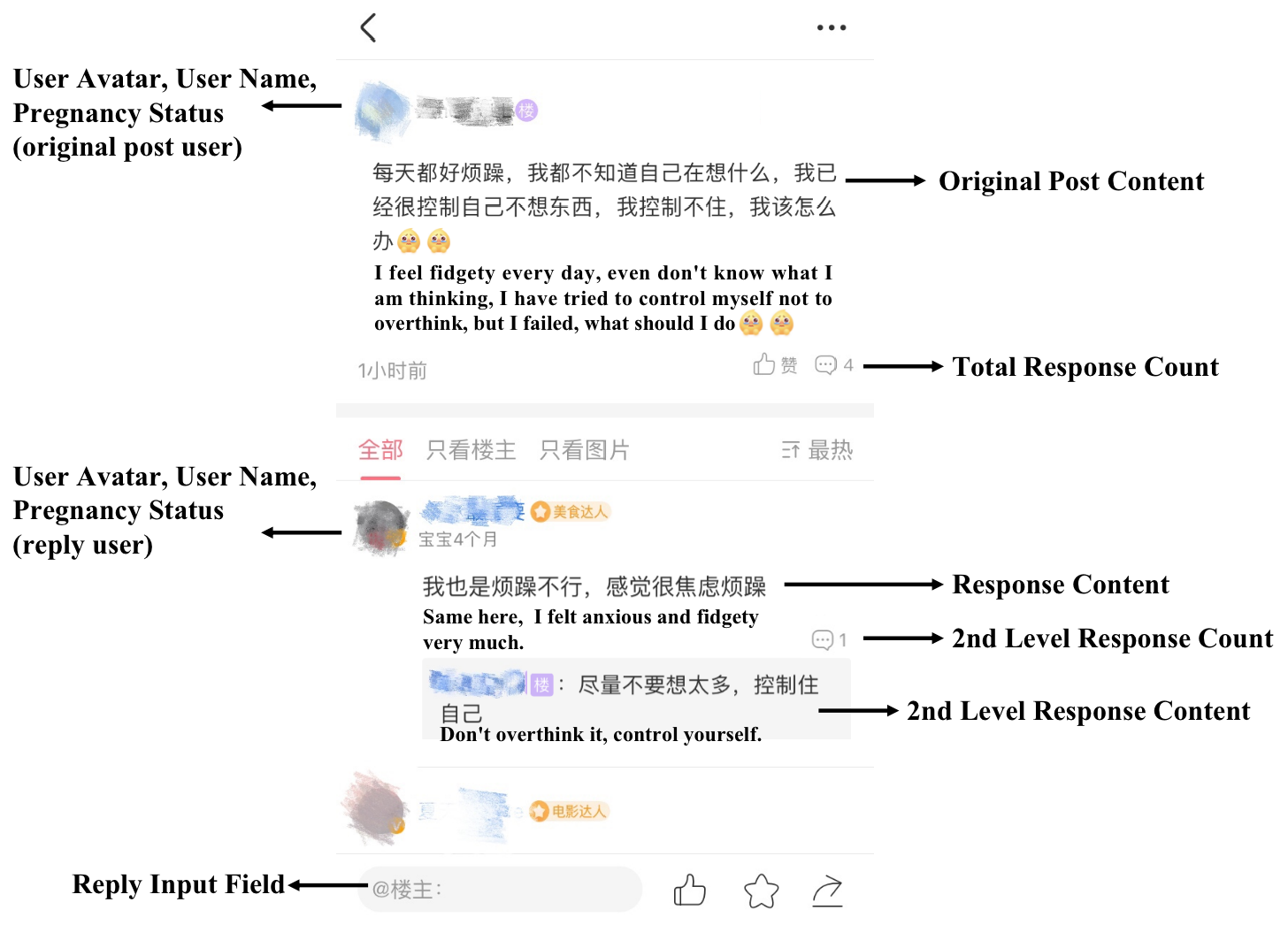}
 \caption{User Interface of YouBaoBao community. It has a original post section to the top, a response section in the middle, and a input field for typing response at the bottom. The content is translated into English in this figure.}~\label{fig:forum_post}
 \Description[A screenshot of a post, including the reply to the post]{The post wrote:"I feel fidgety every day,even don't know what I am thinking. I have tried to control myself not to overthink,but I failed, what should I do." The sentence ends with two crying faces. The first reply wrote:"Same here, I felt anxious and fidgety very much." Then poster replied to the first reply, the poster wrote:"Don't overthink it,control yourself."}
\end{figure}

A \textit{post} or \textit{response} has minimum and maximum word limits---the \textit{content} has to be between 6 and 3000 words. When a post is created, the system can automatically recommend a sub-forum for it according to its content and the user's pregnancy status. Users can also manually modify the recommended sub-forum. Fig.~\ref{fig:forum_post} illustrates a typical post and responses it received (a direct response to the post and a second-level response). The user profile avatar, user name, and pregnancy status are displayed right above the content. The system also shows the \textit{total number} of responses for a post, and of second-level responses for a first-level response.

\subsection{Method}
To better understand what challenges the community faces and where chatbots can support, we first conducted both descriptive statistical analysis and  qualitative content analysis (following the method in ~\cite{Pregnant_community_Gui:2017:ISS:3171581.3134685}) to understand the characteristics of support-seeking activities in this community. Specifically, we explored the following \textbf{research questions}: What types of support do users (e.g., pregnant women) seek in this online community? How do community members respond to those support seeking posts? What issues or barriers exist in communications and interactions in the community?

\subsubsection{Data Collection}
In order to gain a comprehensive view of support-seeking across different stages of pregnancy, we chose five sub-forums with each focusing on one specific pregnancy stage: \textit{``pregnancy preparation'', ``first-trimester'', ``second-trimester'', ``third-trimester'', and ``having a baby less than 6-month-old''}. We collected six months of data from September 2018 to March 2019 through the community's application programming interface (API), and organized them into post-response pairs (N = 220,000). All the data we collected were publicly available but we removed all identifiable personal information (e.g., user names and status) in consideration of research ethics. We also replaced images with a special icon if there were any images in a post content. This data collection method is standard in various HCI~\cite{seekers:Yang:2019:SPW:3290605.3300574,Pregnant_community_Gui:2017:ISS:3171581.3134685} and NLP research literature~\cite{tan2019context,wang2008recovering}. This study is approved by the first author's university Institutional Review Board (IRB).

\revision{As this labeled dataset was later used for model training purposes, two coders removed the malicious posts, such as advertisements, before performing any detailed analysis. We followed iterative coding process and conducted axial coding: two coders first independently coded a group of 200 randomly-selected posts to identify 11 themes and organized these themes into categories. Then, the research team met and discussed the coding schema and differences in coding to resolve all disagreements. After that, the same coders conducted another round of independent coding with a group of 300 randomly-sampled new posts, and discussed newly emerged categories and disagreements in coding. They repeated this process until there was no newly emerged categories or inconsistent codings. In total, they repeated three iterations with each time analyzing 300 new posts (200+3*300=1,100 posts). At the end, we randomly sampled another 2,300 posts for computing the coding reliability score. Based on the agreed coding schema, the two coders independently coded these 2,300 new posts and reached a high inter-rater reliability (Cohen's kappa = 0.86).}
% At the end, two researchers coded 2,300  posts in total with a high inter-rater reliability (Cohen's kappa = 0.86).

\subsection{Result} \label{study1-result}
\subsubsection{Characteristics of Support Seeking and Providing in the Community}
\begin{figure}
\centering
 \includegraphics[width=\columnwidth]{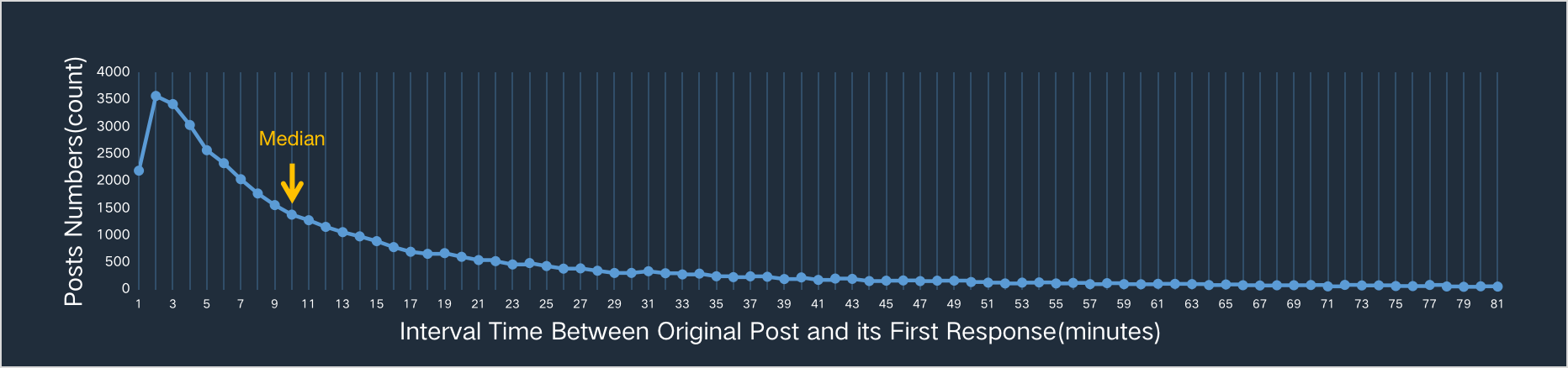}
 \caption{Distribution of Interval Time between a post is published and its first response is published. Median is 10 minutes.}~\label{fig:replyTime}
 \Description{Line graph showing posts numbers from 0 to 4000  in increments of 500 on the Y axis against interval time between original posts and its first response from 1 to 81 minutes in increments of 2 on the X axis. One lines are shown. The X and y of the starting point are 1 and 2192 respectively, and the X and y of the second point are 2 and 3570 respectively. After that, the point showed a downward trend and finally approached 0. The tenth minute position was marked with the median.}
\end{figure}

The descriptive statistical analysis helped us understand the baseline of user behaviors in the community. \revision{These findings later were used to guide the design decisions of the CASS system.} More specifically, we observed that, on average, the users on this community posted 5,000 posts per day, and each post had an average of 6 responses. However, approximately 18\% of the posts did not receive any response. To further investigate what these posts are and how they can be replied, we randomly sampled another 60,000 posts out of the collected 220,000 samples for further analysis. We calculated the \textit{Interval Time} between the original post was published and the first response was made. As shown in Fig.~\ref{fig:replyTime}, the median interval time is 10 minutes. Later, we used this number in Study 3 as a threshold to evaluate whether a post gets timely response after the deployment of our chatbot system.

We also analyzed the time span from the beginning as the original post was published to the end as its  last comment was made. This indicator reflects the retention of a post and its discussion thread in an online community (e.g., how long a post remains ``alive'' to the community members). We name this as the \textit{Lifespan} of a post. We found that the average lifespan for a post thread is about 6 days. We define the lifespan for a post with no response as 0 minute. These numbers were also used in Study 3 as thresholds for the Control Logic Module in step 3 in Fig.~\ref{fig:work_flow}. 

There are some other measurements that we calculated to describe the users' behaviors and the community's characteristics, but the results are  similar to the numbers we will report in Study 3, No-Chatbot group, thus we will report them only in Section~\ref{study3result}.

\subsubsection{Categories of Posts}
The content analysis of posts led to the identification of 11 concepts, which were then grouped into 3 high-level categories, including \textit{Informational-Support Seeking, Emotional-Support Seeking, and Sharing Daily Life}. As shown in Table~\ref{tab:categories}, 27.45\% of the posts were related to seeking emotional support or sharing emotional status, 45.95\% of the posts were about seeking informational support, and 26.60\% of them focused on sharing posters' daily life.

Emotional-Support Seeking refers to posters expressing their frustration and stress (e.g., complaining something happened in their work or life), or sharing positive news (e.g., announcing the confirmation of pregnancy) to seek encouragement or empathy. Informational-Support Seeking refers to seeking information, knowledge, advice, and suggestions from others to manage a situation (e.g., how due dates are calculated). Sharing Daily Life refers to posts where community members post a photo of food or exercise in a gym, or share updates and progress of pregnancy.

The resulting categories are similar to previous findings~\cite{Pregnant_community_Gui:2017:ISS:3171581.3134685, seekers:Yang:2019:SPW:3290605.3300574}, confirming that our research site (YouBaoBao) has similar characteristics as other popular pregnancy-related online communities (such as BabyCenter\footnote{https://www.babycenter.com/} and TheBump\footnote{https://www.thebump.com/}). More importantly, the analyses of posts helped us understand the types and nature of emotional support sought by users. Thus, our primary goal for the chatbot is to provide non-informational support (both emotional-support and sharing dailylife) to the posting users.  %~\cite{Pregnant_community_Gui:2017:ISS:3171581.3134685} has 6 categories from coding 683 posts, but we have 3 categories after coding 2,300 posts. That is because our current work does not aim to address the no-response-post problem for all those categories, and for factual information support we need medical experts from the pregnancy healthcare domain to ensure the quality. Thus, we only focus on the emotional support seeking posts as an illustration. So we do not need to have fine-grained categories about informational support seeking activities. 
% Our ``informational support seeking'' category is a combination of their 3 categories ``advice, formal knowledge, and informal knowledge seeking''. Our ``emotional support seeking and sharing'' is a combination of their ``emotional support and reassurance''. Despite the granularity difference between these two coding schema, our category and concept axis are similar to these existing literature (e.g., our concept of ``Seek personal opinions from peers'' is similar to their ``informal knowledge seeking'' category)~\cite{Pregnant_community_Gui:2017:ISS:3171581.3134685}. 
%We do find a novel category where community members use it to share their own daily life, such as post a photo of food or exercise in a gym. This indicates that community members gain support and contribute to the community at the same time. For our research purpose, we do not need to distinguish the emotional support seeking and the sharing daily life categories, because the responses to these two categories can be overlapped with compliment or empathy and our text generation module can handle them the same way, as long as they are not seeking factorial information. 
Later in Section~\ref{study2-classify}, we will illustrate how we translate these learned contextual knowledge into parameters, and feed the labeled data to train the algorithm models in the system architecture.

% \usepackage{multirow}
\begin{table}[]
\begin{tabular}{|l|l|c|c|}
\hline
\multicolumn{1}{|c|}{Category}                                                                           & \multicolumn{1}{c|}{Concept}                                                                   & N                     & \%                     \\ \hline
\multirow{4}{*}{\begin{tabular}[c]{@{}l@{}}Emotional support seeking\\ and emotion sharing\end{tabular}} & \begin{tabular}[c]{@{}l@{}}Complain about work, life, \\ family, etc.\end{tabular}             & \multirow{4}{*}{631}  & \multirow{4}{*}{27.45} \\ \cline{2-2}
                                                                                                         & \begin{tabular}[c]{@{}l@{}}Have a fantasy conversation\\ with the baby to be born\end{tabular} &                       &                        \\ \cline{2-2}
                                                                                                         & Share happiness                                                                                &                       &                        \\ \cline{2-2}
                                                                                                         & Make a good wish                                                                               &                       &                        \\ \hline
\multirow{3}{*}{Informational support seeking}                                                           & Seek health-related information                                                                & \multirow{3}{*}{1057} & \multirow{3}{*}{45.95} \\ \cline{2-2}
                                                                                                         & \begin{tabular}[c]{@{}l@{}}Seek personal opinions \\ from peers about pregnancy\end{tabular}   &                       &                        \\ \cline{2-2}
                                                                                                         & \begin{tabular}[c]{@{}l@{}}Seek opinions and suggestions\\ on baby-related issues\end{tabular} &                       &                        \\ \hline
\multirow{4}{*}{Sharing daily life}                                                                      & Share food and cuisine                                                                         & \multirow{4}{*}{612}  & \multirow{4}{*}{26.60} \\ \cline{2-2}
                                                                                                         & Share sports and other activities                                                              &                       &                        \\ \cline{2-2}
                                                                                                         & \begin{tabular}[c]{@{}l@{}}Share updates of her body\\ along with pregnancy\end{tabular}       &                       &                        \\ \cline{2-2}
                                                                                                         & \begin{tabular}[c]{@{}l@{}}Share updates and progress\\ of the baby or fetus\end{tabular}      &                       &                        \\ \hline
\end{tabular}
\caption{Axial coding result with 3 Categories and 11 Concepts for original posts. (N = 2,300, Cohen's kappa = 0.86)}
\label{tab:categories}

\end{table}

\section{Study 2: Building a neural-network based chatbot with a scalable and generalizable architecture}
Based on the literature~\cite{terveen2014study,seering2019moderator,Pregnant_community_Gui:2017:ISS:3171581.3134685} and the knowledge gained from Study 1, we learned that one major challenge faced by online health communities is that many posts can not get a timely response (e.g., more than 18\% never got a response); for the posts that have responses, it often takes more than 10 minutes to get the first response. In the YouBaoBao context, this challenge is particularly frustrating considering the pregnant women are already at high stress level. To address this prominent challenge, we built a chatbot system to explore the feasibility of chatbot for providing timely response to support seekers. \revision{Because the responses to the emotional-support-seeking posts and to the sharing-daily-life posts can be overlapped with compliment or empathy thus our text generation module can handle them the same way. We combine these two non-informational support-seeking categories, and train the models to detect them from the informational-support-seeking posts. }

In this section, we will introduce how we build the system, what design decisions we made along the implementation process, and lastly how it performs using common natural language processing (NLP) performance metrics and human judgement in offline evaluation\footnote{The offline evaluation is in contrast to an online evaluation approach. Offline evaluation uses machine learning testing or holdout subset of the data to evaluate the performance of a model or system, whereas online evaluation means deploying the model back to the application environment.}.

\subsection{CASS: Chatbot As a Social Supporter for Online Health Community}
\begin{figure}
\centering
 \includegraphics[width=1.0\columnwidth]{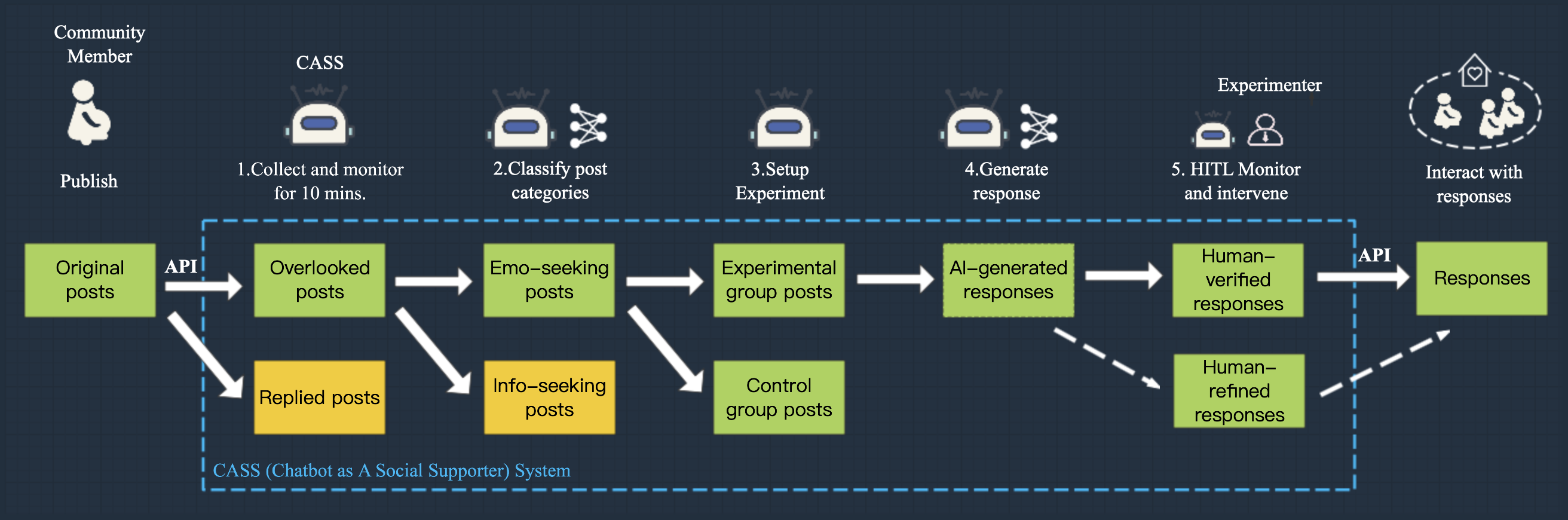}
 \caption{CASS system architecture and workflow. It has 3 modules and 5 steps. Steps 1, 3, and 5 are implemented via the control logic module; step 2 applies the neural network (NN)-based text classification module; and step 4 applies the neural network (NN)-based text generation module.}~\label{fig:work_flow}
 \Description[There are seven stages in the picture.]{Each stage consists of three parts: executor, behavior and object. The objects of execution are posts or replies. The execution object has two colors, green indicates that it is within the scope of study, and yellow indicates that it is not within the scope of study. In the first stage. The executor is a community member whose image is a pregnant woman. This community member's behavior is publishing an original post. The object is original posts. In the second stage, which is step1 mentioned in the text the executor chatbot. The behavior is collecting and monitoring for 10 minutes. The two objects are overlooked posts with green color and replied posts with yellow color.The original posts object points to them respectively, and the arrow pointing to overlooked posts is attached with a text description of the API. In the third stage, which is step2 mentioned in the text the executor is chatbot with neural network. The behavior is classifying posts categories. The two objects are emotion seeking posts with green color and information seeking posts with yellow color. The overlooked posts object points to them respectively. In the fourth stage, which is step3 mentioned in the text, the executor is chatbot. The behavior is setup experiment. The object are experimental group posts with green color and control group posts with green color. They are also pointed to by emotion seeking posts. In the fifth stage, which is step4 mentioned in the text, the executor is chatbot with neural network. The behavior is generating response. The object is AI-based responses with green color. It is pointed to by experimental group posts. In the sixth stage, which is step5 mentioned in the text, the executor is chatbot and human experimenter. The behavior is human-in-the-loop monitoring and intervening. The two objects are human-verified responses with green color which is pointed to by the object of the previous stage with a solid line and human-refined responses with green color which is pointed to by the object of the previous stage with a dotted line. In the last stage, the exectuors are community members. Their behavior is interacting with responses. The object is responses, to which Human-verified responses object points with a solid line and Human-refined responses object points with a dotted line. The arrow pointing by human-refined responses object is attached with a text description of the API. Stages 2 to 6 are framed by a dotted line. The title of this frame is called Chatbot as a social supporter system.}
\end{figure}

In this study, we propose a system architecture that works in the following five steps (Fig.~\ref{fig:work_flow}), and we build a chatbot system--\textbf{CASS}--using this architecture: 
\begin{itemize}
    \item \textbf{Step 1.} Collecting and monitoring new posts \revision{(Input interface: an API URL to fetch data from the community)}; 
    \item \textbf{Step 2.} Classifying posts into emotional-support seeking category and informational-support seeking category using a trained CNN model \revision{(Input: a labeled dataset with posts, responses, and a post's category out of the 3 coded categories)};
    \item \textbf{Step 3.} Randomly dispatching half of the emotional-support seeking posts into a control group and monitor them, this is only for the field experiment in Study 3  \revision{(Parameter: 10 minutes that derived from Study 1, as the threshold for deciding whether the chatbot needs to respond to the overlooked post or not)}; 
    \item \textbf{Step 4.} Generating responses for posts in the experimental group using a trained LSTM model  \revision{(Input: the same input as in step 2 -- the labeled dataset)};
    \item \textbf{Step 5.} A human-in-the-loop verification and intervention user interface, before posting the response back to the forum \revision{(Output interface: an API URL to post response back to the community)}.
\end{itemize}

% Among these steps, Step 3 is specifically designed for the later field experiment in Study 3.
\revision{This chatbot architecture and workflow is quite generic and easy to generalize to other online health forums --- a developer or researcher only needs to alternate the two API URL parameters on how the system reads in original post from the community and how it publishes generated response back to the community, and to feed a labeled dataset as input.} In the following subsection, we present how we use the architecture to build the CASS system with 3 modules: \textbf{a control logic module, a text classification module, and a text generation module}. 

\subsubsection{Control Logic Module}\label{study2logic}
We collectively refer to the posts that can not get a response within 10 minutes as \textbf{overlooked posts}, and the others as replied posts. CASS's control logic module is created to identify and track those posts, dispatch them to the corresponding system modules for text classification or text generation or for human-in-the-loop intervention, and finally publish them back to the community. In Fig.~\ref{fig:work_flow}, the control logic module are responsible for the implementation of \textbf{Step 1, 3, and 5}. 

\textbf{Step 1 and 5} are responsible for collecting posts from the community (i.e., input) and publishing responses to the community (i.e. output), respectively. Step 1 is also responsible to distinguish whether a post is timely-replied (yellow box in Step 1 in  Fig.~\ref{fig:work_flow}; no further interventions were provided for this type of post) or overlooked (green box in Step 1 in  Fig.~\ref{fig:work_flow}). 

In addition to its automated nature of the output step for CASS, we also built a Human-in-the-Loop (\textbf{HITL}) module in Step 5. We hypothesized there may be malfunction of the CASS system as it is fully automated after the model is trained and the system is deployed. Thus, we designed a user interface console, which can be used by a human experimenter to track and monitor each AI-generated response in real-time. Every response generated by the text generation module in Step 4 will show up in the console for 10 seconds before being published to the community. The human experimenter has two options for action: he/she can choose to do nothing then the response is published; or he/she can hit ``Enter'' button to suspend the publishing process. If the latter action is triggered, the human experimenter will be prompted to type in a new sentence in the console, and that new sentence will replace the AI-generated response and then will be sent to the community. 

\revision{In the actual online field experiment in Study 3, two experiment operators carefully monitored the performance of the chatbot in generating responses from 8 a.m. to midnight (16 hours) using the HILT module. But they did not find any inappropriate AI-generated responses that need intervention. This may be because the data collected from YouBaoBao and used to train the models were clean and had nearly no malicious speech corpus. This may also attribute to the friendly nature of the research site~\cite{Pregnant_community_Gui:2017:ISS:3171581.3134685}.}

\textbf{Step 3} is a control logic specifically designed for the field experiment in Study 3. We want to note that step 3 is not needed in a real deployment of the CASS system in the future. This step randomly assigns half of the qualified posts (i.e., overlooked emotional support seeking posts) into an experimental group, and the other half into a control group. For the experimental group posts, CASS generates responses, sends them through HILT step, publishes them on the forum, and keeps tracking for 7 days for the post owner's (i.e. poster) and other community members' (i.e. commenter) actions. For the control group posts, CASS simply tracks the user actions for 7 days without intervention for comparison purpose.

\subsubsection{Text Classification Module}~\label{study2-classify}
Our results from Study 1 suggest that a user typically seek two types of support: informational- and emotional-support. For informational support, posters look for accurate responses rather than diversity in response~\cite{To_stay_or_leave_Wang:2012:SLR:2145204.2145329}. Thus, this type of utilitarian needs can be satisfied by various rule-based algorithms or existing chatbot platforms ( e.g., ~\cite{xu2017new,liao2018all}); this is not the focus of this study. As such, \revision{we build a text classification module based on convolutional neural network (CNN) (\textbf{Step 2} in Fig.\ref{fig:work_flow}) to distinguish informational support seeking posts and non-informational seeking posts (i.e., both emotional support seeking and sharing daily life) so that the chatbot can provide timely response to non-informational support seeking posts only.}

CNN models have been widely used in the computer vision field in recent years~\cite{Krizhevsky2012ImageNet}, and some variations have been extended to text classification tasks \cite{kim2014convolutional}. It normally requires preprocessing each data instance (i.e., posts in this paper) into word vectors with a same dimension, then feed the word vectors into a CNN network; the output of the network is a classification of the input post into one of the three categories. In our study, we do not need to distinguish the detailed categories --- we only need to categorize posts that are informational support seeking, so that our chatbot logic module can ignore those posts. To train this model, we used the labeled data (N=2,300) from Study 1 as training data. Out of the 2,300 labeled posts, 45.95\% of them were seeking informational support. To address the data imbalance issue, we oversampled the data~\cite{barua2012mwmote}, adjusted the ratio of informational support seeking posts to emotional support seeking posts to 1:1 (in total 3,000 posts).

% we first preprocessed the training data and embedded each post to a unified length word vector (600 dimensions)
After the training dataset is prepared, \revision{we first preprocess the training data and encode each post as word vectors using tensorflow ~\cite{tensorflow} (vocabulary size=5000, embedding dimension=64, sequence length=600)}, followed by feeding these vectors into the CNN model. Our CNN model uses a simple structure, including one convolutional layer, two full-connection (FC) layers, one input layer, and one output layer. The input layer has 600 neurons. The convolutional layer consists of three consecutive operations/layers: convolution with kernels, non-linear activation function, and max pooling. The convolutional layer contains 256 kernels, and the FC layer contains 128 units with a dropout rate of 0.5. 

In training, we optimize the model with AdamOptimizer using a learning rate of 0.001, optimized for accuracy. After training, we conduct 5-fold cross-validation. The overall cross-validation accuracy is 0.86 and F1 is 0.87. This accuracy is good enough for our following experiments.

\subsubsection{Text Generation Module}\label{study2-generate}
In \textbf{Step 4}, we build a text generation module. It can be considered as a task that for a given input sequence of words (i.e., post), we need to find another sequence of words (i.e., response) that best suit its input. Thus, we can conceptualize this as a NLP translation task which can be solved using a variation of machine translation networks. To that end, we use OpenNMT\cite{Klein2018OpenNMT}, which is a state-of-the-art open-source toolkit\footnote{https://opennmt.net/} for neural machine translation (NMT) tasks, to build a model to generate responses for a given post. The model is derived from a sequence-to-sequence model with attention mechanism \cite{Luong2015Effective}. \revision{In addition, we build a information retrieval model (IR) with BM25 ranking function~\cite{bm25} as a baseline model.}

The training data is 220,000 post-response pairs from the 5 sub-forums. For each post-response pair, the data contains the post content, a picture icon if the content contains pictures, a post ID,  \# of views, \# of responses, response ID, response user ID, response content, and response timestamp. 

First, we pre-process and embed the input sentence into word vectors (500 dimensions), and then feed them into a Long Short-Term Memory network (LSTM)~\cite{HochreiterLong} as encoder. In the preprocessing step, we first apply the text classification (Section~\ref{study2-classify}) to distinguish and  filter out all the informational support seeking posts and its pair. Then, we manually check the posts to see if there are advertisements (e.g., ``sell infant formula'') or offensive words in the original posts. We do not see any posts using offsensive words but we find many advertisements. Thus, we define a list of stop-words to eliminate advertisement posts. After filtering, 81,000 post-response pairs are left, and we use them as our training dataset.

Second, in the LSTM model training step, attention is applied over the decoder LSTM and combined with the current hidden state at each target time step to produce a prediction of the next word. This prediction is then fed into the decoder LSTM. Specifically, we train a 2-layer deep LSTMs, and each layer has 500 memory cells, using stochastic gradient descent and 0.2 dropout. The batch size is 64, and the global attention translation is conducted using Multilayer Perceptron (MLP). In total, we train 100,000 iterations.

\begin{figure}
\centering
 \includegraphics[width=\columnwidth]{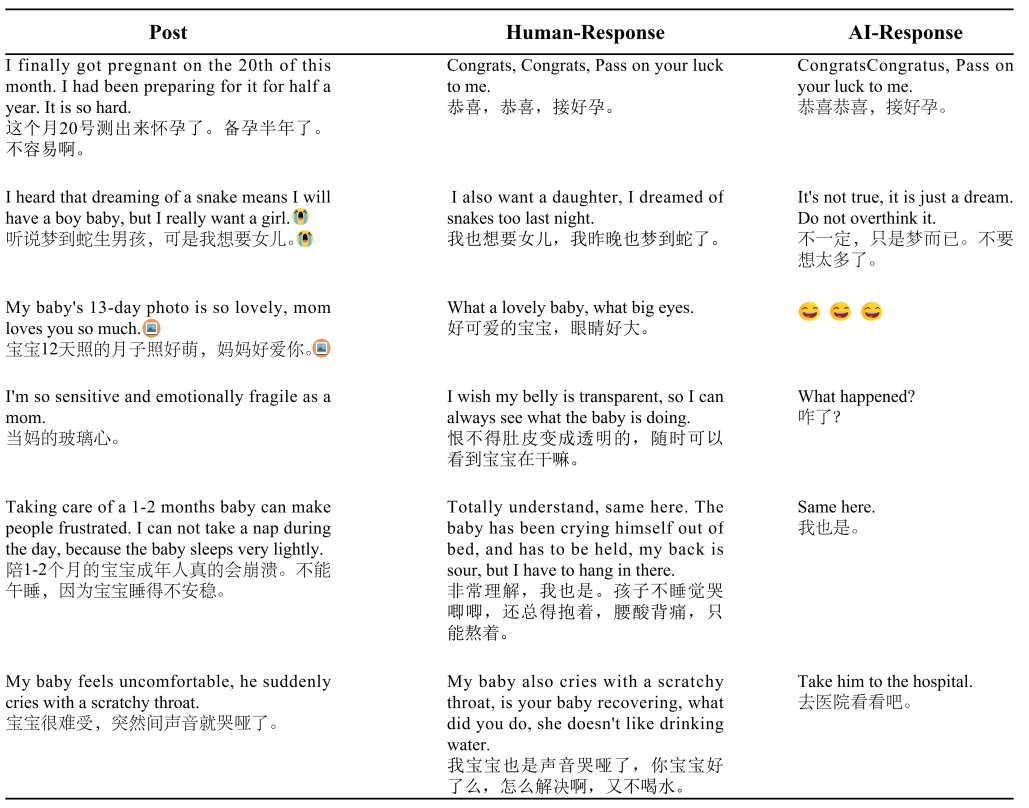}
 \caption{Some examples of human user commented responses as ground truth (Human-Response) and our text generation module generated responses (AI-Response). These sentences are all in Chinese and we translate them into English for this paper. }~\label{fig:dialogue}
 \Description[Seven examples are given in the figure]{First example: the post wrote "I finally got pregnant on the 20th of this month. I had been preparing for it for half a year. It is so hard." Human response is "Congrats, Congrats, Pass on your luck to me." Chatbot replies "CongratsCongrats, Pass on your luck on me." 

Second example: the post wrote "I heard that dreaming of a snake means I will have a boy baby, but I really want a girl (a sob emoji)." Human response is "I also want a daughter, I dreamed of snakes too last night." Chatbot replies "It's not true, it is just a dream. Do not overthink it ." 

Third example: the post wrote "My baby's 13-day photo is so lovely, mom loves you so much (a photo of her daughter)." Human response is "What a lovely baby, what big eyes." Chatbot replies with three grin emoji.

Fourth example: the post wrote "I'm so sensitive and emotionally fragile as a mom." Human response is "I wish my belly is transparent, so I can always see what the baby is doing." Chatbot replies "What happened?" 

Fifth example: the post wrote "Taking care of a 1-2 months baby can make people frustrated. I can not take a nap during the day, because the baby sleeps very lightly.". Human response is "Totally understand, same here. The baby has been crying himself out of bed, and has to be held, my back is sour, but I have to hang in there." Chatbot replies "Same here". 

Sixth example: the post wrote "My baby feels uncomfortable, he suddenly cries with a scratchy throat." Human response is "My baby also cries with a scratchy throat, is your baby recovering, what did you do, she doesn't like drinking water." Chatbot replies "Take him to the hospital." }
\end{figure}

In Fig.~\ref{fig:dialogue}, we illustrate 6 pairs of post-response (originally all in Chinese). The human-response column contains the ground truth data that real human users commented under a post; and the AI-response column contains example responses generated by our text generation module. As illustrated, these generated responses are  vivid and diverse, and some of them even contain emojis.

\subsection{Evaluation}
In this subsection, we present offline evaluation of CASS performance. Offline evaluation refers to the evaluation that is done before deploying the system to let people test in the real world context. There are two types of evaluation approaches: automated performance evaluation and human evaluation. In this study, we adopt both of them to conduct an offline evaluation of our chatbot system. 

Automated performance evaluation is an established mechanism to automatically and quickly evaluate the performance of an AI system in the field of machine learning and AI. It often first reserves a subset of the training data as holdout data, whose both input (i.e., post) and output (i.e., human response) are known. Then, we feed input into the train model which automatically generates a predicted output. We can automatically compare the predicted output with the known human response output to evaluate the model performance.

The human evaluation is a well-known practice to both HCI and machine learning communities. Often, we can define a variate of dimensions and ask human users to rate the AI predicted outputs. Depending on the task, the human graders/coders are not necessarily having to be the actual intended users as long as these graders are capable of completing the task. Thus, many such human evaluations are done by crowd-workers~\cite{xu2017new}.

\subsubsection{Automated Performance Evaluation}
In NLP research, there are many metrics to describe the performance of a machine learning system, with each metric is suitable for different tasks. In our system, the final output is a predicted/generated response based on an input (i.e., a post). This is a Natural Language Generation task which can be evaluated by a widely used metric called \textit{BLEU} score (Bilingual Evaluation Understudy)~\cite{Papineni2002BLEU}. BLEU score represents the similarity between CASS generated response and the human ground truth in the training dataset with a n-gram match. It is a number between 0 and 1; 1 indicates that the generated response is exactly the same as the ground truth. We randomly selected 2,000 post-response pairs and reserved them as holdout data before training. The BLEU score on this holdout dataset was 0.23, which is a fairly acceptable performance score~\cite{DeepProbe,xu2017new}. \revision{As a comparison, the BLEU score of IR-based baseline model is 0.03.}

\subsubsection{Human Evaluation}
BLEU score simply describes the lexical differences of two sentences. It, however, does not reflect the grammar correctness or the semantic meaning. It is possible that CASS generates a much shorter sentence with the same meaning to the original human response, but the BLEU score is low due to the length difference. For example, in the second row in Fig.~\ref{fig:dialogue}, AI-generated response is a valid and even better response to the post, but the BLEU score is low because it is quite different from the human-response. Thus, we recruit human coders to evaluate the generated results. Based on previous literature~\cite{althoff2016large}, and for the research interest of this study, we introduce four dimensions for evaluating AI-generated responses: 
\begin{itemize}
    \item Grammar Correctness. The human coders are asked to evaluate the generated response's grammar correctness with a 5-point Likert-scale ranging from strongly disagree (-2) to strongly agree (+2) that the grammar is correct~\cite{Ritter2011Data}.
    \item Relevance. The response may not be relevant to the given input post so we decided to evaluate the relevance of AI-generated responses. It is also done using a 5-point Likert-scale ranging from strongly disagree (-2) to strongly agree (+2) that the response is relevant to the post's topic ~\cite{Ritter2011Data}.
    \item Willing-to-Reply. A response's content may be correct and relevant, but users in an online community may not feel engaged so they tend to ignore the AI-generated response. In this question, we ask human coders to rate how likely they will comment on the automatically generated response. Again, a 5-point Likert-scale is used with -2 indicating strongly unlikely whereas +2 representing strongly likely. 
    \item Emotional Support. This is a dimension designed specifically for our research context. We are interested in examining the extent to which the human coders perceive that the AI-generated response provides the desired emotional support. Similar to the above dimensions, a 5-point Likert-scale is used.
\end{itemize}

\begin{figure}
\centering
 \includegraphics[width=1.0\columnwidth]{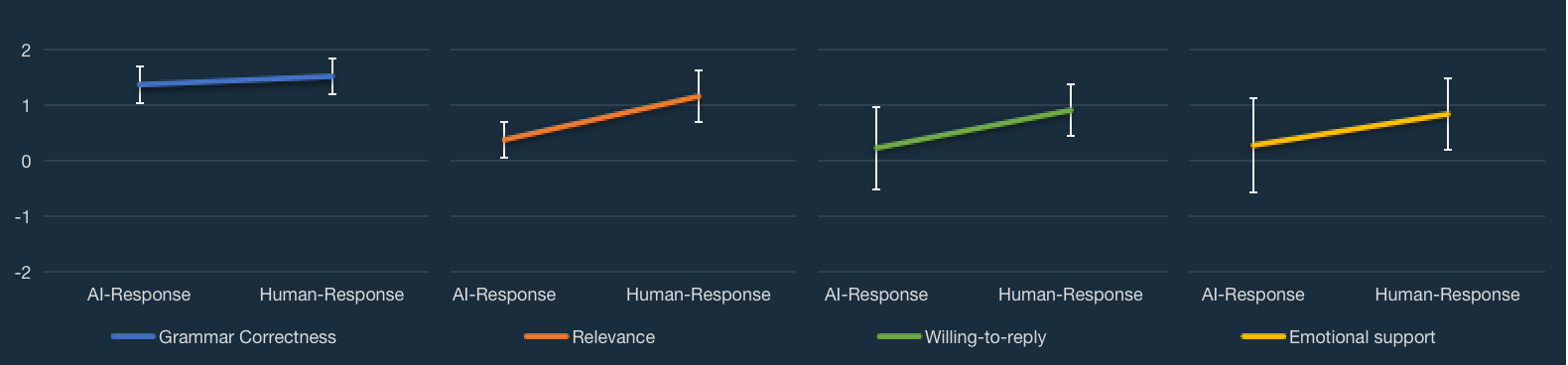}
 \caption{Comparison of AI-generated responses' and human ground-truth responses' quality with 5 human coders on 4 dimensions. We also performed t-test and post-hoc analyses. Grammar correctness: 1.37 vs 1.52, t(199)=-4.393, {$p$} > 0.1; Relevance: 0.38 vs 1.16, t(199)=-12.062, {$p$}<0.01; Willing-to-reply: 0.23 vs 0.91, t(199)=-10.657, {$p$}<0.01; Emotional support: 0.28 vs 0.34, t(199)=-7.186, {$p$}<0.01.}~\label{fig:evaluation}
 \Description{The picture is divided into four parts according to four dimensions. The y-axis is from -2 to 2, which represents the score of likert scale. Each part has AI response and human response corresponding to two points, and there is a line between them. Their scores are described in detail in caption}
\end{figure}

We randomly select 200 post-response pairs from the holdout dataset (the same dataset used in machine evaluation). Each post has both a human response that we collected from the community as a ground truth, and a AI-generated response. Thus, we end up with a total of 400 pairs of post-response. We ask human coders to evaluate both the AI-generated responses and human ground-truth for comparison. We recruit five human coders to rate each of the 400 pairs; each coder provide 4 scores related to grammar correctness, relevance, willing-to-reply, and emotional support dimensions. Intra-class correlations ICC(3,1) for the four dimensions range from 0.74 to 0.99, which indicates a good inter-coder consistency. For each response, it has four dimension scores, and we use the average from the five coder's scores. Then we perform a t-test to compare the AI-generated response and human ground-truth's quality in each of the 4 dimensions (as shown in AI-Response and Human-Response groups in Fig.\ref{fig:evaluation}). 

Fig.\ref{fig:evaluation} shows that AI-response's grammar correctness is rated  high and not that different from human-response (1.37 vs 1.52, t(199)=-4.393, {$p$} > 0.1). However, AI-response's quality is not as good as human-response in topical relevance, willing-to-reply, and emotional support, suggesting that the chatbot should be improved on those dimensions. But building a system to outperform human in writing responses is never the goal of this study. Despite the difference, all the evaluation scores are above 0, meaning that the 5 human coders agree that the AI-generated responses are grammatically correct, of high relevance to the topic, engaging enough for users to reply, and providing a considerate level of emotional support to the poster.

The results of the automated performance evaluation and human evaluation confirm that our chatbot system has a pretty good performance in providing emotional support. Therefore, we deploy the CASS back to the research site to provide timely responses and emotional supports to overlooked posts. We will describe our field deployment and experiment in the next section.

\section{Study 3: Deploying CASS to the study site and evaluating its impact on support seekers and other community members}
In the previous two sections, we presented an exploratory study on the YouBaoBao community to build an contextual understanding of the study site, from which, we identified what types of posts the community members post, and highlighted the challenge that some of their emotional support seeking posts can not get a timely response (Section~\ref{study1-result}). Then, we presented CASS system with a fully automated end-to-end architecture that can read in posts, identify overlooked posts, classify emotional support seeking posts, generate responses for those overlooked posts, and publish the generated response back to the community (Fig.~\ref{fig:work_flow}). Our preliminary evaluations showed that the NN-based models have a satisfying performance score, so the CASS system is ready for deployment. In this section, we present how we deployed the CASS system back to the community following a field experiment research setup, and how we evaluated its impacts on individual members and the community. 
% emotional support to improve community health ~\cite{mcgrath1991time}. 
% This last study concludes our research cycle started with learning knowledge from the community, then building systems with data provided by the community, and finally deploying systems to provide support \textbf{for the community}. 

\subsection{Method}
\textit{Field deployment} is a well-established HCI research practice~\cite{siek2014field}. It can reveal users ``naturalistic usage'' of the introduced new functionality of the system. We follow ~\cite{siek2014field} guideline on how to define the beginning and the end of the field deployment, and its ethical considerations of engaging users. 

Moreover, to quantify the support that CASS brings to individual members as well as to the community, we also adopted an experiment study setup~\cite{gergle2014experimental}. We design a ``control group'' and an ``experiment group'' to measure the difference of its impact between users exposed to it and the ones that are not exposed to it. An experiment design in a real-world deployment has its unique challenges. For example, to ensure the users having the most naturalistic behavior while interacting with the intervention, they should not know they are part of an experiment, when such disguise is not harmful to the users. We follow the \textit{field experiment} research method described in the textbook by Terveen et al. ~\cite{terveen2014study} to ``maximize realism of the context, while still affording some experimental control''.

\subsubsection{Field Experiment Design}
Our research goal is to deploy CASS to provide timely social support for the individual members who seek support. In addition, we are also interested in evaluating its impact on other community members. We deployed the system in August 2019, and conducted the field experiment for 7 days. The reason for deploying the system for 7 days is that the lifespan of a post thread on average is 7 days (as reported in Study 1, Section~\ref{study1-result}). After we deactivated the CASS system, we still kept tracking of the comments and user actives for another 7 days to ensure we gathered a 7-day data for each of the post.

\revision{To evaluate whether CASS helps individual support seekers, our unit of analysis is each individual user who published the original post. We define measurements to capture their online behaviors, and their emotion changes after the post was responded by CASS. To evaluate CASS's impact on other members of the community, the unit of analysis is changed to each individual post and its comments as a thread. We also define measurements (e.g., how many members participated under a post thread) to reflect other community members' behaviors caused by the CASS intervention.}

As shown in Study 1 in Section~\ref{study1-result}, the median \textit{interval time} for a post to get its first response is 10 minutes; and in Study 2 in Section~\ref{study2logic}, we define a post as \textit{overlooked post} if it does not get a human response within 10 minutes. Thus, during the deployment, we design CASS to only track the overlooked non-informational support seeking posts (Section~\ref{study2-classify}). In total, CASS tracks 3,445 overlooked emotional-support-seeking posts during the 7-day field experiment. 

These overlooked posts are then randomly split into two groups: a \textbf{baseline condition}, where we only passively track the original poster's and the other members' activities without CASS intervention; and an \textbf{experiment condition}, where CASS responds to the overlooked post once 10 minutes has passed, and we track activities and measure the poster interactions. \revision{During the 7-day field experiment, 1,717 overlooked posts are assigned to the baseline condition, and another 1,728 overlooked posts are in the experiment condition.}

% there is nearly no activities 7 days after a post is initially published. 
% Thus,in our study, we used 7 days as a threshold to collect all the comments from both the original poster and other community commenters. This gave every post a fair comparison point.

\subsubsection{Disguise the AI Identity}
As instructed in the textbook~\cite{gergle2014experimental}, when possible, participants should not be aware that they are in a field experiment setting. In addition, a number of recent literature~\cite{luo2019machines,Recommend_comment_Morris2018TowardsAA,Jakesch:2019:ACP:3290605.3300469} has highlighted that users' behaviors and perceptions may change if they know or believe they are interacting with an AI system, which will further compromise the experiment results. Therefore, during this experimental study, we disguised CASS as a normal community member with a pseudo user name and a user avatar. We disclosed the chatbot's real identity to all the users who had interacted with it via the community built-in private messages after the study was completed.  This is a common practice in psychology experiments ~\cite{gergle2014experimental} and clinical trial experiments~\cite{dersimonian1986meta}.

\subsubsection{Measurements} \label{study3-measure}
\revision{In this section, we present the measurements to quantify the effectiveness of our chatbot system in providing social support. In McGrath's seminal work, \textit{``Time, Interaction, and Performance''}~\cite{mcgrath1991time}, he proposes a three dimentional framework to describe teamwork: \textit{production, member support, and group well-being}. Grudin argues that a successful CSCW system should be able to provide productivity benefits for not only the individual members, but also for the group dynamics~\cite{grudin20081}. Inspired by these literature, we thus define a set of measurements to quantify CASS's impact on other community members.}
% Community is a large size group, thus we expect CASS can support the YouBaoBao for all the three functions. Below we specify measurements that were used to evaluate the impacts of CASS, which were grouped into three categories --- utilitarian benefit, individual member support, and community well-being. 

The three measurements for determining the CASS's \textit{effectiveness} in generating timely social support include:

\begin{itemize}
    \item \textit{The count of posts with no response.} This variable describes how many posts have no response at all in 7 days (in short \textbf{Post-NoResp-N}). If AI-Response group has a significant smaller number in this measurement, that suggests CASS is effective in addressing no-response posts. When calculating this score, the CASS's responses are not counted, as the chatbot replies to every post in the AI-Response group. 
    
    \item \textit{The time interval between the original post and the 1st response, and the time interval between the 1st response and the 2nd response.} Here we calculate two variables, both time intervals in minutes, to describe how fast a post gets replied (in short \textbf{Post-1Resp-Time}), and how fast the 1st response could attract the 2nd response (in short \textbf{1Resp-2Resp-Time}). If AI-Response group has a smaller time interval measurement, that suggests CASS indeed supports the poster to get a response in a more timely manner.
    
    %the result looks confusing, hide it for now, let's see if it's enough result
    % \item \textit{The time interval between the 1st response and the original poster came back and publish a comment to the 1st response.} Despite that a post may be responded quickly, if the original poster does not find it useful or does not want to interact with this response, her emotional support seeking desire is not completed. This variable (in short \textbf{1Resp-PComm-Time}) is a proxy that suggests the perceived quality of the 1st response to the original poster. The shorter this interval is, the quicker the poster comments on the 1st response, the better quality the 1st response has.
\end{itemize}

Then, we use a set of four measurements to reflect whether the CASS-generated post helps the original support-seeking \textit{individual}. 

\begin{itemize}
    \item \textit{The count of follow-up comments from the original poster.} After the original poster publishes the post and some other users come into the thread and reply to the posts, will the original poster come back to publish follow-up comments and form an active discussion? If so, we count the number of comments from the original poster, in short as \textbf{Poster-Comm-N}. The higher this number is, the more active the poster is. Thus, the support-seeking poster found an emotional outlet under her post and leveraged it to engage more with other community members. If the poster never came back, we count it as 0. 
    
    % result marginal significant, ignore for now
    % \item \textit{The count of follow-up comments specific to the 1st response from the poster herself.} Similar to the previous variable, this measurement looks at a subset of the count: how many comments did the poster publish to directly reply to the first response from others or from CASS? This counts only if the poster's comment is a 2nd level response to the 1st response, exactly as shown in the example in Fig.~\ref{fig:forum_post}, ``Don't overthink it, control yourself''. This measurement, in short \textbf{Poster-Comm-T1R-N}, reflects a quality of the 1st response published either by a human or by CASS in the two conditions. A higher number reflects the poster engages more directly with the human or with CASS in the two conditions. 
    
    \item \textit{The poster's original-emotional states, updated-emotional states, and the difference of these two states.} It is difficult to measure the posters' emotional states without directly asking them, thus we employ the \textit{theory of emotional valence}~\cite{lubis2015study}, and recruit two coders to annotate an original post's emotional valence score as a proxy for the original poster's initial emotional state (in short \textbf{Poster-Emo-Val-Orig}). More specifically, there are three states of emotional valence: \textit{1 for positive (happiness, contentment); 0 for neutral; and -1 for negative (anger, sadness)}. Then, if the original poster publishes a follow-up comment, we use this comment's emotional valence score as a proxy to represent her updated emotional state (in short \textbf{Poster-Emo-Val-Updt}). We also calculate the change of emotional states (in short \textbf{Poster-Emo-Val-Chng}). If an original poster never come back to post a follow-up comment, we exclude this data point. In summary, we have 759 data points in the experiment condition (N=1,728); and 536 data points in the baseline condition (N=1,717). We refer to these data points as AI-Response and Human-Response data points. Inter-rater reliability between the two coders on the emotional valence score of these 1319 data points reaches 0.88 (cohen's kappa), which indicates a high consistency.
\end{itemize}

Lastly, we define three measurements to reflect CASS's impact on other members in the community.
% as community contribution and community commitment are important aspects of a healthy online community~\cite{terveen2014study}.
\begin{itemize}
    \item \textit{The count of total responses to an original post.} This variable reflects the participation level under a post. A higher average number of this variable suggests the community has a higher community contribution level, and the community is healthier (in short \textbf{Post-Resp-N}).
    % \item \textit{The count of total responses to an original post.} This variable reflects the participation level under a post. A higher average number of this variable suggests the community has a higher community contribution level, and the community is healthier (in short \textbf{Post-Resp-N}). We exclude the responses from CASS and the original poster. 这个和上一个二选一
    \item \textit{The count of how many members participated and commented under a post thread.} This variable also reflects the participation level under a post. A higher average number of this variable suggests on average more community members engaged in each post's discussion. It reflects a high community commitment level, and the community is healthier (in short \textbf{Post-Member-N}).
    \item \textit{The time interval between each pair of adjacent responses under a post.} This variable (in short \textbf{Adj-Comm-Time}) represents on average how fast a post thread can get a new comment. If this number is smaller in the AI-Response group, it suggests that CASS not only helps the original poster to get a timely response, but also activates the liveliness of the community (e.g., other members also participate in a post's discussion in a more active fashion).
\end{itemize}

\subsection{Result} \label{study3result}
We organize the result section into three subsections to report that 1) \textit{CASS is effective in providing timely social support to overlooked posts}; 2) \textit{CASS-generated responses can improve the individual support seeker's emotional status}; and 3) \textit{CASS can also positively impact other community members' participation}. 

\subsubsection{CASS is effective in providing timely social support to overlooked posts}
The measurements in this category represent how CASS's functionalities effectively mitigate the primary challenge that a emotional-support-seeking post can not get a timely response. 

\textbf{Post-NoResp-N}: The results show that in the baseline condition, there were 595 out of 1,717 posts never received any response within 7 days of the post being published; in the experiment condition, out of 1,728 posts, this number was 433 (we exclude the response from the original poster, or from the CASS chatbot when calculating this number). A Chi-Square test suggests that the difference is significant ($\chi$(1)=39.65, p<0.01). That is, CASS can effectively generate timely social support messages as response to the support seeking posts and significantly reduce the number of overlooked posts.
% AI-response:433 no human reply, 298 chatbot reply and poster reply to them
% Human-response: 595 no human reply, 433 vs 595 $\chi$(1)=39.65, p<0.01
%与以前的数字不一致原因：以前将comment/reply混淆

\textbf{Post-1Resp-Time \& 1Resp-2Resp-Time}\label{study3result1}: These two variables are used for calculating time intervals between the original post and the 1st response, and the interval between the 1st response and the 2nd response. The results show that \textbf{Post-1Resp-Time} in the baseline condition is 254 minutes on average, and in the experiment condition is 10 minutes. It is not comparable because we simply set the CASS logic to identify and reply to posts in 10 minutes. Thus, a more comparable measure is \textbf{1Resp-2Resp-Time}, which reflects how long it takes for a second response to come in after the 1st response was published. In the baseline condition, the time interval between 1st response and 2nd response is on average 462 minutes, whereas in the experiment condition that number is 349 minutes. An independent sample t-test suggests the difference is significant (
t(1577)=-1.433, p<0.05). This result means that CASS not only posts a timely response to the original post, but also accelerates the time that a following response was posted by another user.
% AI:(349$\pm$1368)
% Human:(462$\pm$1725)
%1577 = 963 +616 -2

In summary, CASS's functionality can help effectively mitigate the primary problem, where some posts can not get a response in a timely manner, by reducing the number of posts with no response (smaller \textbf{Post-NoResp-N}), and speeding up the time that a post is responded (smaller \textbf{1Resp-2Resp-Time}).

%the result looks confusing, hide it for now, let's see if it's enough result
% \textbf{1Resp-PComm-Time}: time interval of poster come back after first reply(第二次数据四维对比.xlsx sheet1)
% AI:(283.79$\pm$1081.90)
% Human:(399.98$\pm$1096.39)
% t(1318)=-1.918,p<0.05
% %1318 = 758+562-2

% after publish:
% AI:(296.51$\pm$1081.84)
% Human:(656.26$\pm$1765.15)
% t(1318)= -4.572,p<0.01

\subsubsection{CASS-generated responses can improve the individual support seeker's emotional valence}
As introduced in section~\ref{study3-measure}, we defined and calculated four measurements to quantify the impact of the CDSS-generated response on an individual support seeker. 
% : \textit{the count of poster's follow-up comments} \textbf{(Poster-Comm-N)}, and \textit{the poster's emotional states and the state change} \textbf{(Poster-Emo-Val-Chng)}.

% IGNORE
% - poster reply first response count v.s (willing-to-reply quality of our AI-generated response) Chi-square 
% 这个可以不要(内容比较.xlsx chatbot sheet and nochatbot sheet)
% 759/1728  536/1122
% 0.44 vs 0.48 $\chi$(1)=4.064, p<0.05

\textbf{Poster-Comm-N}: The results show that original support seekers come back to their post thread and leave 1.36 follow-up comments in the baseline condition (N = 1,717). In contrast, the posters posted on average 1.78 follow-up comments in the experiment condition (N = 1,728). An independent t-test (t(3443) = 2.616, p<0.1) has a marginal significant effect and suggests that CASS indeed made the original support seeker to follow up with more comments, and interact more actively with other community members. These behaviors arguably are beneficial to release their stress. 
% - poster total reply times or response count (poster is open to social) independent sample t-test) (新指标对比.xlsx 重新算host reply times)要这个
% AI:(1.78$\pm$5.12)
% Human:(1.36$\pm$4.16)
% t(3443) = 2.616, p=0.095
% 3443=1728+1717-2

\begin{figure}
\centering
 \includegraphics[width=\columnwidth]{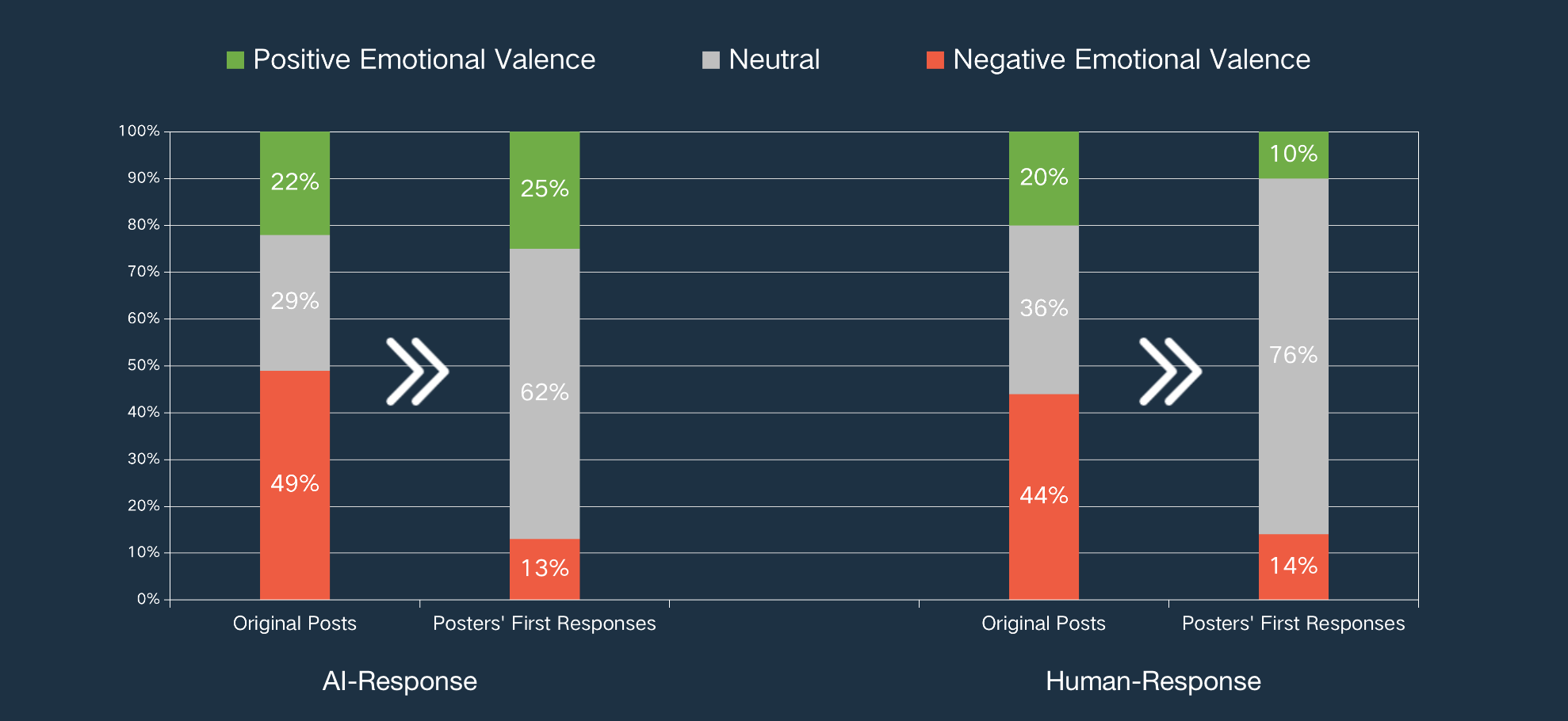}
 \caption{The poster's emotional valence and their changes, as in two proxy variables: the emotional valence score for her original post, and for her first comment to others' response under her own post. }~\label{fig:emotion}
 \Description[This is a bar-chart with four column.]{The y-axis is from 0\% to 100\%. The length of each column is 100\%. Each column contains three color components. Green represents positive emotional valence, gray represents neutral, and red represents negative emotional valence. For AI response, the color proportion of its original posts column is 22\% green, 29\% gray and 49\% red. the posters' first responses column corresponding to 25\% green, 62\% gray and 13\% red.  For human response, the color proportion of its original posts column is 20\% green, 36\% gray and 44\% red. the posters' first responses corresponding to 10\% green, 76\% gray and 14\% red.}
\end{figure}

\textbf{Poster-Emo-Val-Orig}: As shown in Fig.~\ref{fig:emotion}, the results show that for the experiment condition (N = 759), 22\% of the original posts are coded by human coders as having positive emotion valence score, 29\% as neutral, and 49\% have negative emotion valence score. In contrast, in the baseline condition (N = 536), the numbers become 20\%, 36\%, and 44\% for positive,  neutral, and negative emotion scores, respectively. As these post-response pairs are only the ones on which the original support seeker left a follow-up comment; thus, to differentiate them from the entire dataset, we refer to them as the AI-Response group (N = 759) and the Human-Response group (N = 536), respectively.

\textbf{Poster-Emo-Val-Udpt}: In the AI-Response group (N = 759), the original poster's 1st comment is the response she replied to the CASS-generated response. 25\% of these responses are coded as positive emotion, 62\% as neutral emotion, and 13\% as negative emotion. In the Human-Response group (N = 536), human coders rate 10\%, 76\%, and 14\% of the posts expressing positive, neutral, and negative emotion, respectively. 

These results are shown in Fig.~\ref{fig:emotion}. From the chart, we can observe that as for the emotional valence score for the original posts, there is no difference between the AI-Response group and the Human-Response group. But we can see there is a clear difference in the emotional valence score of the poster's 1st response. 

\textbf{Poster-Emo-Val-Chng}: It is interesting to see that the emotional valence score  in both groups changed from the support seeker's original post to her 1st response, in Fig.~\ref{fig:emotion}. It appears that in AI-Response group, many posters' original posts are negative, but then change to neutral (29\% to 62\%), and some neutral ones convert to positive (22\% to 25\%). Many community members' emotional valence become more positive, as shown in the example below:

%Here we list two examples to illustrate how the AI-generated response and Human-generated response influenced the poster's emotional states. The first example is to show that an AI-generated response leads the poster's emotional score to go up:
\begin{quote}
    Poster's original post: \textit{I may have insomnia. I still can't fall asleep and it's already 3:00am in the morning.}
    
    AI-generated response: \textit{Have a good sleep, and don't push yourself too hard.}
    
    Poster's 1st comment: \textit{Wow!!! I want to hug you!}
\end{quote}

%The second example illustrates that an AI-generated response leads the poster's emotional score to go down:

%\begin{quote}
    %Poster's original post: \textit{I hope I can give birth as soon as possible.}
    
    %AI-generated response: \textit{Take it easy. }
    
    %Poster's 1st comment: \textit{It is too difficult [cry][cry][cry].}
%\end{quote}

In contrast, in the Human-Response group, many posters' emotional valence score change from positive to neutral (36\% to 76\%). That suggests, some of the responses posted by another community member may adversely affect a certain number of posters' emotion.

To statistically analyze poster's emotional change, we further calculate the percentage of the posts that have an emotional valence score going up (e.g., neutral to positive), going down (e.g., positive to neutral), or remaining unchanged. In the AI-Response group, 48\% of the posts' scores go up, 38\% remain unchanged, and 14\% go down. In the Human-Response group, the percentages are 38\%, 44\%, and 18\%, respectively. We perform Chi-Square test to compare the percentages across the two groups. 

There are significantly more posters having an increased emotional valence score in the AI-Response group than in the Human-Response group ($\chi$(1)=14.35, p<0.01). Similarly, there are significantly less posters having decreased emotional state score in the AI-Response group than in the Human-Response group ({$\chi$}(1)=5.242, p<0.05). This suggests that the AI-generated response can motivate posters to become more positive, and prevent them from experiencing frustration or a similar negative emotion, when compared to the human-generated response.
% - emotion status and change (figure, 3 Chi-square)
% drop:\[\chi(1)=5.242, p=0.022\]     raise:\[\chi(1)=14.351, p=0.000\]
% nochange:\[\chi(1)=4.619, p=0.032\] 

In summary, our results show that in the AI-Response group, more original support seekers (i.e. posters) revisit and comment under their own posts to interact with other community members. In addition, more posters turn to a positive emotional valence, and less turn to negative emotion, when compared to the posters in the Human-Response group. In this sense, CASS can help individual members use their published post as an emotional outlet to chat with others, and may further improve their emotional states.

\subsubsection{CASS can also positively impact other community members' participation.}
In addition to CASS's utilitarian effectiveness and support for individual members, we are also interested in examining how it may improve the well-being of the community as a whole.

% exclude human and AI group 0 reply:
% 7.63 (34.25) vs 6.3 (35.14)  t(2115)=0.879, p = 0.228 
% AI v.s. Human, no chatbot, no host, N=997, 1120
\textbf{Post-Resp-N}: The results show that in the experiment condition (N = 1,728), there are 7.63 (S.D.=34.25) responses (N = 997) for each post; in contrast, in the baseline condition (N = 1,717), each post has an average of 6.3 responses (N = 1120, S.D. = 35.14). We have excluded the responses published by CASS or by the original poster, so this variable is a pure indicator of other community member's participation. This result suggests that other community members in the experiment condition  have a higher participation level, but an independent sample t-test suggests such difference is not significant (t(2115)=0.879, p = 0.228). 

\textbf{Post-Member-N}: For each post, there are 5 other community members (S.D.=20.97) participated in the thread (N = 997) in the experiment condition. In contrast, about 4 members (S.D.=23.35) participated in each post thread in the baseline condition (N = 1,120). This result suggests that each post in the experiment condition has more community members participated, and it has a higher level of community commitment, when compared to the baseline condition. However, an independent sample t-test suggests such difference is not significant (t(2115)=0.911, p = 0.232). 

\begin{figure}
\centering
 \includegraphics[width=\columnwidth]{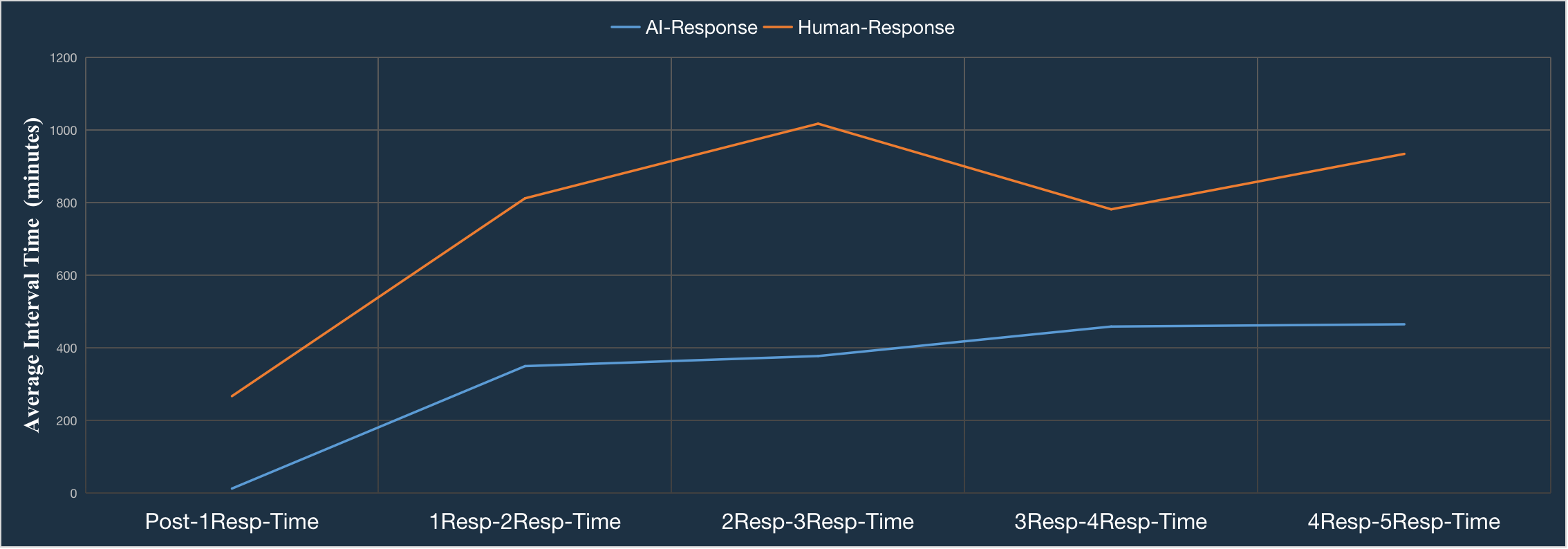}
 \caption{Time intervals for each pair of adjacent responses under a post. In this chart, we only show the first 5 pairs, but this trend persists for the rest pairs. This result suggests the community is more active in the AI-Response group than in the Human-Response group. }~\label{fig:compare}
 \Description{Line graph showing average interval time from 0 to 1,200 minutes in increment of 200 on the Y axis against each pair of adjacent responses under a post on the X axis. Two lines are shown. The line of human response is higher than the line of AI response in each pair.}
\end{figure}

\textbf{Adj-Comm-Time}: We also calculate the time intervals between each pair of adjacent responses in a post thread. In section~\ref{study3result1}, we have reported \textbf{Post-1Resp-Time} is 10 minutes in the experiment condition, and 254 minutes in the the baseline condition; \textbf{1Resp-2Resp-Time} is 349 minutes and 462 minutes, respectively. In addition, we continue to calculate the time intervals up to five follow-up responses. As Fig.~\ref{fig:compare} illustrated, there is a trend that the AI-Response group's time intervals are always shorter than the ones in the Human-Response group. This suggests that CASS not only helps the original poster to get a quicker response, but also energizes other members in the community, and accelerates their participation in a post's discussion.

In summary, we find that in the experiment condition, community members participate more actively in a post thread. Results also suggest that with the presence of our chatbot system, each post has a higher level of community participation, with both more people and more responses, even though difference is not significant.

% 3. Community Well-Being
% - reply per post v.s.  (independent sample t-test)
% 7.18 vs 5.4 p=0.4

% exclude human 0 reply
% 7.18 vs 8.3 p = 0.255

% exclude AI group chatbot and host human
% 8.25 vs 8.38 p=0.874 N = 1295 1122
% % - others' reply per post (no host/no chatbot) v.s.  (independent sample t-test)跟以前的数不一样，以前的分母把 0 reply的去掉了(新指标对比.xlsx 算其他成员的回复数量sheet)
% %这个和上一个二选一
% no host/no chat:
% AI:(4.4$\pm$26.29)
% Human:(4.11$\pm$28.54)
% t(3443)=0.312,p=0.455
% %3443= 1728 + 1717 -2

% exclude human 0 reply:
% 4.4 vs 6.3 p=0.306

% exclude human and AI group 0 reply:
% 7.63 (34.25) vs 6.3 (35.14)  t(2115)=0.879, p = 0.228 
% AI v.s. Human, no chatbot, no host, N=997, 1120

% - unique participants per post v.s.  (independent sample t-test)（新指标对比.xlsx unique user sheet)(all users in replies,if poster did not reply to anyone,do not include poster.)
% not include chatbot:
% AI:(3.86$\pm$16.1)
% Human:(3.22$\pm$19.98)
% t(3443)=1.029(2.647 if include chatbot),p=0.179
% %1728 1717

% exclude two group 0 user
% AI:5.14(18.40) N=1299
% Human:5.03(24.79) N=1100
% t(2397) = 0.117,p=0.488

% - five reply time interval plot (figure) (对比2.xlsx comment faster sheet)

\section{Discussion}

\revision{\subsection{How does Our Work Differ from Existing Literature on Online Healthcare Communities?} }
% Although Gui et el.~\cite{Pregnant_community_Gui:2017:ISS:3171581.3134685} have conducted 

\revision{We start our exploration with a detailed content analysis of the posts on the pregnancy forum.
% , we believe it is necessary for us to explore the characteristics of our research site.
% Results from study 1 show that categories in this PregForm are similar to them. 
Despite the granularity difference between these two coding schema, our category and concept axis echo previous literature~\cite{Pregnant_community_Gui:2017:ISS:3171581.3134685}. 
In ~\cite{Pregnant_community_Gui:2017:ISS:3171581.3134685}, the researchers propose 6 categories from coding 683 online posts; in contrast, we identify 3 top-level and 11 detailed-level concepts after coding 2,300 posts. Our ``informational-support seeking'' category is quite similar to a combination of their 3 categories ``advice'', ``formal knowledge'', and ``informal knowledge seeking''. Our ``emotional support seeking and sharing'' is similar to a combination of their ``emotional support'' and ``reassurance''. }

We find a novel post category that refers to community members sharing their own daily life, such as posting a photo of food or exercise in a gym. This indicates that community members gain support and contribute to the community at the same time. For our research purpose, we do not attempt to distinguish the emotional support seeking category and the sharing daily life category, because the responses to these two categories can be overlapped with compliment or empathy and our text generation module can handle them the same way, as long as they are not seeking factorial information. 

Study 1 suggests approximately 18\% of the posts did not receive any response, signifying an urgent need to address ``post-with-no-response'' issue. Community members engagement is a key factor influencing the response amount. Maintaining the active engagement of members is a key challenge for online communities. To address this challenge, some communities employed incentives schemes to motivate member participation, such as gamification ~\cite{gamification_stackoverflow} or monetary reward~\cite{Wu2017Money}. While they are effective to some extend, such solutions come with a relatively high implementation and maintenance cost~\cite{ren2012encouraging}. With the advances in AI and chatbot technology, some platforms have started experimenting with the use of chatbot service to more closely connect members, for instance, by recommending members to reach out to others~\cite{Peng2019github} or by suggesting further actions based on participation status~\cite{grembot_peng}. 

Our work takes a different route --- builing a chatbot to act as an active member to boost the community engagement level. This result sounds similar to the catfish effect\footnote{https://en.wikipedia.org/wiki/Catfish\_effect} --- an active catfish in a group of sardines in a fish tank can stimulus the sardines to be more active and live longer during the transition. CASS is an ``active member'', which generates diverse responses to reply to other members' posts. The existence of CASS energizes other members to post more responses to the support-seeking post. We speculate this is one plausible explanation to interpret our results in Study 3. 
% that CASS can attract other members to engage in the thread discussion.

\revision{\subsection{The Neural-Network-based Chatbot Architecture is a Promising Solution for Building CSCW Systems to Support Online Communities.}}
To develop a successful AI system to provide social support for an online community, a chatbot needs to meet high creterias with respect to its post classification and response generation capabilities. From the original support seeker's perspective, they need authentic and helpful responses in a timely manner, otherwise it may even negatively affect their mood. From the other community member's perspective, if a post's first response has a high quality and high relevancy to the post, they may follow the lead and be more likely to provide their high quality comments to the post thread. 

Therefore, CASS needs to be able to generate authentic, diverse, and high-quality responses. However, existing rule-based and IR-based chatbots cannot sufficiently meet such needs~\cite{peng2019survey}. NN-based chatbot can be a promising solution in this case, but it may be affected by the noisy-label issue in training data. We avoid this potential risk with three pre-processing methods. First, our research site is a friendly community. When we clean up the data, we only see some advertisement (e,g., which brand of milk powder is better for the baby) and do not find any offensive or hatred language. Second, we build a CNN-based classification model to filter out informational-seeking posts, as CASS is not designed to generate responses that are relevant to medical information. Thirdly, we design a user interface console with human-in-the-loop AI design, which can be used by a human experimenter to track and monitor each AI-generated response in real-time. From data collection to response generation and publication, we employ rigorous methods to avoid the potential risk of such NN-based chatbot architecture.

However, an online community is fluid and its norms and topics may evolve along with time~\cite{wang2003building}. AI-generated responses rely on the style and topics in its training dataset. Thus, the NN-based chatbot may need to be re-trained with the latest community data to keep up-to-date with its popular memes and language norms after a period of deployment. \revision{This process may sound labor-intensive, but it is much easier than re-training and refining a rule-based chatbot system.}

\subsection{Human Behavior May Change in a Human-AI Co-Existing Online Community in the Future.}

Previous work that studied the human interaction with chatbots often focused only on the impact of a chatbot on an individual user. Little is known about how chatbot system may affect a group or a community. Our results provide evidence that the NN-based AI system can provide timely social support to individual members, as well as boost up other community members' engagement level. 

Existing literature suggests that chatbots could play a significant role in boosting the popularity of post content~\cite{TwitterBotGilani:2017:BIT:3041021.3054255, Peng2019github}. Seering et al. ~\cite{seering_village10.1145/3313831.3376708} also find the existence of the chatbot sparked both human-chatbot interactions and human-human interactions. \revision{The results of our Study 3 suggest that CASS can help individual members to use their published post as an emotional outlet and interact more with other community members, and ultimately improve their emotional states. These results extend prior work and provide valuable insights into the design of future chatbot systems for online community.}

In addition, we also find that CASS can promote other team members to engage more with a post. For example, results from study 3 show that other community members who did not interact with CASS directly have also been affected by CASS. That is, CASS made them become more engaged and more active in replying to those support-seeking posts. We speculate that our research site is a Chinese online community, thus Chinese culture~\cite{bond1996handbook} may contribute to this effect.  For example, people may not want to be the first person to express their opinion in a public discourse~\cite{bond1996handbook}. Being the first commenter seems to require more courage and a stronger desire to express. If the post has already been replied by someone (in this case by the charbot), then the psychological pressure to reply will be less. So when CASS replied to the post, other members may be more willing to participate in the discussion with lower pressure.
For HCI and AI researchers, this could be a new research direction that how different cultures interplay with the user behaviors when people and AI systems co-exist in an online community.

\revision{
More importantly, with more and more chatbots and AI-based moderation applications being built and adopted into various online communities, we argue that in the future it will be common for online communities to have a human-AI co-existence ecosystem. To support this new paradigm, we need to extend the current human-in-the-loop AI design philosophy \cite{cranshaw2017calendar} and the mixed-initiative design approach~\cite{horvitz1999principles}. An AI system with high accuracy is not enough. In addition, the AI system should also be designed to cooperate with the primary target users and the other users in the community. An \textbf{cooperative AI} system should seamlessly fit into the context and bring some benefits to the community (e.g., supports stressed community members and increases community engagement). And the users who interact with it should feel comfortable to have the AI system in a community, instead of feeling threatened or intimidated. This ``Human-AI Collaboration'' future~\cite{aidoctor} is the ultimate research goal that we are aiming for. 
}

\subsection{Limitation and Future Directions}
\revision{One limitation of our work is that the CASS system was designed to handle the emotional-support-seeking posts only, and neglected those informational support seeking posts. It is because the expected responses are different: for emotional support responses, users prefer diverse and lively content, and it is time-sensitive; whereas responses for informational support seeking posts require accurate and factorial answers, thus the diversity is not a critical concern~\cite{Pregnant_community_Gui:2017:ISS:3171581.3134685}. }That being said, NN-based approaches can also help with such task, but it can not be the same seq-to-seq architecture we used for emotional support. Open domain Q\&A ~\cite{wang2018r} is a promising NLP technique that can parse a whole textbook and automatically generate answers for any given question about the book. With this technique, chatbot can reply to users' informational support seeking questions by generating answers from medical textbooks, wikipedia or other authentic data sources. The CASS system can still follow the architecture to have 5 Steps and 3 Modules. Only inside the response generation module, CASS needs a hybrid architecture to use the corresponding NLP technique for the incoming posts to generate responses. Theoretically, it is easy to do. But it requires careful design considerations, solid data preparation, and rigorous model training and experimentation work. We will explore this research direction in our future work.

% We did not disclose CASS identity before experiment ended, because human perception may be biased towards an AI response. Many research ~\cite{luo2019machines,Recommend_comment_Morris2018TowardsAA,Jakesch:2019:ACP:3290605.3300469,Human-Bot_social_QA_Murgia:2016:MHI:2851581.2892311,Botivist:Savage:2016:BCV:2818048.2819985} suggests that people may have negative bias towards the response if they know that it is from a chatbot. We believe that the identity of chatbot will affect the results of the experiment, attracting too much attention~\cite{seering_village10.1145/3313831.3376708}
% or making the responses biased~\cite{Jakesch:2019:ACP:3290605.3300469}. Thus, we only disclosed the chatbot's real identity to all the users after the study was completed. If the future, we would like to replicate others' experiment in this community and examine the effect of whether to disclosure of chatbot's profile or not on community member's trust and interaction behaviors.

We did not disclose CASS identity before or during the experiment for two main reasons: Firstly, as suggested by previous literature~\cite{luo2019machines,Recommend_comment_Morris2018TowardsAA,Jakesch:2019:ACP:3290605.3300469,Human-Bot_social_QA_Murgia:2016:MHI:2851581.2892311,Botivist:Savage:2016:BCV:2818048.2819985}, we believe that the identity of chatbot will affect the results of the experiment, attracting too much attention~\cite{seering_village10.1145/3313831.3376708}
or making the responses biased~\cite{Jakesch:2019:ACP:3290605.3300469}. For the purpose of this study, we want to avoid that. \revision{Secondly, the potential impact of disclosing chatbot's identity on community users' behaviors was not the scope of this study. We are planning to thoroughly examine this topic in our future work. Whether disclosing the chatbot identity is related to an important design question: the appropriateness of languages for a chatbot. In our study, the appropriateness of the generated language is evaluated based on the accuracy and whether it answers the given input. However, we may anticipate that some of the words may be perceived appropriate if they are posted by pregnant human members but inappropriate if they are posted by an AI algorithm. For example, in the Fig~\ref{fig:dialogue} example, one pregnant woman complains about her baby causing her some sleeping issues, and our chatbot answered ``same here''. The answer is totally appropriate if it is from another human member of the community, but some people may perceive it weird as the AI is suggesting ``her baby'' also caused ``her'' similar troubles. Thus, there may need a novel dimension for characterizing the generated language's AI-appropriateness.

}

Another limitation of our chatbot is that in the field experiment, we limited our chatbot's engagement with the poster to only one round. It is not because our chatbot can not deal with multiple round conversations, rather, it is due to the semi-controlled experiment design consideration: if our chatbot can reply to poster for an arbitrary number of rounds, it introduces more confounding variables to the major behavioral measurements of interest. In the future, we can extend the CASS capability to have multi-round dialogue skills~\cite{nan2018state}, and engage with community members in a more natural manner.

\revision{Our work is under the assumption that for these support-seeking posts, the sooner they get answers, the better it is for the support seeker. However, based on Figure~\ref{fig:forum_post}, it is clear that most posts either quickly receive a response, or never receive a response. The introduction of the proposed chatbot may significantly change such community dynamic. Maybe this change will have some unexpected negative implications. Further study is required to fully understand its impact in a longer-term deployment.}

\revision{Our deployment in this particular healthcare community may be a limitation for generalizing the empirical findings, for example, those findings specific to the pregnant women community may not apply to other online communities, or the Chinese cultural context may hinder the generalizability to other cultures. However, we believe the scalable and generalizable chatbot architecture can be easily customized for other online community contexts. Thus, we welcome other researchers to join our effort, together we can replicate and extend this study design in other community contexts.
}

\section{conclusion}
\revision{In this paper, we present a comprehensive research project consisting of three studies, in which we developed and deployed a chatbot system to automatically generate and post responses to emotional support seeking posts in an online health community for pregnant women. Our studies show that the neural-network-based chatbot architecture is a promising solution to build chatbots to generate timely responses for posts with no replies. In addition, we present evidence regarding the chatbot's positive impact on influencing the support seeker's emotional status, and encouraging other community members to be more engaging.} 
% More importantly, we demonstrated an action research approach that the development of novel AI systems using data provided by a community can also provide benefits to individual community members as well as for the community as a whole.

%\begin{acks}
%xxx
%\end{acks}

\bibliographystyle{ACM-Reference-Format}
\bibliography{main}


\begin{thebibliography}{134}

%%% the Web address.

\ifx \showCODEN    \undefined \def \showCODEN     #1{\unskip}     \fi
\ifx \showDOI      \undefined \def \showDOI       #1{#1}\fi
\ifx \showISBNx    \undefined \def \showISBNx     #1{\unskip}     \fi
\ifx \showISBNxiii \undefined \def \showISBNxiii  #1{\unskip}     \fi
\ifx \showISSN     \undefined \def \showISSN      #1{\unskip}     \fi
\ifx \showLCCN     \undefined \def \showLCCN      #1{\unskip}     \fi
\ifx \shownote     \undefined \def \shownote      #1{#1}          \fi
\ifx \showarticletitle \undefined \def \showarticletitle #1{#1}   \fi
\ifx \showURL      \undefined \def \showURL       {\relax}        \fi
% invisible to TeX
\providecommand\bibfield[2]{#2}
\providecommand\bibinfo[2]{#2}
\providecommand\natexlab[1]{#1}
\providecommand\showeprint[2][]{arXiv:#2}

\bibitem[\protect\citeauthoryear{Abadi, Barham, Chen, Chen, Davis, Dean, Devin,
  Ghemawat, Irving, Isard, Kudlur, Levenberg, Monga, Moore, Murray, Steiner,
  Tucker, Vasudevan, Warden, Wicke, Yu, and Zheng}{Abadi et~al\mbox{.}}{2016}]%
        {tensorflow}
\bibfield{author}{\bibinfo{person}{Mart{\'\i}n Abadi}, \bibinfo{person}{Paul
  Barham}, \bibinfo{person}{Jianmin Chen}, \bibinfo{person}{Zhifeng Chen},
  \bibinfo{person}{Andy Davis}, \bibinfo{person}{Jeffrey Dean},
  \bibinfo{person}{Matthieu Devin}, \bibinfo{person}{Sanjay Ghemawat},
  \bibinfo{person}{Geoffrey Irving}, \bibinfo{person}{Michael Isard},
  \bibinfo{person}{Manjunath Kudlur}, \bibinfo{person}{Josh Levenberg},
  \bibinfo{person}{Rajat Monga}, \bibinfo{person}{Sherry Moore},
  \bibinfo{person}{Derek~G. Murray}, \bibinfo{person}{Benoit Steiner},
  \bibinfo{person}{Paul Tucker}, \bibinfo{person}{Vijay Vasudevan},
  \bibinfo{person}{Pete Warden}, \bibinfo{person}{Martin Wicke},
  \bibinfo{person}{Yuan Yu}, {and} \bibinfo{person}{Xiaoqiang Zheng}.}
  \bibinfo{year}{2016}\natexlab{}.
\newblock \showarticletitle{TensorFlow: A System for Large-Scale Machine
  Learning}. In \bibinfo{booktitle}{\emph{12th {USENIX} Symposium on Operating
  Systems Design and Implementation ({OSDI} 16)}}. \bibinfo{publisher}{{USENIX}
  Association}, \bibinfo{address}{Savannah, GA}, \bibinfo{pages}{265--283}.
\newblock
\showISBNx{978-1-931971-33-1}
\urldef\tempurl%
\url{https://www.usenix.org/conference/osdi16/technical-sessions/presentation/abadi}
\showURL{%
\tempurl}


\bibitem[\protect\citeauthoryear{Abokhodair, Yoo, and McDonald}{Abokhodair
  et~al\mbox{.}}{2015}]%
        {dissecting10.1145/2675133.2675208}
\bibfield{author}{\bibinfo{person}{Norah Abokhodair}, \bibinfo{person}{Daisy
  Yoo}, {and} \bibinfo{person}{David~W. McDonald}.}
  \bibinfo{year}{2015}\natexlab{}.
\newblock \showarticletitle{Dissecting a Social Botnet: Growth, Content and
  Influence in Twitter}. In \bibinfo{booktitle}{\emph{Proceedings of the 18th
  ACM Conference on Computer Supported Cooperative Work \& Social Computing}}
  \emph{(\bibinfo{series}{CSCW ’15})}. \bibinfo{publisher}{Association for
  Computing Machinery}, \bibinfo{address}{New York, NY, USA},
  \bibinfo{pages}{839–851}.
\newblock
\showISBNx{9781450329224}
\urldef\tempurl%
\url{https://doi.org/10.1145/2675133.2675208}
\showDOI{\tempurl}


\bibitem[\protect\citeauthoryear{Ackerman and McDonald}{Ackerman and
  McDonald}{1996}]%
        {ackerman1996answer}
\bibfield{author}{\bibinfo{person}{Mark~S Ackerman} {and}
  \bibinfo{person}{David~W McDonald}.} \bibinfo{year}{1996}\natexlab{}.
\newblock \showarticletitle{Answer Garden 2: merging organizational memory with
  collaborative help}. In \bibinfo{booktitle}{\emph{Proceedings of the 1996 ACM
  conference on Computer supported cooperative work}}.
  \bibinfo{pages}{97--105}.
\newblock


\bibitem[\protect\citeauthoryear{Almeida, Comber, and Balaam}{Almeida
  et~al\mbox{.}}{2016}]%
        {hci_and_intimate10.1145/2858036.2858187}
\bibfield{author}{\bibinfo{person}{Teresa Almeida}, \bibinfo{person}{Rob
  Comber}, {and} \bibinfo{person}{Madeline Balaam}.}
  \bibinfo{year}{2016}\natexlab{}.
\newblock \showarticletitle{HCI and Intimate Care as an Agenda for Change in
  Women’s Health}. In \bibinfo{booktitle}{\emph{Proceedings of the 2016 CHI
  Conference on Human Factors in Computing Systems}}
  \emph{(\bibinfo{series}{CHI ’16})}. \bibinfo{publisher}{Association for
  Computing Machinery}, \bibinfo{address}{New York, NY, USA},
  \bibinfo{pages}{2599–2611}.
\newblock
\showISBNx{9781450333627}
\urldef\tempurl%
\url{https://doi.org/10.1145/2858036.2858187}
\showDOI{\tempurl}


\bibitem[\protect\citeauthoryear{Althoff, Clark, and Leskovec}{Althoff
  et~al\mbox{.}}{2016}]%
        {althoff2016large}
\bibfield{author}{\bibinfo{person}{Tim Althoff}, \bibinfo{person}{Kevin Clark},
  {and} \bibinfo{person}{Jure Leskovec}.} \bibinfo{year}{2016}\natexlab{}.
\newblock \showarticletitle{Large-scale analysis of counseling conversations:
  An application of natural language processing to mental health}.
\newblock \bibinfo{journal}{\emph{Transactions of the Association for
  Computational Linguistics}}  \bibinfo{volume}{4} (\bibinfo{year}{2016}),
  \bibinfo{pages}{463--476}.
\newblock


\bibitem[\protect\citeauthoryear{Amershi, Weld, Vorvoreanu, Fourney, Nushi,
  Collisson, Suh, Iqbal, Bennett, Inkpen, et~al\mbox{.}}{Amershi
  et~al\mbox{.}}{2019}]%
        {amershi2019guidelines}
\bibfield{author}{\bibinfo{person}{Saleema Amershi}, \bibinfo{person}{Dan
  Weld}, \bibinfo{person}{Mihaela Vorvoreanu}, \bibinfo{person}{Adam Fourney},
  \bibinfo{person}{Besmira Nushi}, \bibinfo{person}{Penny Collisson},
  \bibinfo{person}{Jina Suh}, \bibinfo{person}{Shamsi Iqbal},
  \bibinfo{person}{Paul~N Bennett}, \bibinfo{person}{Kori Inkpen},
  {et~al\mbox{.}}} \bibinfo{year}{2019}\natexlab{}.
\newblock \showarticletitle{Guidelines for human-AI interaction}. In
  \bibinfo{booktitle}{\emph{Proceedings of the 2019 chi conference on human
  factors in computing systems}}. \bibinfo{pages}{1--13}.
\newblock


\bibitem[\protect\citeauthoryear{Ammari, Schoenebeck, and Romero}{Ammari
  et~al\mbox{.}}{2018}]%
        {Pseudonymous_10.1145/3173574.3174063}
\bibfield{author}{\bibinfo{person}{Tawfiq Ammari}, \bibinfo{person}{Sarita
  Schoenebeck}, {and} \bibinfo{person}{Daniel~M. Romero}.}
  \bibinfo{year}{2018}\natexlab{}.
\newblock \showarticletitle{Pseudonymous Parents: Comparing Parenting Roles and
  Identities on the Mommit and Daddit Subreddits}. In
  \bibinfo{booktitle}{\emph{Proceedings of the 2018 CHI Conference on Human
  Factors in Computing Systems}} \emph{(\bibinfo{series}{CHI ’18})}.
  \bibinfo{publisher}{Association for Computing Machinery},
  \bibinfo{address}{New York, NY, USA}, \bibinfo{pages}{1–13}.
\newblock
\showISBNx{9781450356206}
\urldef\tempurl%
\url{https://doi.org/10.1145/3173574.3174063}
\showDOI{\tempurl}


\bibitem[\protect\citeauthoryear{Arguello, Butler, Joyce, Kraut, Ling,
  Ros{\'e}, and Wang}{Arguello et~al\mbox{.}}{2006}]%
        {talk_to_me:Arguello:2006:TMF:1124772.1124916}
\bibfield{author}{\bibinfo{person}{Jaime Arguello}, \bibinfo{person}{Brian~S.
  Butler}, \bibinfo{person}{Elisabeth Joyce}, \bibinfo{person}{Robert Kraut},
  \bibinfo{person}{Kimberly~S. Ling}, \bibinfo{person}{Carolyn Ros{\'e}}, {and}
  \bibinfo{person}{Xiaoqing Wang}.} \bibinfo{year}{2006}\natexlab{}.
\newblock \showarticletitle{Talk to Me: Foundations for Successful
  Individual-group Interactions in Online Communities}. In
  \bibinfo{booktitle}{\emph{Proceedings of the SIGCHI Conference on Human
  Factors in Computing Systems}} \emph{(\bibinfo{series}{CHI '06})}.
  \bibinfo{publisher}{ACM}, \bibinfo{address}{New York, NY, USA},
  \bibinfo{pages}{959--968}.
\newblock
\showISBNx{1-59593-372-7}
\urldef\tempurl%
\url{https://doi.org/10.1145/1124772.1124916}
\showDOI{\tempurl}


\bibitem[\protect\citeauthoryear{Bakhshi, Shamma, and Gilbert}{Bakhshi
  et~al\mbox{.}}{2014}]%
        {face_attract_Bakhshi:2014:FEU:2556288.2557403}
\bibfield{author}{\bibinfo{person}{Saeideh Bakhshi}, \bibinfo{person}{David~A.
  Shamma}, {and} \bibinfo{person}{Eric Gilbert}.}
  \bibinfo{year}{2014}\natexlab{}.
\newblock \showarticletitle{Faces Engage Us: Photos with Faces Attract More
  Likes and Comments on Instagram}. In \bibinfo{booktitle}{\emph{Proceedings of
  the SIGCHI Conference on Human Factors in Computing Systems}}
  \emph{(\bibinfo{series}{CHI '14})}. \bibinfo{publisher}{ACM},
  \bibinfo{address}{New York, NY, USA}, \bibinfo{pages}{965--974}.
\newblock
\showISBNx{978-1-4503-2473-1}
\urldef\tempurl%
\url{https://doi.org/10.1145/2556288.2557403}
\showDOI{\tempurl}


\bibitem[\protect\citeauthoryear{Bambina}{Bambina}{2007}]%
        {Social_support_definition:Bambina:2007:OSS:1534606}
\bibfield{author}{\bibinfo{person}{Antonina~D. Bambina}.}
  \bibinfo{year}{2007}\natexlab{}.
\newblock \bibinfo{booktitle}{\emph{Online Social Support: The Interplay of
  Social Networks and Computer-Mediated Communication}}.
\newblock \bibinfo{publisher}{Cambria Press}.
\newblock
\showISBNx{1934043257, 9781934043257}


\bibitem[\protect\citeauthoryear{Banti, Mauri, Oppo, Borri, Rambelli,
  Ramacciotti, Montagnani, Camilleri, Cortopassi, Rucci, et~al\mbox{.}}{Banti
  et~al\mbox{.}}{2011}]%
        {Banti2011From}
\bibfield{author}{\bibinfo{person}{S Banti}, \bibinfo{person}{Mauro Mauri},
  \bibinfo{person}{A Oppo}, \bibinfo{person}{C Borri}, \bibinfo{person}{C
  Rambelli}, \bibinfo{person}{D Ramacciotti}, \bibinfo{person}{M~S Montagnani},
  \bibinfo{person}{V Camilleri}, \bibinfo{person}{S Cortopassi},
  \bibinfo{person}{Paola Rucci}, {et~al\mbox{.}}}
  \bibinfo{year}{2011}\natexlab{}.
\newblock \showarticletitle{From the third month of pregnancy to 1 year
  postpartum. Prevalence, incidence, recurrence, and new onset of depression.
  Results from the Perinatal Depression-Research \& Screening Unit study}.
\newblock \bibinfo{journal}{\emph{Comprehensive Psychiatry}}
  \bibinfo{volume}{52}, \bibinfo{number}{4} (\bibinfo{year}{2011}),
  \bibinfo{pages}{343--351}.
\newblock


\bibitem[\protect\citeauthoryear{Barry, Doherty, Marcano~Belisario, Car,
  Morrison, and Doherty}{Barry et~al\mbox{.}}{2017}]%
        {depression_pregnancy:Barry:2017:MMM:3025453.3025918}
\bibfield{author}{\bibinfo{person}{Marguerite Barry}, \bibinfo{person}{Kevin
  Doherty}, \bibinfo{person}{Jose Marcano~Belisario}, \bibinfo{person}{Josip
  Car}, \bibinfo{person}{Cecily Morrison}, {and} \bibinfo{person}{Gavin
  Doherty}.} \bibinfo{year}{2017}\natexlab{}.
\newblock \showarticletitle{mHealth for Maternal Mental Health: Everyday Wisdom
  in Ethical Design}. In \bibinfo{booktitle}{\emph{Proceedings of the 2017 CHI
  Conference on Human Factors in Computing Systems}}
  \emph{(\bibinfo{series}{CHI '17})}. \bibinfo{publisher}{ACM},
  \bibinfo{address}{New York, NY, USA}, \bibinfo{pages}{2708--2756}.
\newblock
\showISBNx{978-1-4503-4655-9}
\urldef\tempurl%
\url{https://doi.org/10.1145/3025453.3025918}
\showDOI{\tempurl}


\bibitem[\protect\citeauthoryear{Barua, Islam, Yao, and Murase}{Barua
  et~al\mbox{.}}{2012}]%
        {barua2012mwmote}
\bibfield{author}{\bibinfo{person}{Sukarna Barua}, \bibinfo{person}{Md~Monirul
  Islam}, \bibinfo{person}{Xin Yao}, {and} \bibinfo{person}{Kazuyuki Murase}.}
  \bibinfo{year}{2012}\natexlab{}.
\newblock \showarticletitle{MWMOTE--majority weighted minority oversampling
  technique for imbalanced data set learning}.
\newblock \bibinfo{journal}{\emph{IEEE Transactions on knowledge and data
  engineering}} \bibinfo{volume}{26}, \bibinfo{number}{2}
  (\bibinfo{year}{2012}), \bibinfo{pages}{405--425}.
\newblock


\bibitem[\protect\citeauthoryear{Bazarova, Choi, Schwanda~Sosik, Cosley, and
  Whitlock}{Bazarova et~al\mbox{.}}{2015}]%
        {social_sharing_10.1145/2675133.2675297}
\bibfield{author}{\bibinfo{person}{Natalya~N. Bazarova},
  \bibinfo{person}{Yoon~Hyung Choi}, \bibinfo{person}{Victoria Schwanda~Sosik},
  \bibinfo{person}{Dan Cosley}, {and} \bibinfo{person}{Janis Whitlock}.}
  \bibinfo{year}{2015}\natexlab{}.
\newblock \showarticletitle{Social Sharing of Emotions on Facebook: Channel
  Differences, Satisfaction, and Replies}. In
  \bibinfo{booktitle}{\emph{Proceedings of the 18th ACM Conference on Computer
  Supported Cooperative Work \& Social Computing}} \emph{(\bibinfo{series}{CSCW
  ’15})}. \bibinfo{publisher}{Association for Computing Machinery},
  \bibinfo{address}{New York, NY, USA}, \bibinfo{pages}{154–164}.
\newblock
\showISBNx{9781450329224}
\urldef\tempurl%
\url{https://doi.org/10.1145/2675133.2675297}
\showDOI{\tempurl}


\bibitem[\protect\citeauthoryear{Biyani, Caragea, Mitra, and Yen}{Biyani
  et~al\mbox{.}}{2014}]%
        {biyani2014identifying}
\bibfield{author}{\bibinfo{person}{Prakhar Biyani}, \bibinfo{person}{Cornelia
  Caragea}, \bibinfo{person}{Prasenjit Mitra}, {and} \bibinfo{person}{John
  Yen}.} \bibinfo{year}{2014}\natexlab{}.
\newblock \showarticletitle{Identifying emotional and informational support in
  online health communities}. In \bibinfo{booktitle}{\emph{Proceedings of
  COLING 2014, the 25th International Conference on Computational Linguistics:
  Technical Papers}}. \bibinfo{pages}{827--836}.
\newblock


\bibitem[\protect\citeauthoryear{Bond}{Bond}{1996}]%
        {bond1996handbook}
\bibfield{author}{\bibinfo{person}{Michael~Harris Bond}.}
  \bibinfo{year}{1996}\natexlab{}.
\newblock \bibinfo{booktitle}{\emph{The handbook of Chinese psychology}}.
\newblock \bibinfo{publisher}{Oxford University Press Hong Kong}.
\newblock


\bibitem[\protect\citeauthoryear{Bratteteig and Verne}{Bratteteig and
  Verne}{2018}]%
        {bratteteig2018does}
\bibfield{author}{\bibinfo{person}{Tone Bratteteig} {and} \bibinfo{person}{Guri
  Verne}.} \bibinfo{year}{2018}\natexlab{}.
\newblock \showarticletitle{Does AI make PD obsolete? exploring challenges from
  artificial intelligence to participatory design}. In
  \bibinfo{booktitle}{\emph{Proceedings of the 15th Participatory Design
  Conference: Short Papers, Situated Actions, Workshops and Tutorial-Volume
  2}}. \bibinfo{pages}{1--5}.
\newblock


\bibitem[\protect\citeauthoryear{Candello, Pinhanez, and Figueiredo}{Candello
  et~al\mbox{.}}{2017}]%
        {Typefaces10.1145/3025453.3025919}
\bibfield{author}{\bibinfo{person}{Heloisa Candello}, \bibinfo{person}{Claudio
  Pinhanez}, {and} \bibinfo{person}{Flavio Figueiredo}.}
  \bibinfo{year}{2017}\natexlab{}.
\newblock \showarticletitle{Typefaces and the Perception of Humanness in
  Natural Language Chatbots}. In \bibinfo{booktitle}{\emph{Proceedings of the
  2017 CHI Conference on Human Factors in Computing Systems}}
  \emph{(\bibinfo{series}{CHI ’17})}. \bibinfo{publisher}{Association for
  Computing Machinery}, \bibinfo{address}{New York, NY, USA},
  \bibinfo{pages}{3476–3487}.
\newblock
\showISBNx{9781450346559}
\urldef\tempurl%
\url{https://doi.org/10.1145/3025453.3025919}
\showDOI{\tempurl}


\bibitem[\protect\citeauthoryear{Cavusoglu, Li, and Huang}{Cavusoglu
  et~al\mbox{.}}{2015}]%
        {gamification_stackoverflow}
\bibfield{author}{\bibinfo{person}{Huseyin Cavusoglu}, \bibinfo{person}{Zhuolun
  Li}, {and} \bibinfo{person}{Ke-Wei Huang}.} \bibinfo{year}{2015}\natexlab{}.
\newblock \showarticletitle{Can Gamification Motivate Voluntary Contributions?
  The Case of StackOverflow Q\&A Community}. In
  \bibinfo{booktitle}{\emph{Proceedings of the 18th ACM Conference Companion on
  Computer Supported Cooperative Work \& Social Computing}}
  \emph{(\bibinfo{series}{CSCW’15 Companion})}.
  \bibinfo{publisher}{Association for Computing Machinery},
  \bibinfo{address}{New York, NY, USA}, \bibinfo{pages}{171–174}.
\newblock
\showISBNx{9781450329460}
\urldef\tempurl%
\url{https://doi.org/10.1145/2685553.2698999}
\showDOI{\tempurl}


\bibitem[\protect\citeauthoryear{Chancellor, Hu, and De~Choudhury}{Chancellor
  et~al\mbox{.}}{2018}]%
        {Norms_matter:Chancellor:2018:NMC:3173574.3174240}
\bibfield{author}{\bibinfo{person}{Stevie Chancellor}, \bibinfo{person}{Andrea
  Hu}, {and} \bibinfo{person}{Munmun De~Choudhury}.}
  \bibinfo{year}{2018}\natexlab{}.
\newblock \showarticletitle{Norms Matter: Contrasting Social Support Around
  Behavior Change in Online Weight Loss Communities}. In
  \bibinfo{booktitle}{\emph{Proceedings of the 2018 CHI Conference on Human
  Factors in Computing Systems}} \emph{(\bibinfo{series}{CHI '18})}.
  \bibinfo{publisher}{ACM}, \bibinfo{address}{New York, NY, USA}, Article
  \bibinfo{articleno}{666}, \bibinfo{numpages}{14}~pages.
\newblock
\showISBNx{978-1-4503-5620-6}
\urldef\tempurl%
\url{https://doi.org/10.1145/3173574.3174240}
\showDOI{\tempurl}


\bibitem[\protect\citeauthoryear{Cranshaw, Elwany, Newman, Kocielnik, Yu, Soni,
  Teevan, and Monroy-Hern{\'a}ndez}{Cranshaw et~al\mbox{.}}{2017}]%
        {cranshaw2017calendar}
\bibfield{author}{\bibinfo{person}{Justin Cranshaw}, \bibinfo{person}{Emad
  Elwany}, \bibinfo{person}{Todd Newman}, \bibinfo{person}{Rafal Kocielnik},
  \bibinfo{person}{Bowen Yu}, \bibinfo{person}{Sandeep Soni},
  \bibinfo{person}{Jaime Teevan}, {and} \bibinfo{person}{Andr{\'e}s
  Monroy-Hern{\'a}ndez}.} \bibinfo{year}{2017}\natexlab{}.
\newblock \showarticletitle{Calendar. help: Designing a workflow-based
  scheduling agent with humans in the loop}. In
  \bibinfo{booktitle}{\emph{Proceedings of the 2017 CHI Conference on Human
  Factors in Computing Systems}}. \bibinfo{pages}{2382--2393}.
\newblock


\bibitem[\protect\citeauthoryear{Cutrona and Suhr}{Cutrona and Suhr}{1994}]%
        {cutrona1994social}
\bibfield{author}{\bibinfo{person}{Carolyn~E Cutrona} {and}
  \bibinfo{person}{Julie~A Suhr}.} \bibinfo{year}{1994}\natexlab{}.
\newblock \showarticletitle{Social support communication in the context of
  marriage: an analysis of couples' supportive interactions.}
\newblock  (\bibinfo{year}{1994}).
\newblock


\bibitem[\protect\citeauthoryear{De~Choudhury, Counts, and
  Horvitz}{De~Choudhury et~al\mbox{.}}{2013}]%
        {predicting_postpartum10.1145/2470654.2466447}
\bibfield{author}{\bibinfo{person}{Munmun De~Choudhury}, \bibinfo{person}{Scott
  Counts}, {and} \bibinfo{person}{Eric Horvitz}.}
  \bibinfo{year}{2013}\natexlab{}.
\newblock \showarticletitle{Predicting Postpartum Changes in Emotion and
  Behavior via Social Media}. In \bibinfo{booktitle}{\emph{Proceedings of the
  SIGCHI Conference on Human Factors in Computing Systems}}
  \emph{(\bibinfo{series}{CHI ’13})}. \bibinfo{publisher}{Association for
  Computing Machinery}, \bibinfo{address}{New York, NY, USA},
  \bibinfo{pages}{3267–3276}.
\newblock
\showISBNx{9781450318990}
\urldef\tempurl%
\url{https://doi.org/10.1145/2470654.2466447}
\showDOI{\tempurl}


\bibitem[\protect\citeauthoryear{De~Choudhury, Counts, Horvitz, and
  Hoff}{De~Choudhury et~al\mbox{.}}{2014}]%
        {characterizing_predicting10.1145/2531602.2531675}
\bibfield{author}{\bibinfo{person}{Munmun De~Choudhury}, \bibinfo{person}{Scott
  Counts}, \bibinfo{person}{Eric~J. Horvitz}, {and} \bibinfo{person}{Aaron
  Hoff}.} \bibinfo{year}{2014}\natexlab{}.
\newblock \showarticletitle{Characterizing and Predicting Postpartum Depression
  from Shared Facebook Data}. In \bibinfo{booktitle}{\emph{Proceedings of the
  17th ACM Conference on Computer Supported Cooperative Work \& Social
  Computing}} \emph{(\bibinfo{series}{CSCW ’14})}.
  \bibinfo{publisher}{Association for Computing Machinery},
  \bibinfo{address}{New York, NY, USA}, \bibinfo{pages}{626–638}.
\newblock
\showISBNx{9781450325400}
\urldef\tempurl%
\url{https://doi.org/10.1145/2531602.2531675}
\showDOI{\tempurl}


\bibitem[\protect\citeauthoryear{De~Choudhury and De}{De~Choudhury and
  De}{2014}]%
        {de2014mental}
\bibfield{author}{\bibinfo{person}{Munmun De~Choudhury} {and}
  \bibinfo{person}{Sushovan De}.} \bibinfo{year}{2014}\natexlab{}.
\newblock \showarticletitle{Mental health discourse on reddit: Self-disclosure,
  social support, and anonymity}. In \bibinfo{booktitle}{\emph{Eighth
  international AAAI conference on weblogs and social media}}.
\newblock


\bibitem[\protect\citeauthoryear{DerSimonian and Laird}{DerSimonian and
  Laird}{1986}]%
        {dersimonian1986meta}
\bibfield{author}{\bibinfo{person}{Rebecca DerSimonian} {and}
  \bibinfo{person}{Nan Laird}.} \bibinfo{year}{1986}\natexlab{}.
\newblock \showarticletitle{Meta-analysis in clinical trials}.
\newblock \bibinfo{journal}{\emph{Controlled clinical trials}}
  \bibinfo{volume}{7}, \bibinfo{number}{3} (\bibinfo{year}{1986}),
  \bibinfo{pages}{177--188}.
\newblock


\bibitem[\protect\citeauthoryear{Donath}{Donath}{2002}]%
        {donath2002identity}
\bibfield{author}{\bibinfo{person}{Judith~S Donath}.}
  \bibinfo{year}{2002}\natexlab{}.
\newblock \showarticletitle{Identity and deception in the virtual community}.
\newblock In \bibinfo{booktitle}{\emph{Communities in cyberspace}}.
  \bibinfo{publisher}{Routledge}, \bibinfo{pages}{37--68}.
\newblock


\bibitem[\protect\citeauthoryear{Drentea and Moren-Cross}{Drentea and
  Moren-Cross}{2005}]%
        {social_capital_doi:10.1111/j.1467-9566.2005.00464.x}
\bibfield{author}{\bibinfo{person}{Patricia Drentea} {and}
  \bibinfo{person}{Jennifer~L. Moren-Cross}.} \bibinfo{year}{2005}\natexlab{}.
\newblock \showarticletitle{Social capital and social support on the web: the
  case of an internet mother site}.
\newblock \bibinfo{journal}{\emph{Sociology of Health \& Illness}}
  \bibinfo{volume}{27}, \bibinfo{number}{7} (\bibinfo{year}{2005}),
  \bibinfo{pages}{920--943}.
\newblock
\urldef\tempurl%
\url{https://doi.org/10.1111/j.1467-9566.2005.00464.x}
\showDOI{\tempurl}
\showeprint{https://onlinelibrary.wiley.com/doi/pdf/10.1111/j.1467-9566.2005.00464.x}


\bibitem[\protect\citeauthoryear{Evans, Donelle, and Hume-Loveland}{Evans
  et~al\mbox{.}}{2012}]%
        {social_support_EVANS2012405}
\bibfield{author}{\bibinfo{person}{Marilyn Evans}, \bibinfo{person}{Lorie
  Donelle}, {and} \bibinfo{person}{Laurie Hume-Loveland}.}
  \bibinfo{year}{2012}\natexlab{}.
\newblock \showarticletitle{Social support and online postpartum depression
  discussion groups: A content analysis}.
\newblock \bibinfo{journal}{\emph{Patient Education and Counseling}}
  \bibinfo{volume}{87}, \bibinfo{number}{3} (\bibinfo{year}{2012}),
  \bibinfo{pages}{405 -- 410}.
\newblock
\showISSN{0738-3991}
\urldef\tempurl%
\url{https://doi.org/10.1016/j.pec.2011.09.011}
\showDOI{\tempurl}


\bibitem[\protect\citeauthoryear{Fan, Chao, Zhang, Wang, Li, and Tian}{Fan
  et~al\mbox{.}}{2020}]%
        {doctorbot}
\bibfield{author}{\bibinfo{person}{Xiangmin Fan}, \bibinfo{person}{Daren Chao},
  \bibinfo{person}{Zhan Zhang}, \bibinfo{person}{Dakuo Wang},
  \bibinfo{person}{Xiaohua Li}, {and} \bibinfo{person}{Feng Tian}.}
  \bibinfo{year}{2020}\natexlab{}.
\newblock \showarticletitle{Utilization of Self-Diagnosis Health Chatbots in
  Real-World Settings: Case Study}.
\newblock \bibinfo{journal}{\emph{Journal of Medical Internet Research}}
  \bibinfo{volume}{22} (\bibinfo{year}{2020}).
\newblock


\bibitem[\protect\citeauthoryear{Fitzpatrick, Darcy, and Vierhile}{Fitzpatrick
  et~al\mbox{.}}{2017}]%
        {woebot}
\bibfield{author}{\bibinfo{person}{Kathleen~Kara Fitzpatrick},
  \bibinfo{person}{Alison Darcy}, {and} \bibinfo{person}{Molly Vierhile}.}
  \bibinfo{year}{2017}\natexlab{}.
\newblock \showarticletitle{Delivering Cognitive Behavior Therapy to Young
  Adults With Symptoms of Depression and Anxiety Using a Fully Automated
  Conversational Agent (Woebot): A Randomized Controlled Trial}.
\newblock \bibinfo{journal}{\emph{JMIR Mental Health}}  \bibinfo{volume}{4}
  (\bibinfo{date}{06} \bibinfo{year}{2017}), \bibinfo{pages}{e19}.
\newblock
\urldef\tempurl%
\url{https://doi.org/10.2196/mental.7785}
\showDOI{\tempurl}


\bibitem[\protect\citeauthoryear{Geiger}{Geiger}{2018}]%
        {DBLP:journals/corr/abs-1810-09590}
\bibfield{author}{\bibinfo{person}{R.~Stuart Geiger}.}
  \bibinfo{year}{2018}\natexlab{}.
\newblock \showarticletitle{The Lives of Bots}.
\newblock \bibinfo{journal}{\emph{CoRR}}  \bibinfo{volume}{abs/1810.09590}
  (\bibinfo{year}{2018}).
\newblock
\showeprint[arxiv]{1810.09590}
\urldef\tempurl%
\url{http://arxiv.org/abs/1810.09590}
\showURL{%
\tempurl}


\bibitem[\protect\citeauthoryear{Gergle and Tan}{Gergle and Tan}{2014}]%
        {gergle2014experimental}
\bibfield{author}{\bibinfo{person}{Darren Gergle} {and}
  \bibinfo{person}{Desney~S Tan}.} \bibinfo{year}{2014}\natexlab{}.
\newblock \showarticletitle{Experimental research in HCI}.
\newblock In \bibinfo{booktitle}{\emph{Ways of Knowing in HCI}}.
  \bibinfo{publisher}{Springer}, \bibinfo{pages}{191--227}.
\newblock


\bibitem[\protect\citeauthoryear{Gilani, Farahbakhsh, and Crowcroft}{Gilani
  et~al\mbox{.}}{2017}]%
        {TwitterBotGilani:2017:BIT:3041021.3054255}
\bibfield{author}{\bibinfo{person}{Zafar Gilani}, \bibinfo{person}{Reza
  Farahbakhsh}, {and} \bibinfo{person}{Jon Crowcroft}.}
  \bibinfo{year}{2017}\natexlab{}.
\newblock \showarticletitle{Do Bots Impact Twitter Activity?}. In
  \bibinfo{booktitle}{\emph{Proceedings of the 26th International Conference on
  World Wide Web Companion}} \emph{(\bibinfo{series}{WWW '17 Companion})}.
  \bibinfo{publisher}{International World Wide Web Conferences Steering
  Committee}, \bibinfo{address}{Republic and Canton of Geneva, Switzerland},
  \bibinfo{pages}{781--782}.
\newblock
\showISBNx{978-1-4503-4914-7}
\urldef\tempurl%
\url{https://doi.org/10.1145/3041021.3054255}
\showDOI{\tempurl}


\bibitem[\protect\citeauthoryear{Grudin}{Grudin}{2008}]%
        {grudin20081}
\bibfield{author}{\bibinfo{person}{Jonathan Grudin}.}
  \bibinfo{year}{2008}\natexlab{}.
\newblock \showarticletitle{1 McGrath and the Behaviors of Groups (BOGs)}.
\newblock  (\bibinfo{year}{2008}).
\newblock


\bibitem[\protect\citeauthoryear{Grudin}{Grudin}{2017}]%
        {grudin2017tool}
\bibfield{author}{\bibinfo{person}{Jonathan Grudin}.}
  \bibinfo{year}{2017}\natexlab{}.
\newblock \showarticletitle{From tool to partner: The evolution of
  human-computer interaction}.
\newblock \bibinfo{journal}{\emph{Synthesis Lectures on Human-Centered
  Interaction}} \bibinfo{volume}{10}, \bibinfo{number}{1}
  (\bibinfo{year}{2017}), \bibinfo{pages}{i--183}.
\newblock


\bibitem[\protect\citeauthoryear{Gui, Chen, Kou, Pine, and Chen}{Gui
  et~al\mbox{.}}{2017}]%
        {Pregnant_community_Gui:2017:ISS:3171581.3134685}
\bibfield{author}{\bibinfo{person}{Xinning Gui}, \bibinfo{person}{Yu Chen},
  \bibinfo{person}{Yubo Kou}, \bibinfo{person}{Katie Pine}, {and}
  \bibinfo{person}{Yunan Chen}.} \bibinfo{year}{2017}\natexlab{}.
\newblock \showarticletitle{Investigating Support Seeking from Peers for
  Pregnancy in Online Health Communities}.
\newblock \bibinfo{journal}{\emph{Proc. ACM Hum.-Comput. Interact.}}
  \bibinfo{volume}{1}, \bibinfo{number}{CSCW}, Article \bibinfo{articleno}{50}
  (\bibinfo{date}{Dec.} \bibinfo{year}{2017}), \bibinfo{numpages}{19}~pages.
\newblock
\showISSN{2573-0142}
\urldef\tempurl%
\url{https://doi.org/10.1145/3134685}
\showDOI{\tempurl}


\bibitem[\protect\citeauthoryear{Halfaker and Geiger}{Halfaker and
  Geiger}{2019}]%
        {halfaker2019ores}
\bibfield{author}{\bibinfo{person}{Aaron Halfaker} {and}
  \bibinfo{person}{R~Stuart Geiger}.} \bibinfo{year}{2019}\natexlab{}.
\newblock \showarticletitle{ORES: Lowering Barriers with Participatory Machine
  Learning in Wikipedia}.
\newblock \bibinfo{journal}{\emph{arXiv preprint arXiv:1909.05189}}
  (\bibinfo{year}{2019}).
\newblock


\bibitem[\protect\citeauthoryear{Halfaker, Geiger, and Terveen}{Halfaker
  et~al\mbox{.}}{2014}]%
        {halfaker2014snuggle}
\bibfield{author}{\bibinfo{person}{Aaron Halfaker}, \bibinfo{person}{R~Stuart
  Geiger}, {and} \bibinfo{person}{Loren~G Terveen}.}
  \bibinfo{year}{2014}\natexlab{}.
\newblock \showarticletitle{Snuggle: Designing for efficient socialization and
  ideological critique}. In \bibinfo{booktitle}{\emph{Proceedings of the SIGCHI
  conference on human factors in computing systems}}.
  \bibinfo{pages}{311--320}.
\newblock


\bibitem[\protect\citeauthoryear{Harper, Raban, Rafaeli, and Konstan}{Harper
  et~al\mbox{.}}{2008}]%
        {predictors_of_answer10.1145/1357054.1357191}
\bibfield{author}{\bibinfo{person}{F.~Maxwell Harper}, \bibinfo{person}{Daphne
  Raban}, \bibinfo{person}{Sheizaf Rafaeli}, {and} \bibinfo{person}{Joseph~A.
  Konstan}.} \bibinfo{year}{2008}\natexlab{}.
\newblock \showarticletitle{Predictors of Answer Quality in Online Q\&A Sites}.
  In \bibinfo{booktitle}{\emph{Proceedings of the SIGCHI Conference on Human
  Factors in Computing Systems}} \emph{(\bibinfo{series}{CHI ’08})}.
  \bibinfo{publisher}{Association for Computing Machinery},
  \bibinfo{address}{New York, NY, USA}, \bibinfo{pages}{865–874}.
\newblock
\showISBNx{9781605580111}
\urldef\tempurl%
\url{https://doi.org/10.1145/1357054.1357191}
\showDOI{\tempurl}


\bibitem[\protect\citeauthoryear{Hochreiter and Schmidhuber}{Hochreiter and
  Schmidhuber}{[n.d.]}]%
        {HochreiterLong}
\bibfield{author}{\bibinfo{person}{Sepp Hochreiter} {and}
  \bibinfo{person}{Jürgen Schmidhuber}.} \bibinfo{year}{[n.d.]}\natexlab{}.
\newblock \showarticletitle{Long Short-Term Memory}.
\newblock \bibinfo{journal}{\emph{Neural Computation}} \bibinfo{volume}{9},
  \bibinfo{number}{8} (\bibinfo{year}{[n.\,d.]}), \bibinfo{pages}{1735--1780}.
\newblock


\bibitem[\protect\citeauthoryear{Holtz, Smock, and Reyes-Gastelum}{Holtz
  et~al\mbox{.}}{2015}]%
        {connected_motherhood_doi:10.1089/tmj.2014.0118}
\bibfield{author}{\bibinfo{person}{Bree Holtz}, \bibinfo{person}{Andrew Smock},
  {and} \bibinfo{person}{David Reyes-Gastelum}.}
  \bibinfo{year}{2015}\natexlab{}.
\newblock \showarticletitle{Connected Motherhood: Social Support for Moms and
  Moms-to-Be on Facebook}.
\newblock \bibinfo{journal}{\emph{Telemedicine and e-Health}}
  \bibinfo{volume}{21}, \bibinfo{number}{5} (\bibinfo{year}{2015}),
  \bibinfo{pages}{415--421}.
\newblock
\urldef\tempurl%
\url{https://doi.org/10.1089/tmj.2014.0118}
\showDOI{\tempurl}
\showeprint{https://doi.org/10.1089/tmj.2014.0118}
\newblock
\shownote{PMID: 25665177.}


\bibitem[\protect\citeauthoryear{Horvitz}{Horvitz}{1999}]%
        {horvitz1999principles}
\bibfield{author}{\bibinfo{person}{Eric Horvitz}.}
  \bibinfo{year}{1999}\natexlab{}.
\newblock \showarticletitle{Principles of mixed-initiative user interfaces}. In
  \bibinfo{booktitle}{\emph{Proceedings of the SIGCHI conference on Human
  Factors in Computing Systems}}. \bibinfo{pages}{159--166}.
\newblock


\bibitem[\protect\citeauthoryear{Hu, Xu, Liu, You, Guo, Sinha, Luo, and
  Akkiraju}{Hu et~al\mbox{.}}{2018}]%
        {hu2018touch}
\bibfield{author}{\bibinfo{person}{Tianran Hu}, \bibinfo{person}{Anbang Xu},
  \bibinfo{person}{Zhe Liu}, \bibinfo{person}{Quanzeng You},
  \bibinfo{person}{Yufan Guo}, \bibinfo{person}{Vibha Sinha},
  \bibinfo{person}{Jiebo Luo}, {and} \bibinfo{person}{Rama Akkiraju}.}
  \bibinfo{year}{2018}\natexlab{}.
\newblock \showarticletitle{Touch your heart: a tone-aware chatbot for customer
  care on social media}. In \bibinfo{booktitle}{\emph{Proceedings of the 2018
  CHI Conference on Human Factors in Computing Systems}}.
  \bibinfo{pages}{1--12}.
\newblock


\bibitem[\protect\citeauthoryear{Huh and Ackerman}{Huh and Ackerman}{2012}]%
        {collaborative_10.1145/2145204.2145331}
\bibfield{author}{\bibinfo{person}{Jina Huh} {and} \bibinfo{person}{Mark~S.
  Ackerman}.} \bibinfo{year}{2012}\natexlab{}.
\newblock \showarticletitle{Collaborative Help in Chronic Disease Management:
  Supporting Individualized Problems}. In \bibinfo{booktitle}{\emph{Proceedings
  of the ACM 2012 Conference on Computer Supported Cooperative Work}}
  \emph{(\bibinfo{series}{CSCW ’12})}. \bibinfo{publisher}{Association for
  Computing Machinery}, \bibinfo{address}{New York, NY, USA},
  \bibinfo{pages}{853–862}.
\newblock
\showISBNx{9781450310864}
\urldef\tempurl%
\url{https://doi.org/10.1145/2145204.2145331}
\showDOI{\tempurl}


\bibitem[\protect\citeauthoryear{IBM}{IBM}{2017}]%
        {customer_engagement_tone}
\bibfield{author}{\bibinfo{person}{IBM}.} \bibinfo{year}{2017}\natexlab{}.
\newblock \bibinfo{title}{Tone Analyzer for Customer Engagement}.
\newblock
  \bibinfo{howpublished}{\url{https://www.ibm.com/blogs/bluemix/2017/04/tone-analyzer-customer-engagement-7-new-tones-help-understand-customers-feeling/}}.
\newblock
\newblock
\shownote{Accessed: 2019-04-03.}


\bibitem[\protect\citeauthoryear{IBM}{IBM}{2018}]%
        {IBMWatson}
\bibfield{author}{\bibinfo{person}{IBM}.} \bibinfo{year}{2018}\natexlab{}.
\newblock \bibinfo{title}{IBM Watson Assistant}.
\newblock
\newblock
\newblock
\shownote{Retrieved: 2018-12-10.
  \url{https://assistant-us-south.watsonplatform.net/us-south/b3a5bd9b-9ea9-4be8-9ec7-145f04f69453/home}.}


\bibitem[\protect\citeauthoryear{Ishibe, Albitar, Jilani, Goldin, Marti, and
  Caporaso}{Ishibe et~al\mbox{.}}{2006}]%
        {Ishibe2006Incidence}
\bibfield{author}{\bibinfo{person}{N Ishibe}, \bibinfo{person}{M Albitar},
  \bibinfo{person}{I.~B. Jilani}, \bibinfo{person}{L.~R. Goldin},
  \bibinfo{person}{G.~E. Marti}, {and} \bibinfo{person}{N.~E. Caporaso}.}
  \bibinfo{year}{2006}\natexlab{}.
\newblock \showarticletitle{Incidence of postpartum depression in Trabzon
  province and risk factors at gestation}.
\newblock \bibinfo{journal}{\emph{Turk psikiyatri dergisi = Turkish journal of
  psychiatry}} \bibinfo{volume}{17}, \bibinfo{number}{4}
  (\bibinfo{year}{2006}), \bibinfo{pages}{243--251}.
\newblock


\bibitem[\protect\citeauthoryear{Jakesch, French, Ma, Hancock, and
  Naaman}{Jakesch et~al\mbox{.}}{2019}]%
        {Jakesch:2019:ACP:3290605.3300469}
\bibfield{author}{\bibinfo{person}{Maurice Jakesch}, \bibinfo{person}{Megan
  French}, \bibinfo{person}{Xiao Ma}, \bibinfo{person}{Jeffrey~T. Hancock},
  {and} \bibinfo{person}{Mor Naaman}.} \bibinfo{year}{2019}\natexlab{}.
\newblock \showarticletitle{AI-Mediated Communication: How the Perception That
  Profile Text Was Written by AI Affects Trustworthiness}. In
  \bibinfo{booktitle}{\emph{Proceedings of the 2019 CHI Conference on Human
  Factors in Computing Systems}} \emph{(\bibinfo{series}{CHI '19})}.
  \bibinfo{publisher}{ACM}, \bibinfo{address}{New York, NY, USA}, Article
  \bibinfo{articleno}{239}, \bibinfo{numpages}{13}~pages.
\newblock
\showISBNx{978-1-4503-5970-2}
\urldef\tempurl%
\url{https://doi.org/10.1145/3290605.3300469}
\showDOI{\tempurl}


\bibitem[\protect\citeauthoryear{Johnson, Lin, Li, Hall, Halfaker,
  Sch\"{o}ning, and Hecht}{Johnson et~al\mbox{.}}{2016}]%
        {Wikipedia:Johnson:2016:HRP:2858036.2858123}
\bibfield{author}{\bibinfo{person}{Isaac~L. Johnson}, \bibinfo{person}{Yilun
  Lin}, \bibinfo{person}{Toby Jia-Jun Li}, \bibinfo{person}{Andrew Hall},
  \bibinfo{person}{Aaron Halfaker}, \bibinfo{person}{Johannes Sch\"{o}ning},
  {and} \bibinfo{person}{Brent Hecht}.} \bibinfo{year}{2016}\natexlab{}.
\newblock \showarticletitle{Not at Home on the Range: Peer Production and the
  Urban/Rural Divide}. In \bibinfo{booktitle}{\emph{Proceedings of the 2016 CHI
  Conference on Human Factors in Computing Systems}}
  \emph{(\bibinfo{series}{CHI '16})}. \bibinfo{publisher}{ACM},
  \bibinfo{address}{New York, NY, USA}, \bibinfo{pages}{13--25}.
\newblock
\showISBNx{978-1-4503-3362-7}
\urldef\tempurl%
\url{https://doi.org/10.1145/2858036.2858123}
\showDOI{\tempurl}


\bibitem[\protect\citeauthoryear{Kim}{Kim}{2014}]%
        {kim2014convolutional}
\bibfield{author}{\bibinfo{person}{Yoon Kim}.} \bibinfo{year}{2014}\natexlab{}.
\newblock \showarticletitle{Convolutional neural networks for sentence
  classification}.
\newblock \bibinfo{journal}{\emph{arXiv preprint arXiv:1408.5882}}
  (\bibinfo{year}{2014}).
\newblock


\bibitem[\protect\citeauthoryear{Klein, Kim, Deng, Nguyen, and Rush}{Klein
  et~al\mbox{.}}{2018}]%
        {Klein2018OpenNMT}
\bibfield{author}{\bibinfo{person}{Guillaume Klein}, \bibinfo{person}{Yoon
  Kim}, \bibinfo{person}{Yuntian Deng}, \bibinfo{person}{Vincent Nguyen}, {and}
  \bibinfo{person}{Alexander~M. Rush}.} \bibinfo{year}{2018}\natexlab{}.
\newblock \showarticletitle{OpenNMT: Neural Machine Translation Toolkit}.
\newblock  (\bibinfo{year}{2018}).
\newblock


\bibitem[\protect\citeauthoryear{Kraut, Burke, Riedl, and Resnick}{Kraut
  et~al\mbox{.}}{2010}]%
        {kraut2010dealing}
\bibfield{author}{\bibinfo{person}{Robert Kraut}, \bibinfo{person}{Moira
  Burke}, \bibinfo{person}{John Riedl}, {and} \bibinfo{person}{Paul Resnick}.}
  \bibinfo{year}{2010}\natexlab{}.
\newblock \showarticletitle{Dealing with newcomers}.
\newblock \bibinfo{journal}{\emph{Evidencebased Social Design Mining the Social
  Sciences to Build Online Communities}}  \bibinfo{volume}{1}
  (\bibinfo{year}{2010}), \bibinfo{pages}{42}.
\newblock


\bibitem[\protect\citeauthoryear{Kraut and Resnick}{Kraut and Resnick}{2011}]%
        {kraut2011encouraging}
\bibfield{author}{\bibinfo{person}{Robert~E Kraut} {and} \bibinfo{person}{Paul
  Resnick}.} \bibinfo{year}{2011}\natexlab{}.
\newblock \showarticletitle{Encouraging contribution to online communities}.
\newblock \bibinfo{journal}{\emph{Building successful online communities:
  Evidence-based social design}} (\bibinfo{year}{2011}),
  \bibinfo{pages}{21--76}.
\newblock


\bibitem[\protect\citeauthoryear{Kraut and Resnick}{Kraut and Resnick}{2012}]%
        {kraut2012building}
\bibfield{author}{\bibinfo{person}{Robert~E Kraut} {and} \bibinfo{person}{Paul
  Resnick}.} \bibinfo{year}{2012}\natexlab{}.
\newblock \bibinfo{booktitle}{\emph{Building successful online communities:
  Evidence-based social design}}.
\newblock \bibinfo{publisher}{Mit Press}.
\newblock


\bibitem[\protect\citeauthoryear{Krizhevsky, Sutskever, and Hinton}{Krizhevsky
  et~al\mbox{.}}{2012}]%
        {Krizhevsky2012ImageNet}
\bibfield{author}{\bibinfo{person}{Alex Krizhevsky}, \bibinfo{person}{I.
  Sutskever}, {and} \bibinfo{person}{G. Hinton}.}
  \bibinfo{year}{2012}\natexlab{}.
\newblock \showarticletitle{ImageNet Classification with Deep Convolutional
  Neural Networks}.
\newblock \bibinfo{journal}{\emph{Advances in neural information processing
  systems}} \bibinfo{volume}{25}, \bibinfo{number}{2} (\bibinfo{year}{2012}).
\newblock


\bibitem[\protect\citeauthoryear{Kumar and Anderson}{Kumar and
  Anderson}{2015}]%
        {mobile_phone_10.1145/2702123.2702258}
\bibfield{author}{\bibinfo{person}{Neha Kumar} {and}
  \bibinfo{person}{Richard~J. Anderson}.} \bibinfo{year}{2015}\natexlab{}.
\newblock \showarticletitle{Mobile Phones for Maternal Health in Rural India}.
  In \bibinfo{booktitle}{\emph{Proceedings of the 33rd Annual ACM Conference on
  Human Factors in Computing Systems}} \emph{(\bibinfo{series}{CHI ’15})}.
  \bibinfo{publisher}{Association for Computing Machinery},
  \bibinfo{address}{New York, NY, USA}, \bibinfo{pages}{427–436}.
\newblock
\showISBNx{9781450331456}
\urldef\tempurl%
\url{https://doi.org/10.1145/2702123.2702258}
\showDOI{\tempurl}


\bibitem[\protect\citeauthoryear{Lee, Ackermans, van As, Chang, Lucas, and
  IJsselsteijn}{Lee et~al\mbox{.}}{2019a}]%
        {Care_for_bot_Lee:2019:CVC:3290605.3300932}
\bibfield{author}{\bibinfo{person}{Minha Lee}, \bibinfo{person}{Sander
  Ackermans}, \bibinfo{person}{Nena van As}, \bibinfo{person}{Hanwen Chang},
  \bibinfo{person}{Enzo Lucas}, {and} \bibinfo{person}{Wijnand IJsselsteijn}.}
  \bibinfo{year}{2019}\natexlab{a}.
\newblock \showarticletitle{Caring for Vincent: A Chatbot for Self-Compassion}.
  In \bibinfo{booktitle}{\emph{Proceedings of the 2019 CHI Conference on Human
  Factors in Computing Systems}} \emph{(\bibinfo{series}{CHI '19})}.
  \bibinfo{publisher}{ACM}, \bibinfo{address}{New York, NY, USA}, Article
  \bibinfo{articleno}{702}, \bibinfo{numpages}{13}~pages.
\newblock
\showISBNx{978-1-4503-5970-2}
\urldef\tempurl%
\url{https://doi.org/10.1145/3290605.3300932}
\showDOI{\tempurl}


\bibitem[\protect\citeauthoryear{Lee, Kim, and Lizarondo}{Lee
  et~al\mbox{.}}{2017}]%
        {lee2017human}
\bibfield{author}{\bibinfo{person}{Min~Kyung Lee}, \bibinfo{person}{Ji~Tae
  Kim}, {and} \bibinfo{person}{Leah Lizarondo}.}
  \bibinfo{year}{2017}\natexlab{}.
\newblock \showarticletitle{A human-centered approach to algorithmic services:
  Considerations for fair and motivating smart community service management
  that allocates donations to non-profit organizations}. In
  \bibinfo{booktitle}{\emph{Proceedings of the 2017 CHI Conference on Human
  Factors in Computing Systems}}. \bibinfo{pages}{3365--3376}.
\newblock


\bibitem[\protect\citeauthoryear{Lee, Kusbit, Kahng, Kim, Yuan, Chan, See,
  Noothigattu, Lee, Psomas, et~al\mbox{.}}{Lee et~al\mbox{.}}{2019b}]%
        {lee2019webuildai}
\bibfield{author}{\bibinfo{person}{Min~Kyung Lee}, \bibinfo{person}{Daniel
  Kusbit}, \bibinfo{person}{Anson Kahng}, \bibinfo{person}{Ji~Tae Kim},
  \bibinfo{person}{Xinran Yuan}, \bibinfo{person}{Allissa Chan},
  \bibinfo{person}{Daniel See}, \bibinfo{person}{Ritesh Noothigattu},
  \bibinfo{person}{Siheon Lee}, \bibinfo{person}{Alexandros Psomas},
  {et~al\mbox{.}}} \bibinfo{year}{2019}\natexlab{b}.
\newblock \showarticletitle{WeBuildAI: Participatory framework for algorithmic
  governance}.
\newblock \bibinfo{journal}{\emph{Proceedings of the ACM on Human-Computer
  Interaction}} \bibinfo{volume}{3}, \bibinfo{number}{CSCW}
  (\bibinfo{year}{2019}), \bibinfo{pages}{1--35}.
\newblock


\bibitem[\protect\citeauthoryear{Lee, Kim, Yi, Sung, and Gerla}{Lee
  et~al\mbox{.}}{2013}]%
        {lee2013analyzing}
\bibfield{author}{\bibinfo{person}{Uichin Lee}, \bibinfo{person}{Jihyoung Kim},
  \bibinfo{person}{Eunhee Yi}, \bibinfo{person}{Juyup Sung}, {and}
  \bibinfo{person}{Mario Gerla}.} \bibinfo{year}{2013}\natexlab{}.
\newblock \showarticletitle{Analyzing crowd workers in mobile pay-for-answer
  q\&a}. In \bibinfo{booktitle}{\emph{Proceedings of the SIGCHI Conference on
  Human Factors in Computing Systems}}. \bibinfo{pages}{533--542}.
\newblock


\bibitem[\protect\citeauthoryear{Leigh and Milgrom}{Leigh and Milgrom}{2008}]%
        {Depressed_women_Leigh2008RiskFF}
\bibfield{author}{\bibinfo{person}{Bronwyn Leigh} {and}
  \bibinfo{person}{Jeannette Milgrom}.} \bibinfo{year}{2008}\natexlab{}.
\newblock \showarticletitle{Risk factors for antenatal depression, postnatal
  depression and parenting stress}.
\newblock \bibinfo{journal}{\emph{BMC Psychiatry}}  \bibinfo{volume}{8}
  (\bibinfo{year}{2008}), \bibinfo{pages}{24 -- 24}.
\newblock


\bibitem[\protect\citeauthoryear{Li, Monroe, Ritter, Galley, Gao, and
  Jurafsky}{Li et~al\mbox{.}}{2016}]%
        {li2016deep}
\bibfield{author}{\bibinfo{person}{Jiwei Li}, \bibinfo{person}{Will Monroe},
  \bibinfo{person}{Alan Ritter}, \bibinfo{person}{Michel Galley},
  \bibinfo{person}{Jianfeng Gao}, {and} \bibinfo{person}{Dan Jurafsky}.}
  \bibinfo{year}{2016}\natexlab{}.
\newblock \showarticletitle{Deep reinforcement learning for dialogue
  generation}.
\newblock \bibinfo{journal}{\emph{arXiv preprint arXiv:1606.01541}}
  (\bibinfo{year}{2016}).
\newblock


\bibitem[\protect\citeauthoryear{Liao, Mas-ud Hussain, Chandar, Davis,
  Khazaeni, Crasso, Wang, Muller, Shami, and Geyer}{Liao et~al\mbox{.}}{2018}]%
        {liao2018all}
\bibfield{author}{\bibinfo{person}{Q~Vera Liao}, \bibinfo{person}{Muhammed
  Mas-ud Hussain}, \bibinfo{person}{Praveen Chandar}, \bibinfo{person}{Matthew
  Davis}, \bibinfo{person}{Yasaman Khazaeni}, \bibinfo{person}{Marco~Patricio
  Crasso}, \bibinfo{person}{Dakuo Wang}, \bibinfo{person}{Michael Muller},
  \bibinfo{person}{N~Sadat Shami}, {and} \bibinfo{person}{Werner Geyer}.}
  \bibinfo{year}{2018}\natexlab{}.
\newblock \showarticletitle{All Work and No Play?}. In
  \bibinfo{booktitle}{\emph{Proceedings of the 2018 CHI Conference on Human
  Factors in Computing Systems}}. \bibinfo{pages}{1--13}.
\newblock


\bibitem[\protect\citeauthoryear{Logie, Weinberg, Harper, and Konstan}{Logie
  et~al\mbox{.}}{2011}]%
        {Logie2011AskedAA}
\bibfield{author}{\bibinfo{person}{John Logie}, \bibinfo{person}{Joseph
  Weinberg}, \bibinfo{person}{F.~Maxwell Harper}, {and}
  \bibinfo{person}{Joseph~A. Konstan}.} \bibinfo{year}{2011}\natexlab{}.
\newblock \showarticletitle{Asked and Answered: On Qualities and Quantities of
  Answers in Online Q\&A Sites}. In \bibinfo{booktitle}{\emph{The Social Mobile
  Web}}.
\newblock


\bibitem[\protect\citeauthoryear{Long, Vines, Sutton, Brooker, Feltwell,
  Kirman, Barnett, and Lawson}{Long et~al\mbox{.}}{2017}]%
        {Could_you_10.1145/3025453.3025830}
\bibfield{author}{\bibinfo{person}{Kiel Long}, \bibinfo{person}{John Vines},
  \bibinfo{person}{Selina Sutton}, \bibinfo{person}{Phillip Brooker},
  \bibinfo{person}{Tom Feltwell}, \bibinfo{person}{Ben Kirman},
  \bibinfo{person}{Julie Barnett}, {and} \bibinfo{person}{Shaun Lawson}.}
  \bibinfo{year}{2017}\natexlab{}.
\newblock \showarticletitle{“Could You Define That in Bot Terms”?
  Requesting, Creating and Using Bots on Reddit}. In
  \bibinfo{booktitle}{\emph{Proceedings of the 2017 CHI Conference on Human
  Factors in Computing Systems}} \emph{(\bibinfo{series}{CHI ’17})}.
  \bibinfo{publisher}{Association for Computing Machinery},
  \bibinfo{address}{New York, NY, USA}, \bibinfo{pages}{3488–3500}.
\newblock
\showISBNx{9781450346559}
\urldef\tempurl%
\url{https://doi.org/10.1145/3025453.3025830}
\showDOI{\tempurl}


\bibitem[\protect\citeauthoryear{Lubis, Sakti, Neubig, Yoshino, Toda, and
  Nakamura}{Lubis et~al\mbox{.}}{2015}]%
        {lubis2015study}
\bibfield{author}{\bibinfo{person}{Nurul Lubis}, \bibinfo{person}{Sakriani
  Sakti}, \bibinfo{person}{Graham Neubig}, \bibinfo{person}{Koichiro Yoshino},
  \bibinfo{person}{Tomoki Toda}, {and} \bibinfo{person}{Satoshi Nakamura}.}
  \bibinfo{year}{2015}\natexlab{}.
\newblock \showarticletitle{A study of social-affective communication:
  Automatic prediction of emotion triggers and responses in television talk
  shows}. In \bibinfo{booktitle}{\emph{2015 IEEE Workshop on Automatic Speech
  Recognition and Understanding (ASRU)}}. IEEE, \bibinfo{pages}{777--783}.
\newblock


\bibitem[\protect\citeauthoryear{Luo, Tong, Fang, and Qu}{Luo
  et~al\mbox{.}}{2019}]%
        {luo2019machines}
\bibfield{author}{\bibinfo{person}{Xueming Luo}, \bibinfo{person}{Siliang
  Tong}, \bibinfo{person}{Zheng Fang}, {and} \bibinfo{person}{Zhe Qu}.}
  \bibinfo{year}{2019}\natexlab{}.
\newblock \showarticletitle{Machines Versus Humans: The Impact of AI Chatbot
  Disclosure on Customer Purchases}.
\newblock \bibinfo{journal}{\emph{Luo, X, Tong S, Fang Z, Qu}}
  \bibinfo{number}{2019} (\bibinfo{year}{2019}).
\newblock


\bibitem[\protect\citeauthoryear{Luong, Pham, and Manning}{Luong
  et~al\mbox{.}}{2015}]%
        {Luong2015Effective}
\bibfield{author}{\bibinfo{person}{Minh~Thang Luong}, \bibinfo{person}{Hieu
  Pham}, {and} \bibinfo{person}{Christopher~D. Manning}.}
  \bibinfo{year}{2015}\natexlab{}.
\newblock \showarticletitle{Effective Approaches to Attention-based Neural
  Machine Translation}.
\newblock \bibinfo{journal}{\emph{Computer Science}} (\bibinfo{year}{2015}).
\newblock


\bibitem[\protect\citeauthoryear{Mamykina, Manoim, Mittal, Hripcsak, and
  Hartmann}{Mamykina et~al\mbox{.}}{2011}]%
        {mamykina2011design}
\bibfield{author}{\bibinfo{person}{Lena Mamykina}, \bibinfo{person}{Bella
  Manoim}, \bibinfo{person}{Manas Mittal}, \bibinfo{person}{George Hripcsak},
  {and} \bibinfo{person}{Bj{\"o}rn Hartmann}.} \bibinfo{year}{2011}\natexlab{}.
\newblock \showarticletitle{Design lessons from the fastest q\&a site in the
  west}. In \bibinfo{booktitle}{\emph{Proceedings of the SIGCHI conference on
  Human factors in computing systems}}. \bibinfo{pages}{2857--2866}.
\newblock


\bibitem[\protect\citeauthoryear{Mamykina, Nakikj, and Elhadad}{Mamykina
  et~al\mbox{.}}{2015}]%
        {collective_10.1145/2702123.2702566}
\bibfield{author}{\bibinfo{person}{Lena Mamykina}, \bibinfo{person}{Drashko
  Nakikj}, {and} \bibinfo{person}{Noemie Elhadad}.}
  \bibinfo{year}{2015}\natexlab{}.
\newblock \showarticletitle{Collective Sensemaking in Online Health Forums}. In
  \bibinfo{booktitle}{\emph{Proceedings of the 33rd Annual ACM Conference on
  Human Factors in Computing Systems}} \emph{(\bibinfo{series}{CHI ’15})}.
  \bibinfo{publisher}{Association for Computing Machinery},
  \bibinfo{address}{New York, NY, USA}, \bibinfo{pages}{3217–3226}.
\newblock
\showISBNx{9781450331456}
\urldef\tempurl%
\url{https://doi.org/10.1145/2702123.2702566}
\showDOI{\tempurl}


\bibitem[\protect\citeauthoryear{Mayfield, Wen, Golant, and
  Penstein~Ros\'{e}}{Mayfield et~al\mbox{.}}{2012}]%
        {discovering10.1145/2389176.2389216}
\bibfield{author}{\bibinfo{person}{Elijah Mayfield}, \bibinfo{person}{Miaomiao
  Wen}, \bibinfo{person}{Mitch Golant}, {and} \bibinfo{person}{Carolyn
  Penstein~Ros\'{e}}.} \bibinfo{year}{2012}\natexlab{}.
\newblock \showarticletitle{Discovering Habits of Effective Online Support
  Group Chatrooms}. In \bibinfo{booktitle}{\emph{Proceedings of the 17th ACM
  International Conference on Supporting Group Work}}
  \emph{(\bibinfo{series}{GROUP ’12})}. \bibinfo{publisher}{Association for
  Computing Machinery}, \bibinfo{address}{New York, NY, USA},
  \bibinfo{pages}{263–272}.
\newblock
\showISBNx{9781450314862}
\urldef\tempurl%
\url{https://doi.org/10.1145/2389176.2389216}
\showDOI{\tempurl}


\bibitem[\protect\citeauthoryear{McGrath}{McGrath}{1991}]%
        {mcgrath1991time}
\bibfield{author}{\bibinfo{person}{Joseph~E McGrath}.}
  \bibinfo{year}{1991}\natexlab{}.
\newblock \showarticletitle{Time, interaction, and performance (TIP) A Theory
  of Groups}.
\newblock \bibinfo{journal}{\emph{Small group research}} \bibinfo{volume}{22},
  \bibinfo{number}{2} (\bibinfo{year}{1991}), \bibinfo{pages}{147--174}.
\newblock


\bibitem[\protect\citeauthoryear{Microsoft}{Microsoft}{2018}]%
        {MicrosoftLUIS}
\bibfield{author}{\bibinfo{person}{Microsoft}.}
  \bibinfo{year}{2018}\natexlab{}.
\newblock \bibinfo{title}{Cognitive Services: Language Understanding (LUIS)}.
\newblock
\newblock
\newblock
\shownote{Retrieved: 2018-12-10. \url{https://www.luis.ai/home}.}


\bibitem[\protect\citeauthoryear{Morris, Kouddous, Kshirsagar, and
  Schueller}{Morris et~al\mbox{.}}{2018}]%
        {Recommend_comment_Morris2018TowardsAA}
\bibfield{author}{\bibinfo{person}{Robert~R. Morris}, \bibinfo{person}{Kareem
  Kouddous}, \bibinfo{person}{Rohan Kshirsagar}, {and}
  \bibinfo{person}{Stephen~Matthew Schueller}.}
  \bibinfo{year}{2018}\natexlab{}.
\newblock \showarticletitle{Towards an Artificially Empathic Conversational
  Agent for Mental Health Applications: System Design and User Perceptions}. In
  \bibinfo{booktitle}{\emph{Journal of medical Internet research}}.
\newblock


\bibitem[\protect\citeauthoryear{Murgia, Janssens, Demeyer, and
  Vasilescu}{Murgia et~al\mbox{.}}{2016}]%
        {Human-Bot_social_QA_Murgia:2016:MHI:2851581.2892311}
\bibfield{author}{\bibinfo{person}{Alessandro Murgia}, \bibinfo{person}{Daan
  Janssens}, \bibinfo{person}{Serge Demeyer}, {and} \bibinfo{person}{Bogdan
  Vasilescu}.} \bibinfo{year}{2016}\natexlab{}.
\newblock \showarticletitle{Among the Machines: Human-Bot Interaction on Social
  Q\&A Websites}. In \bibinfo{booktitle}{\emph{Proceedings of the 2016 CHI
  Conference Extended Abstracts on Human Factors in Computing Systems}}
  \emph{(\bibinfo{series}{CHI EA '16})}. \bibinfo{publisher}{ACM},
  \bibinfo{address}{New York, NY, USA}, \bibinfo{pages}{1272--1279}.
\newblock
\showISBNx{978-1-4503-4082-3}
\urldef\tempurl%
\url{https://doi.org/10.1145/2851581.2892311}
\showDOI{\tempurl}


\bibitem[\protect\citeauthoryear{Nam, Ackerman, and Adamic}{Nam
  et~al\mbox{.}}{2009}]%
        {nam2009questions}
\bibfield{author}{\bibinfo{person}{Kevin~Kyung Nam}, \bibinfo{person}{Mark~S
  Ackerman}, {and} \bibinfo{person}{Lada~A Adamic}.}
  \bibinfo{year}{2009}\natexlab{}.
\newblock \showarticletitle{Questions in, knowledge in? A study of Naver's
  question answering community}. In \bibinfo{booktitle}{\emph{Proceedings of
  the SIGCHI conference on human factors in computing systems}}.
  \bibinfo{pages}{779--788}.
\newblock


\bibitem[\protect\citeauthoryear{Nan and Haofen}{Nan and Haofen}{2018}]%
        {nan2018state}
\bibfield{author}{\bibinfo{person}{QIU Nan} {and} \bibinfo{person}{WANG
  Haofen}.} \bibinfo{year}{2018}\natexlab{}.
\newblock \bibinfo{title}{State machine based context-sensitive system for
  managing multi-round dialog}.
\newblock
\newblock
\newblock
\shownote{US Patent App. 15/694,917.}


\bibitem[\protect\citeauthoryear{O'Leary, Bhattacharya, Munson, Wobbrock, and
  Pratt}{O'Leary et~al\mbox{.}}{2017}]%
  {Design_Opportunities_for_Mental_Health_Peer_Support_TechnologiesO'Leary:2017:DOM:2998181.2998349}
\bibfield{author}{\bibinfo{person}{Kathleen O'Leary}, \bibinfo{person}{Arpita
  Bhattacharya}, \bibinfo{person}{Sean~A. Munson}, \bibinfo{person}{Jacob~O.
  Wobbrock}, {and} \bibinfo{person}{Wanda Pratt}.}
  \bibinfo{year}{2017}\natexlab{}.
\newblock \showarticletitle{Design Opportunities for Mental Health Peer Support
  Technologies}. In \bibinfo{booktitle}{\emph{Proceedings of the 2017 ACM
  Conference on Computer Supported Cooperative Work and Social Computing}}
  \emph{(\bibinfo{series}{CSCW '17})}. \bibinfo{publisher}{ACM},
  \bibinfo{address}{New York, NY, USA}, \bibinfo{pages}{1470--1484}.
\newblock
\showISBNx{978-1-4503-4335-0}
\urldef\tempurl%
\url{https://doi.org/10.1145/2998181.2998349}
\showDOI{\tempurl}


\bibitem[\protect\citeauthoryear{Papineni, Roukos, Ward, and Zhu}{Papineni
  et~al\mbox{.}}{2002}]%
        {Papineni2002BLEU}
\bibfield{author}{\bibinfo{person}{Kishore Papineni}, \bibinfo{person}{Salim
  Roukos}, \bibinfo{person}{Todd Ward}, {and} \bibinfo{person}{Wei-Jing Zhu}.}
  \bibinfo{year}{2002}\natexlab{}.
\newblock \showarticletitle{BLEU: a method for automatic evaluation of machine
  translation}. In \bibinfo{booktitle}{\emph{Proceedings of the 40th annual
  meeting of the Association for Computational Linguistics}}.
  \bibinfo{pages}{311--318}.
\newblock


\bibitem[\protect\citeauthoryear{Peng, Kim, and Ma}{Peng et~al\mbox{.}}{2019}]%
        {grembot_peng}
\bibfield{author}{\bibinfo{person}{Zhenhui Peng}, \bibinfo{person}{Taewook
  Kim}, {and} \bibinfo{person}{Xiaojuan Ma}.} \bibinfo{year}{2019}\natexlab{}.
\newblock \showarticletitle{GremoBot: Exploring Emotion Regulation in Group
  Chat}. In \bibinfo{booktitle}{\emph{Conference Companion Publication of the
  2019 on Computer Supported Cooperative Work and Social Computing}}
  \emph{(\bibinfo{series}{CSCW ’19})}. \bibinfo{publisher}{Association for
  Computing Machinery}, \bibinfo{address}{New York, NY, USA},
  \bibinfo{pages}{335–340}.
\newblock
\showISBNx{9781450366922}
\urldef\tempurl%
\url{https://doi.org/10.1145/3311957.3359472}
\showDOI{\tempurl}


\bibitem[\protect\citeauthoryear{Peng and Ma}{Peng and Ma}{2019a}]%
        {Peng2019github}
\bibfield{author}{\bibinfo{person}{Zhenhui Peng} {and}
  \bibinfo{person}{Xiaojuan Ma}.} \bibinfo{year}{2019}\natexlab{a}.
\newblock \showarticletitle{Exploring how software developers work with mention
  bot in GitHub}.
\newblock \bibinfo{journal}{\emph{CCF Transactions on Pervasive Computing and
  Interaction}} (\bibinfo{date}{05 Sep} \bibinfo{year}{2019}).
\newblock
\showISSN{2524-5228}
\urldef\tempurl%
\url{https://doi.org/10.1007/s42486-019-00013-2}
\showDOI{\tempurl}


\bibitem[\protect\citeauthoryear{Peng and Ma}{Peng and Ma}{2019b}]%
        {peng2019survey}
\bibfield{author}{\bibinfo{person}{Zhenhui Peng} {and}
  \bibinfo{person}{Xiaojuan Ma}.} \bibinfo{year}{2019}\natexlab{b}.
\newblock \showarticletitle{A survey on construction and enhancement methods in
  service chatbots design}.
\newblock \bibinfo{journal}{\emph{CCF Transactions on Pervasive Computing and
  Interaction}} \bibinfo{volume}{1}, \bibinfo{number}{3}
  (\bibinfo{year}{2019}), \bibinfo{pages}{204--223}.
\newblock


\bibitem[\protect\citeauthoryear{Perrier, Dell, DeRenzi, Anderson, Kinuthia,
  Unger, and John-Stewart}{Perrier et~al\mbox{.}}{2015}]%
        {engaging_10.1145/2702123.2702124}
\bibfield{author}{\bibinfo{person}{Trevor Perrier}, \bibinfo{person}{Nicola
  Dell}, \bibinfo{person}{Brian DeRenzi}, \bibinfo{person}{Richard Anderson},
  \bibinfo{person}{John Kinuthia}, \bibinfo{person}{Jennifer Unger}, {and}
  \bibinfo{person}{Grace John-Stewart}.} \bibinfo{year}{2015}\natexlab{}.
\newblock \showarticletitle{Engaging Pregnant Women in Kenya with a Hybrid
  Computer-Human SMS Communication System}. In
  \bibinfo{booktitle}{\emph{Proceedings of the 33rd Annual ACM Conference on
  Human Factors in Computing Systems}} \emph{(\bibinfo{series}{CHI ’15})}.
  \bibinfo{publisher}{Association for Computing Machinery},
  \bibinfo{address}{New York, NY, USA}, \bibinfo{pages}{1429–1438}.
\newblock
\showISBNx{9781450331456}
\urldef\tempurl%
\url{https://doi.org/10.1145/2702123.2702124}
\showDOI{\tempurl}


\bibitem[\protect\citeauthoryear{Pettman}{Pettman}{2009}]%
        {pettman2009love}
\bibfield{author}{\bibinfo{person}{Dominic Pettman}.}
  \bibinfo{year}{2009}\natexlab{}.
\newblock \showarticletitle{Love in the Time of Tamagotchi}.
\newblock \bibinfo{journal}{\emph{Theory, culture \& society}}
  \bibinfo{volume}{26}, \bibinfo{number}{2-3} (\bibinfo{year}{2009}),
  \bibinfo{pages}{189--208}.
\newblock


\bibitem[\protect\citeauthoryear{Ren, Harper, Drenner, Terveen, Kiesler, Riedl,
  and Kraut}{Ren et~al\mbox{.}}{2012a}]%
        {ren2012building}
\bibfield{author}{\bibinfo{person}{Yuqing Ren}, \bibinfo{person}{F~Maxwell
  Harper}, \bibinfo{person}{Sara Drenner}, \bibinfo{person}{Loren Terveen},
  \bibinfo{person}{Sara Kiesler}, \bibinfo{person}{John Riedl}, {and}
  \bibinfo{person}{Robert~E Kraut}.} \bibinfo{year}{2012}\natexlab{a}.
\newblock \showarticletitle{BUILDING MEMBER ATTACHMENT IN ONLINE COMMUNITIES:
  APPLYING THEORIES OF GROUP IDENTITY AND INTERPERSONAL BONDS}.
\newblock \bibinfo{journal}{\emph{MIS Quarterly}} \bibinfo{volume}{36},
  \bibinfo{number}{3} (\bibinfo{year}{2012}), \bibinfo{pages}{841--864}.
\newblock


\bibitem[\protect\citeauthoryear{Ren, Kraut, Kiesler, and Resnick}{Ren
  et~al\mbox{.}}{2012b}]%
        {ren2012encouraging}
\bibfield{author}{\bibinfo{person}{Yuqing Ren}, \bibinfo{person}{Robert Kraut},
  \bibinfo{person}{Sara Kiesler}, {and} \bibinfo{person}{Paul Resnick}.}
  \bibinfo{year}{2012}\natexlab{b}.
\newblock \showarticletitle{Encouraging commitment in online communities}.
\newblock \bibinfo{journal}{\emph{Building successful online communities:
  Evidence-based social design}} (\bibinfo{year}{2012}),
  \bibinfo{pages}{77--124}.
\newblock


\bibitem[\protect\citeauthoryear{Resnick, Konstan, Chen, and Kraut}{Resnick
  et~al\mbox{.}}{2012}]%
        {resnick2012starting}
\bibfield{author}{\bibinfo{person}{Paul Resnick}, \bibinfo{person}{Joseph
  Konstan}, \bibinfo{person}{Yan Chen}, {and} \bibinfo{person}{Robert~E
  Kraut}.} \bibinfo{year}{2012}\natexlab{}.
\newblock \showarticletitle{Starting new online communities}.
\newblock \bibinfo{journal}{\emph{Building successful online communities:
  Evidence-based social design}}  \bibinfo{volume}{231} (\bibinfo{year}{2012}).
\newblock


\bibitem[\protect\citeauthoryear{Ritter, Cherry, and Dolan}{Ritter
  et~al\mbox{.}}{2011}]%
        {Ritter2011Data}
\bibfield{author}{\bibinfo{person}{Alan Ritter}, \bibinfo{person}{Colin
  Cherry}, {and} \bibinfo{person}{William~B. Dolan}.}
  \bibinfo{year}{2011}\natexlab{}.
\newblock \showarticletitle{Data-Driven Response Generation in Social Media}.
  In \bibinfo{booktitle}{\emph{Conference on Empirical Methods in Natural
  Language Processing}}.
\newblock


\bibitem[\protect\citeauthoryear{Robert, Pierce, Marquis, Kim, and
  Alahmad}{Robert et~al\mbox{.}}{2020}]%
        {robert2020designing}
\bibfield{author}{\bibinfo{person}{Lionel~P Robert}, \bibinfo{person}{Casey
  Pierce}, \bibinfo{person}{Liz Marquis}, \bibinfo{person}{Sangmi Kim}, {and}
  \bibinfo{person}{Rasha Alahmad}.} \bibinfo{year}{2020}\natexlab{}.
\newblock \showarticletitle{Designing fair AI for managing employees in
  organizations: a review, critique, and design agenda}.
\newblock \bibinfo{journal}{\emph{Human--Computer Interaction}}
  (\bibinfo{year}{2020}), \bibinfo{pages}{1--31}.
\newblock


\bibitem[\protect\citeauthoryear{Robertson, Walker, Jones, Hancock-Beaulieu,
  Gatford, et~al\mbox{.}}{Robertson et~al\mbox{.}}{1995}]%
        {bm25}
\bibfield{author}{\bibinfo{person}{Stephen~E Robertson}, \bibinfo{person}{Steve
  Walker}, \bibinfo{person}{Susan Jones}, \bibinfo{person}{Micheline~M
  Hancock-Beaulieu}, \bibinfo{person}{Mike Gatford}, {et~al\mbox{.}}}
  \bibinfo{year}{1995}\natexlab{}.
\newblock \showarticletitle{Okapi at TREC-3}.
\newblock \bibinfo{journal}{\emph{Nist Special Publication Sp}}
  \bibinfo{volume}{109} (\bibinfo{year}{1995}), \bibinfo{pages}{109}.
\newblock


\bibitem[\protect\citeauthoryear{Savage, Monroy-Hernandez, and
  H\"{o}llerer}{Savage et~al\mbox{.}}{2016}]%
        {Botivist:Savage:2016:BCV:2818048.2819985}
\bibfield{author}{\bibinfo{person}{Saiph Savage}, \bibinfo{person}{Andres
  Monroy-Hernandez}, {and} \bibinfo{person}{Tobias H\"{o}llerer}.}
  \bibinfo{year}{2016}\natexlab{}.
\newblock \showarticletitle{Botivist: Calling Volunteers to Action Using Online
  Bots}. In \bibinfo{booktitle}{\emph{Proceedings of the 19th ACM Conference on
  Computer-Supported Cooperative Work \& Social Computing}}
  \emph{(\bibinfo{series}{CSCW '16})}. \bibinfo{publisher}{ACM},
  \bibinfo{address}{New York, NY, USA}, \bibinfo{pages}{813--822}.
\newblock
\showISBNx{978-1-4503-3592-8}
\urldef\tempurl%
\url{https://doi.org/10.1145/2818048.2819985}
\showDOI{\tempurl}


\bibitem[\protect\citeauthoryear{Schlesinger, O'Hara, and Taylor}{Schlesinger
  et~al\mbox{.}}{2018}]%
        {Race:Schlesinger:2018:LTR:3173574.3173889}
\bibfield{author}{\bibinfo{person}{Ari Schlesinger}, \bibinfo{person}{Kenton~P.
  O'Hara}, {and} \bibinfo{person}{Alex~S. Taylor}.}
  \bibinfo{year}{2018}\natexlab{}.
\newblock \showarticletitle{Let's Talk About Race: Identity, Chatbots, and AI}.
  In \bibinfo{booktitle}{\emph{Proceedings of the 2018 CHI Conference on Human
  Factors in Computing Systems}} \emph{(\bibinfo{series}{CHI '18})}.
  \bibinfo{publisher}{ACM}, \bibinfo{address}{New York, NY, USA}, Article
  \bibinfo{articleno}{315}, \bibinfo{numpages}{14}~pages.
\newblock
\showISBNx{978-1-4503-5620-6}
\urldef\tempurl%
\url{https://doi.org/10.1145/3173574.3173889}
\showDOI{\tempurl}


\bibitem[\protect\citeauthoryear{Schroeder, Wilkes, Rowan, Toledo, Paradiso,
  Czerwinski, Mark, and Linehan}{Schroeder et~al\mbox{.}}{2018}]%
        {DBT:Schroeder:2018:PSC:3173574.3173972}
\bibfield{author}{\bibinfo{person}{Jessica Schroeder}, \bibinfo{person}{Chelsey
  Wilkes}, \bibinfo{person}{Kael Rowan}, \bibinfo{person}{Arturo Toledo},
  \bibinfo{person}{Ann Paradiso}, \bibinfo{person}{Mary Czerwinski},
  \bibinfo{person}{Gloria Mark}, {and} \bibinfo{person}{Marsha~M. Linehan}.}
  \bibinfo{year}{2018}\natexlab{}.
\newblock \showarticletitle{Pocket Skills: A Conversational Mobile Web App To
  Support Dialectical Behavioral Therapy}. In
  \bibinfo{booktitle}{\emph{Proceedings of the 2018 CHI Conference on Human
  Factors in Computing Systems}} \emph{(\bibinfo{series}{CHI '18})}.
  \bibinfo{publisher}{ACM}, \bibinfo{address}{New York, NY, USA}, Article
  \bibinfo{articleno}{398}, \bibinfo{numpages}{15}~pages.
\newblock
\showISBNx{978-1-4503-5620-6}
\urldef\tempurl%
\url{https://doi.org/10.1145/3173574.3173972}
\showDOI{\tempurl}


\bibitem[\protect\citeauthoryear{Seering, Kraut, and Dabbish}{Seering
  et~al\mbox{.}}{2017}]%
        {shaping_pro_10.1145/2998181.2998277}
\bibfield{author}{\bibinfo{person}{Joseph Seering}, \bibinfo{person}{Robert
  Kraut}, {and} \bibinfo{person}{Laura Dabbish}.}
  \bibinfo{year}{2017}\natexlab{}.
\newblock \showarticletitle{Shaping Pro and Anti-Social Behavior on Twitch
  Through Moderation and Example-Setting}. In
  \bibinfo{booktitle}{\emph{Proceedings of the 2017 ACM Conference on Computer
  Supported Cooperative Work and Social Computing}}
  \emph{(\bibinfo{series}{CSCW ’17})}. \bibinfo{publisher}{Association for
  Computing Machinery}, \bibinfo{address}{New York, NY, USA},
  \bibinfo{pages}{111–125}.
\newblock
\showISBNx{9781450343350}
\urldef\tempurl%
\url{https://doi.org/10.1145/2998181.2998277}
\showDOI{\tempurl}


\bibitem[\protect\citeauthoryear{Seering, Luria, Ye, Kaufman, and
  Hammer}{Seering et~al\mbox{.}}{2020}]%
        {seering_village10.1145/3313831.3376708}
\bibfield{author}{\bibinfo{person}{Joseph Seering}, \bibinfo{person}{Michal
  Luria}, \bibinfo{person}{Connie Ye}, \bibinfo{person}{Geoff Kaufman}, {and}
  \bibinfo{person}{Jessica Hammer}.} \bibinfo{year}{2020}\natexlab{}.
\newblock \showarticletitle{It Takes a Village: Integrating an Adaptive Chatbot
  into an Online Gaming Community}. In \bibinfo{booktitle}{\emph{Proceedings of
  the 2020 CHI Conference on Human Factors in Computing Systems}}
  \emph{(\bibinfo{series}{CHI ’20})}. \bibinfo{publisher}{Association for
  Computing Machinery}, \bibinfo{address}{New York, NY, USA},
  \bibinfo{pages}{1–13}.
\newblock
\showISBNx{9781450367080}
\urldef\tempurl%
\url{https://doi.org/10.1145/3313831.3376708}
\showDOI{\tempurl}


\bibitem[\protect\citeauthoryear{Seering, Wang, Yoon, and Kaufman}{Seering
  et~al\mbox{.}}{2019}]%
        {seering2019moderator}
\bibfield{author}{\bibinfo{person}{Joseph Seering}, \bibinfo{person}{Tony
  Wang}, \bibinfo{person}{Jina Yoon}, {and} \bibinfo{person}{Geoff Kaufman}.}
  \bibinfo{year}{2019}\natexlab{}.
\newblock \showarticletitle{Moderator engagement and community development in
  the age of algorithms}.
\newblock \bibinfo{journal}{\emph{New Media \& Society}} \bibinfo{volume}{21},
  \bibinfo{number}{7} (\bibinfo{year}{2019}), \bibinfo{pages}{1417--1443}.
\newblock


\bibitem[\protect\citeauthoryear{Shamekhi, Liao, Wang, Bellamy, and
  Erickson}{Shamekhi et~al\mbox{.}}{2018}]%
        {Face_value:Shamekhi:2018:FVE:3173574.3173965}
\bibfield{author}{\bibinfo{person}{Ameneh Shamekhi}, \bibinfo{person}{Q.~Vera
  Liao}, \bibinfo{person}{Dakuo Wang}, \bibinfo{person}{Rachel K.~E. Bellamy},
  {and} \bibinfo{person}{Thomas Erickson}.} \bibinfo{year}{2018}\natexlab{}.
\newblock \showarticletitle{Face Value? Exploring the Effects of Embodiment for
  a Group Facilitation Agent}. In \bibinfo{booktitle}{\emph{Proceedings of the
  2018 CHI Conference on Human Factors in Computing Systems}}
  \emph{(\bibinfo{series}{CHI '18})}. \bibinfo{publisher}{ACM},
  \bibinfo{address}{New York, NY, USA}, Article \bibinfo{articleno}{391},
  \bibinfo{numpages}{13}~pages.
\newblock
\showISBNx{978-1-4503-5620-6}
\urldef\tempurl%
\url{https://doi.org/10.1145/3173574.3173965}
\showDOI{\tempurl}


\bibitem[\protect\citeauthoryear{Sharma and De~Choudhury}{Sharma and
  De~Choudhury}{2018}]%
        {linguisticSharma:2018:MHS:3173574.3174215}
\bibfield{author}{\bibinfo{person}{Eva Sharma} {and} \bibinfo{person}{Munmun
  De~Choudhury}.} \bibinfo{year}{2018}\natexlab{}.
\newblock \showarticletitle{Mental Health Support and Its Relationship to
  Linguistic Accommodation in Online Communities}. In
  \bibinfo{booktitle}{\emph{Proceedings of the 2018 CHI Conference on Human
  Factors in Computing Systems}} \emph{(\bibinfo{series}{CHI '18})}.
  \bibinfo{publisher}{ACM}, \bibinfo{address}{New York, NY, USA}, Article
  \bibinfo{articleno}{641}, \bibinfo{numpages}{13}~pages.
\newblock
\showISBNx{978-1-4503-5620-6}
\urldef\tempurl%
\url{https://doi.org/10.1145/3173574.3174215}
\showDOI{\tempurl}


\bibitem[\protect\citeauthoryear{Siek, Hayes, Newman, and Tang}{Siek
  et~al\mbox{.}}{2014}]%
        {siek2014field}
\bibfield{author}{\bibinfo{person}{Katie~A Siek}, \bibinfo{person}{Gillian~R
  Hayes}, \bibinfo{person}{Mark~W Newman}, {and} \bibinfo{person}{John~C
  Tang}.} \bibinfo{year}{2014}\natexlab{}.
\newblock \showarticletitle{Field deployments: Knowing from using in context}.
\newblock In \bibinfo{booktitle}{\emph{Ways of Knowing in HCI}}.
  \bibinfo{publisher}{Springer}, \bibinfo{pages}{119--142}.
\newblock


\bibitem[\protect\citeauthoryear{Tan, Wang, Gao, Wang, Potdar, Guo, Chang, and
  Yu}{Tan et~al\mbox{.}}{2019}]%
        {tan2019context}
\bibfield{author}{\bibinfo{person}{Ming Tan}, \bibinfo{person}{Dakuo Wang},
  \bibinfo{person}{Yupeng Gao}, \bibinfo{person}{Haoyu Wang},
  \bibinfo{person}{Saloni Potdar}, \bibinfo{person}{Xiaoxiao Guo},
  \bibinfo{person}{Shiyu Chang}, {and} \bibinfo{person}{Mo Yu}.}
  \bibinfo{year}{2019}\natexlab{}.
\newblock \showarticletitle{Context-Aware Conversation Thread Detection in
  Multi-Party Chat}. In \bibinfo{booktitle}{\emph{Proceedings of the 2019
  Conference on Empirical Methods in Natural Language Processing and the 9th
  International Joint Conference on Natural Language Processing
  (EMNLP-IJCNLP)}}. \bibinfo{pages}{6457--6462}.
\newblock


\bibitem[\protect\citeauthoryear{Terveen, Konstan, and Lampe}{Terveen
  et~al\mbox{.}}{2014}]%
        {terveen2014study}
\bibfield{author}{\bibinfo{person}{Loren Terveen}, \bibinfo{person}{Joseph~A
  Konstan}, {and} \bibinfo{person}{Cliff Lampe}.}
  \bibinfo{year}{2014}\natexlab{}.
\newblock \showarticletitle{Study, Build, Repeat: Using Online Communities as a
  Research Platform}.
\newblock In \bibinfo{booktitle}{\emph{Ways of Knowing in HCI}}.
  \bibinfo{publisher}{Springer}, \bibinfo{pages}{95--117}.
\newblock


\bibitem[\protect\citeauthoryear{Toxtli, Monroy-Hern\'{a}ndez, and
  Cranshaw}{Toxtli et~al\mbox{.}}{2018}]%
        {TaskManagement:Toxtli:2018:UCT:3173574.3173632}
\bibfield{author}{\bibinfo{person}{Carlos Toxtli}, \bibinfo{person}{Andr{\'e}s
  Monroy-Hern\'{a}ndez}, {and} \bibinfo{person}{Justin Cranshaw}.}
  \bibinfo{year}{2018}\natexlab{}.
\newblock \showarticletitle{Understanding Chatbot-mediated Task Management}. In
  \bibinfo{booktitle}{\emph{Proceedings of the 2018 CHI Conference on Human
  Factors in Computing Systems}} \emph{(\bibinfo{series}{CHI '18})}.
  \bibinfo{publisher}{ACM}, \bibinfo{address}{New York, NY, USA}, Article
  \bibinfo{articleno}{58}, \bibinfo{numpages}{6}~pages.
\newblock
\showISBNx{978-1-4503-5620-6}
\urldef\tempurl%
\url{https://doi.org/10.1145/3173574.3173632}
\showDOI{\tempurl}


\bibitem[\protect\citeauthoryear{van Uden-Kraan, Drossaert, Taal, Shaw, Seydel,
  and van~de Laar}{van Uden-Kraan et~al\mbox{.}}{2008}]%
        {Emotional_support_UdenKraan2012BreastC}
\bibfield{author}{\bibinfo{person}{Cornelia~F. van Uden-Kraan},
  \bibinfo{person}{Constance H.~C. Drossaert}, \bibinfo{person}{Erik Taal},
  \bibinfo{person}{Bret~R. Shaw}, \bibinfo{person}{Erwin~R. Seydel}, {and}
  \bibinfo{person}{Mart A. F.~J. van~de Laar}.}
  \bibinfo{year}{2008}\natexlab{}.
\newblock \showarticletitle{Empowering Processes and Outcomes of Participation
  in Online Support Groups for Patients With Breast Cancer, Arthritis, or
  Fibromyalgia}.
\newblock \bibinfo{journal}{\emph{Qualitative Health Research}}
  \bibinfo{volume}{18}, \bibinfo{number}{3} (\bibinfo{year}{2008}),
  \bibinfo{pages}{405--417}.
\newblock
\urldef\tempurl%
\url{https://doi.org/10.1177/1049732307313429}
\showDOI{\tempurl}
\showeprint{https://doi.org/10.1177/1049732307313429}
\newblock
\shownote{PMID: 18235163.}


\bibitem[\protect\citeauthoryear{Walther and Boyd}{Walther and Boyd}{2002}]%
        {social_support_article}
\bibfield{author}{\bibinfo{person}{Joseph Walther} {and} \bibinfo{person}{Saxon
  Boyd}.} \bibinfo{year}{2002}\natexlab{}.
\newblock \showarticletitle{Attraction to computer-mediated social support}.
\newblock \bibinfo{journal}{\emph{Communication Technology and Society:
  Audience Adoption and Uses}} (\bibinfo{date}{01} \bibinfo{year}{2002}),
  \bibinfo{pages}{153--188}.
\newblock


\bibitem[\protect\citeauthoryear{Wang, Andres, Weisz, Oduor, and Dugan}{Wang
  et~al\mbox{.}}{2021a}]%
        {autods}
\bibfield{author}{\bibinfo{person}{Dakuo Wang}, \bibinfo{person}{Josh Andres},
  \bibinfo{person}{Justin Weisz}, \bibinfo{person}{Erick Oduor}, {and}
  \bibinfo{person}{Casey Dugan}.} \bibinfo{year}{2021}\natexlab{a}.
\newblock \showarticletitle{AutoDS: Towards Human-Centered Automation of Data
  Science}. In \bibinfo{booktitle}{\emph{Proceedings of the CHI 2021}}.
\newblock


\bibitem[\protect\citeauthoryear{Wang, Hou, Luo, and Pan}{Wang
  et~al\mbox{.}}{2016}]%
        {wang2016answerer}
\bibfield{author}{\bibinfo{person}{Dakuo Wang}, \bibinfo{person}{Youyang Hou},
  \bibinfo{person}{Lin Luo}, {and} \bibinfo{person}{Yingxin Pan}.}
  \bibinfo{year}{2016}\natexlab{}.
\newblock \showarticletitle{Answerer engagement in an enterprise social
  question \& answering system}.
\newblock \bibinfo{journal}{\emph{IConference 2016 Proceedings}}
  (\bibinfo{year}{2016}).
\newblock


\bibitem[\protect\citeauthoryear{Wang, Wang, Zhang, Wang, Zhu, Gao, Fan, and
  Tian}{Wang et~al\mbox{.}}{2021b}]%
        {aidoctor}
\bibfield{author}{\bibinfo{person}{Dakuo Wang}, \bibinfo{person}{Liuping Wang},
  \bibinfo{person}{Zhan Zhang}, \bibinfo{person}{Ding Wang},
  \bibinfo{person}{Haiyi Zhu}, \bibinfo{person}{Yvonne Gao},
  \bibinfo{person}{Xiangmin Fan}, {and} \bibinfo{person}{Feng Tian}.}
  \bibinfo{year}{2021}\natexlab{b}.
\newblock \showarticletitle{Brilliant AI Doctor in Rural China: Tensions and
  Challenges in AI-Powered CDSS Deployment}. In
  \bibinfo{booktitle}{\emph{Proceedings of the CHI 2021}}.
\newblock


\bibitem[\protect\citeauthoryear{Wang, Weisz, Muller, Ram, Geyer, Dugan,
  Tausczik, Samulowitz, and Gray}{Wang et~al\mbox{.}}{2019}]%
        {wang2019humanai}
\bibfield{author}{\bibinfo{person}{Dakuo Wang}, \bibinfo{person}{Justin~D.
  Weisz}, \bibinfo{person}{Michael Muller}, \bibinfo{person}{Parikshit Ram},
  \bibinfo{person}{Werner Geyer}, \bibinfo{person}{Casey Dugan},
  \bibinfo{person}{Yla Tausczik}, \bibinfo{person}{Horst Samulowitz}, {and}
  \bibinfo{person}{Alexander Gray}.} \bibinfo{year}{2019}\natexlab{}.
\newblock \showarticletitle{Human-AI Collaboration in Data Science: Exploring
  Data Scientists' Perceptions of Automated AI}.
\newblock \bibinfo{journal}{\emph{To appear in Computer Supported Cooperative
  Work (CSCW)}} (\bibinfo{year}{2019}).
\newblock


\bibitem[\protect\citeauthoryear{Wang, Sierra, and Folger}{Wang
  et~al\mbox{.}}{2003}]%
        {wang2003building}
\bibfield{author}{\bibinfo{person}{Minjuan Wang}, \bibinfo{person}{Christina
  Sierra}, {and} \bibinfo{person}{Terre Folger}.}
  \bibinfo{year}{2003}\natexlab{}.
\newblock \showarticletitle{Building a dynamic online learning community among
  adult learners}.
\newblock \bibinfo{journal}{\emph{Educational Media International}}
  \bibinfo{volume}{40}, \bibinfo{number}{1-2} (\bibinfo{year}{2003}),
  \bibinfo{pages}{49--62}.
\newblock


\bibitem[\protect\citeauthoryear{Wang, Yu, Guo, Wang, Klinger, Zhang, Chang,
  Tesauro, Zhou, and Jiang}{Wang et~al\mbox{.}}{2018}]%
        {wang2018r}
\bibfield{author}{\bibinfo{person}{Shuohang Wang}, \bibinfo{person}{Mo Yu},
  \bibinfo{person}{Xiaoxiao Guo}, \bibinfo{person}{Zhiguo Wang},
  \bibinfo{person}{Tim Klinger}, \bibinfo{person}{Wei Zhang},
  \bibinfo{person}{Shiyu Chang}, \bibinfo{person}{Gerry Tesauro},
  \bibinfo{person}{Bowen Zhou}, {and} \bibinfo{person}{Jing Jiang}.}
  \bibinfo{year}{2018}\natexlab{}.
\newblock \showarticletitle{R 3: Reinforced ranker-reader for open-domain
  question answering}. In \bibinfo{booktitle}{\emph{Thirty-Second AAAI
  Conference on Artificial Intelligence}}.
\newblock


\bibitem[\protect\citeauthoryear{Wang, Joshi, Cohen, and Ros{\'e}}{Wang
  et~al\mbox{.}}{2008}]%
        {wang2008recovering}
\bibfield{author}{\bibinfo{person}{Yi-Chia Wang}, \bibinfo{person}{Mahesh
  Joshi}, \bibinfo{person}{William~W Cohen}, {and}
  \bibinfo{person}{Carolyn~Penstein Ros{\'e}}.}
  \bibinfo{year}{2008}\natexlab{}.
\newblock \showarticletitle{Recovering Implicit Thread Structure in Newsgroup
  Style Conversations.}. In \bibinfo{booktitle}{\emph{ICWSM}}.
\newblock


\bibitem[\protect\citeauthoryear{Wang, Kraut, and Levine}{Wang
  et~al\mbox{.}}{2015}]%
        {no_reply_community:article}
\bibfield{author}{\bibinfo{person}{Yi-Chia Wang}, \bibinfo{person}{Robert
  Kraut}, {and} \bibinfo{person}{John Levine}.}
  \bibinfo{year}{2015}\natexlab{}.
\newblock \showarticletitle{Eliciting and Receiving Online Support: Using
  Computer-Aided Content Analysis to Examine the Dynamics of Online Social
  Support}.
\newblock \bibinfo{journal}{\emph{Journal of medical Internet research}}
  \bibinfo{volume}{17} (\bibinfo{date}{04} \bibinfo{year}{2015}),
  \bibinfo{pages}{e99}.
\newblock
\urldef\tempurl%
\url{https://doi.org/10.2196/jmir.3558}
\showDOI{\tempurl}


\bibitem[\protect\citeauthoryear{Wang, Kraut, and Levine}{Wang
  et~al\mbox{.}}{2012}]%
        {To_stay_or_leave_Wang:2012:SLR:2145204.2145329}
\bibfield{author}{\bibinfo{person}{Yi-Chia Wang}, \bibinfo{person}{Robert
  Kraut}, {and} \bibinfo{person}{John~M. Levine}.}
  \bibinfo{year}{2012}\natexlab{}.
\newblock \showarticletitle{To Stay or Leave?: The Relationship of Emotional
  and Informational Support to Commitment in Online Health Support Groups}. In
  \bibinfo{booktitle}{\emph{Proceedings of the ACM 2012 Conference on Computer
  Supported Cooperative Work}} \emph{(\bibinfo{series}{CSCW '12})}.
  \bibinfo{publisher}{ACM}, \bibinfo{address}{New York, NY, USA},
  \bibinfo{pages}{833--842}.
\newblock
\showISBNx{978-1-4503-1086-4}
\urldef\tempurl%
\url{https://doi.org/10.1145/2145204.2145329}
\showDOI{\tempurl}


\bibitem[\protect\citeauthoryear{Weizenbaum}{Weizenbaum}{1966}]%
        {ELIZA:Weizenbaum:1966:ECP:365153.365168}
\bibfield{author}{\bibinfo{person}{Joseph Weizenbaum}.}
  \bibinfo{year}{1966}\natexlab{}.
\newblock \showarticletitle{ELIZA\&Mdash;a Computer Program for the Study of
  Natural Language Communication Between Man and Machine}.
\newblock \bibinfo{journal}{\emph{Commun. ACM}} \bibinfo{volume}{9},
  \bibinfo{number}{1} (\bibinfo{date}{Jan.} \bibinfo{year}{1966}),
  \bibinfo{pages}{36--45}.
\newblock
\showISSN{0001-0782}
\urldef\tempurl%
\url{https://doi.org/10.1145/365153.365168}
\showDOI{\tempurl}


\bibitem[\protect\citeauthoryear{Wen and Ros{\'e}}{Wen and Ros{\'e}}{2012}]%
        {wen2012understanding}
\bibfield{author}{\bibinfo{person}{Miaomiao Wen} {and}
  \bibinfo{person}{Carolyn~Penstein Ros{\'e}}.}
  \bibinfo{year}{2012}\natexlab{}.
\newblock \showarticletitle{Understanding participant behavior trajectories in
  online health support groups using automatic extraction methods}. In
  \bibinfo{booktitle}{\emph{Proceedings of the 17th ACM international
  conference on Supporting group work}}. \bibinfo{pages}{179--188}.
\newblock


\bibitem[\protect\citeauthoryear{Wolf, Miller, and Grodzinsky}{Wolf
  et~al\mbox{.}}{2017}]%
        {wolf2017we}
\bibfield{author}{\bibinfo{person}{Marty~J Wolf}, \bibinfo{person}{K Miller},
  {and} \bibinfo{person}{Frances~S Grodzinsky}.}
  \bibinfo{year}{2017}\natexlab{}.
\newblock \showarticletitle{Why we should have seen that coming: comments on
  Microsoft's tay" experiment," and wider implications}.
\newblock \bibinfo{journal}{\emph{ACM SIGCAS Computers and Society}}
  \bibinfo{volume}{47}, \bibinfo{number}{3} (\bibinfo{year}{2017}),
  \bibinfo{pages}{54--64}.
\newblock


\bibitem[\protect\citeauthoryear{Wong-Villacres, Kumar, and
  DiSalvo}{Wong-Villacres et~al\mbox{.}}{2019}]%
        {Actor_Network10.1145/3290605.3300914}
\bibfield{author}{\bibinfo{person}{Marisol Wong-Villacres},
  \bibinfo{person}{Neha Kumar}, {and} \bibinfo{person}{Betsy DiSalvo}.}
  \bibinfo{year}{2019}\natexlab{}.
\newblock \showarticletitle{The Parenting Actor-Network of Latino Immigrants in
  the United States}. In \bibinfo{booktitle}{\emph{Proceedings of the 2019 CHI
  Conference on Human Factors in Computing Systems}}
  \emph{(\bibinfo{series}{CHI ’19})}. \bibinfo{publisher}{Association for
  Computing Machinery}, \bibinfo{address}{New York, NY, USA},
  \bibinfo{pages}{1–12}.
\newblock
\showISBNx{9781450359702}
\urldef\tempurl%
\url{https://doi.org/10.1145/3290605.3300914}
\showDOI{\tempurl}


\bibitem[\protect\citeauthoryear{Woodruff, Fox, Rousso-Schindler, and
  Warshaw}{Woodruff et~al\mbox{.}}{2018}]%
        {woodruff2018qualitative}
\bibfield{author}{\bibinfo{person}{Allison Woodruff}, \bibinfo{person}{Sarah~E
  Fox}, \bibinfo{person}{Steven Rousso-Schindler}, {and}
  \bibinfo{person}{Jeffrey Warshaw}.} \bibinfo{year}{2018}\natexlab{}.
\newblock \showarticletitle{A qualitative exploration of perceptions of
  algorithmic fairness}. In \bibinfo{booktitle}{\emph{Proceedings of the 2018
  chi conference on human factors in computing systems}}.
  \bibinfo{pages}{1--14}.
\newblock


\bibitem[\protect\citeauthoryear{Wu and Ma}{Wu and Ma}{2017}]%
        {Wu2017Money}
\bibfield{author}{\bibinfo{person}{Ziming Wu} {and} \bibinfo{person}{Xiaojuan
  Ma}.} \bibinfo{year}{2017}\natexlab{}.
\newblock \showarticletitle{Money as a Social Currency to Manage Group
  Dynamics: Red Packet Gifting in Chinese Online Communities}. In
  \bibinfo{booktitle}{\emph{Chi Conference Extended Abstracts on Human Factors
  in Computing Systems}}.
\newblock


\bibitem[\protect\citeauthoryear{Xiang, Zhang, and Liang}{Xiang
  et~al\mbox{.}}{2020}]%
        {xiang2020sedentary}
\bibfield{author}{\bibinfo{person}{Mi Xiang}, \bibinfo{person}{Zhiruo Zhang},
  {and} \bibinfo{person}{Huigang Liang}.} \bibinfo{year}{2020}\natexlab{}.
\newblock \showarticletitle{Sedentary behavior relates to mental distress of
  pregnant women differently across trimesters: An observational study in
  China}.
\newblock \bibinfo{journal}{\emph{Journal of affective disorders}}
  \bibinfo{volume}{260} (\bibinfo{year}{2020}), \bibinfo{pages}{187--193}.
\newblock


\bibitem[\protect\citeauthoryear{Xu, Liu, Guo, Sinha, and Akkiraju}{Xu
  et~al\mbox{.}}{2017}]%
        {xu2017new}
\bibfield{author}{\bibinfo{person}{Anbang Xu}, \bibinfo{person}{Zhe Liu},
  \bibinfo{person}{Yufan Guo}, \bibinfo{person}{Vibha Sinha}, {and}
  \bibinfo{person}{Rama Akkiraju}.} \bibinfo{year}{2017}\natexlab{}.
\newblock \showarticletitle{A new chatbot for customer service on social
  media}. In \bibinfo{booktitle}{\emph{Proceedings of the 2017 CHI Conference
  on Human Factors in Computing Systems}}. \bibinfo{pages}{3506--3510}.
\newblock


\bibitem[\protect\citeauthoryear{Xu, Wang, Collins, Lee, and Warschauer}{Xu
  et~al\mbox{.}}{[n.d.]}]%
        {xu161same}
\bibfield{author}{\bibinfo{person}{Ying Xu}, \bibinfo{person}{Dakuo Wang},
  \bibinfo{person}{Penelope Collins}, \bibinfo{person}{Hyelim Lee}, {and}
  \bibinfo{person}{Mark Warschauer}.} \bibinfo{year}{[n.d.]}\natexlab{}.
\newblock \showarticletitle{Same benefits, different communication patterns:
  Comparing Children's reading with a conversational agent vs. a human
  partner}.
\newblock \bibinfo{journal}{\emph{Computers \& Education}}
  \bibinfo{volume}{161} (\bibinfo{year}{[n.\,d.]}), \bibinfo{pages}{104059}.
\newblock


\bibitem[\protect\citeauthoryear{Yang, Kraut, and Levine}{Yang
  et~al\mbox{.}}{2017a}]%
        {Commitment10.1145/3025453.3026008}
\bibfield{author}{\bibinfo{person}{Diyi Yang}, \bibinfo{person}{Robert Kraut},
  {and} \bibinfo{person}{John~M. Levine}.} \bibinfo{year}{2017}\natexlab{a}.
\newblock \showarticletitle{Commitment of Newcomers and Old-Timers to Online
  Health Support Communities}. In \bibinfo{booktitle}{\emph{Proceedings of the
  2017 CHI Conference on Human Factors in Computing Systems}}
  \emph{(\bibinfo{series}{CHI ’17})}. \bibinfo{publisher}{Association for
  Computing Machinery}, \bibinfo{address}{New York, NY, USA},
  \bibinfo{pages}{6363–6375}.
\newblock
\showISBNx{9781450346559}
\urldef\tempurl%
\url{https://doi.org/10.1145/3025453.3026008}
\showDOI{\tempurl}


\bibitem[\protect\citeauthoryear{Yang, Kraut, Smith, Mayfield, and
  Jurafsky}{Yang et~al\mbox{.}}{2019a}]%
        {seekers:Yang:2019:SPW:3290605.3300574}
\bibfield{author}{\bibinfo{person}{Diyi Yang}, \bibinfo{person}{Robert~E.
  Kraut}, \bibinfo{person}{Tenbroeck Smith}, \bibinfo{person}{Elijah Mayfield},
  {and} \bibinfo{person}{Dan Jurafsky}.} \bibinfo{year}{2019}\natexlab{a}.
\newblock \showarticletitle{Seekers, Providers, Welcomers, and Storytellers:
  Modeling Social Roles in Online Health Communities}. In
  \bibinfo{booktitle}{\emph{Proceedings of the 2019 CHI Conference on Human
  Factors in Computing Systems}} \emph{(\bibinfo{series}{CHI '19})}.
  \bibinfo{publisher}{ACM}, \bibinfo{address}{New York, NY, USA}, Article
  \bibinfo{articleno}{344}, \bibinfo{numpages}{14}~pages.
\newblock
\showISBNx{978-1-4503-5970-2}
\urldef\tempurl%
\url{https://doi.org/10.1145/3290605.3300574}
\showDOI{\tempurl}


\bibitem[\protect\citeauthoryear{Yang, Yao, and Kraut}{Yang
  et~al\mbox{.}}{2017b}]%
        {yang2017self}
\bibfield{author}{\bibinfo{person}{Diyi Yang}, \bibinfo{person}{Zheng Yao},
  {and} \bibinfo{person}{Robert Kraut}.} \bibinfo{year}{2017}\natexlab{b}.
\newblock \showarticletitle{Self-disclosure and channel difference in online
  health support groups}. In \bibinfo{booktitle}{\emph{Eleventh International
  AAAI Conference on Web and Social Media}}.
\newblock


\bibitem[\protect\citeauthoryear{Yang, Yao, Seering, and Kraut}{Yang
  et~al\mbox{.}}{2019b}]%
        {Channel:Yang:2019:CMS:3290605.3300261}
\bibfield{author}{\bibinfo{person}{Diyi Yang}, \bibinfo{person}{Zheng Yao},
  \bibinfo{person}{Joseph Seering}, {and} \bibinfo{person}{Robert Kraut}.}
  \bibinfo{year}{2019}\natexlab{b}.
\newblock \showarticletitle{The Channel Matters: Self-disclosure, Reciprocity
  and Social Support in Online Cancer Support Groups}. In
  \bibinfo{booktitle}{\emph{Proceedings of the 2019 CHI Conference on Human
  Factors in Computing Systems}} \emph{(\bibinfo{series}{CHI '19})}.
  \bibinfo{publisher}{ACM}, \bibinfo{address}{New York, NY, USA}, Article
  \bibinfo{articleno}{31}, \bibinfo{numpages}{15}~pages.
\newblock
\showISBNx{978-1-4503-5970-2}
\urldef\tempurl%
\url{https://doi.org/10.1145/3290605.3300261}
\showDOI{\tempurl}


\bibitem[\protect\citeauthoryear{Yazici, Sati~Kirkan, Akcali~Aslan, Aydin, and
  Yazici}{Yazici et~al\mbox{.}}{2015}]%
        {pregnancy_depressionarticle}
\bibfield{author}{\bibinfo{person}{Esra Yazici}, \bibinfo{person}{Tulay
  Sati~Kirkan}, \bibinfo{person}{Puren Akcali~Aslan}, \bibinfo{person}{Nazan
  Aydin}, {and} \bibinfo{person}{A Yazici}.} \bibinfo{year}{2015}\natexlab{}.
\newblock \showarticletitle{Untreated depression in the first trimester of
  pregnancy leads to postpartum depression: High rates from a natural follow-up
  study}.
\newblock \bibinfo{journal}{\emph{Neuropsychiatric Disease and Treatment}}
  \bibinfo{volume}{11} (\bibinfo{date}{02} \bibinfo{year}{2015}),
  \bibinfo{pages}{405--11}.
\newblock
\urldef\tempurl%
\url{https://doi.org/10.2147/ndt.s77194}
\showDOI{\tempurl}


\bibitem[\protect\citeauthoryear{Yin, Chang, and Zhang}{Yin
  et~al\mbox{.}}{2017}]%
        {DeepProbe}
\bibfield{author}{\bibinfo{person}{Zi Yin}, \bibinfo{person}{Keng-hao Chang},
  {and} \bibinfo{person}{Ruofei Zhang}.} \bibinfo{year}{2017}\natexlab{}.
\newblock \showarticletitle{DeepProbe: Information Directed Sequence
  Understanding and Chatbot Design via Recurrent Neural Networks}. In
  \bibinfo{booktitle}{\emph{Proceedings of the 23rd ACM SIGKDD International
  Conference on Knowledge Discovery and Data Mining}}
  \emph{(\bibinfo{series}{KDD ’17})}. \bibinfo{publisher}{Association for
  Computing Machinery}, \bibinfo{address}{New York, NY, USA},
  \bibinfo{pages}{2131–2139}.
\newblock
\showISBNx{9781450348874}
\urldef\tempurl%
\url{https://doi.org/10.1145/3097983.3098148}
\showDOI{\tempurl}


\bibitem[\protect\citeauthoryear{Zhang and Cranshaw}{Zhang and
  Cranshaw}{2018}]%
        {MakeSenseGroupChat:Zhang:2018:MSG:3290265.3274465}
\bibfield{author}{\bibinfo{person}{Amy~X. Zhang} {and} \bibinfo{person}{Justin
  Cranshaw}.} \bibinfo{year}{2018}\natexlab{}.
\newblock \showarticletitle{Making Sense of Group Chat Through Collaborative
  Tagging and Summarization}.
\newblock \bibinfo{journal}{\emph{Proc. ACM Hum.-Comput. Interact.}}
  \bibinfo{volume}{2}, \bibinfo{number}{CSCW}, Article \bibinfo{articleno}{196}
  (\bibinfo{date}{Nov.} \bibinfo{year}{2018}), \bibinfo{numpages}{27}~pages.
\newblock
\showISSN{2573-0142}
\urldef\tempurl%
\url{https://doi.org/10.1145/3274465}
\showDOI{\tempurl}


\bibitem[\protect\citeauthoryear{Zhou, Gao, Li, and Shum}{Zhou
  et~al\mbox{.}}{2020}]%
        {zhou2020design}
\bibfield{author}{\bibinfo{person}{Li Zhou}, \bibinfo{person}{Jianfeng Gao},
  \bibinfo{person}{Di Li}, {and} \bibinfo{person}{Heung-Yeung Shum}.}
  \bibinfo{year}{2020}\natexlab{}.
\newblock \showarticletitle{The design and implementation of xiaoice, an
  empathetic social chatbot}.
\newblock \bibinfo{journal}{\emph{Computational Linguistics}}
  \bibinfo{volume}{46}, \bibinfo{number}{1} (\bibinfo{year}{2020}),
  \bibinfo{pages}{53--93}.
\newblock


\bibitem[\protect\citeauthoryear{Zhu, Kraut, and Kittur}{Zhu
  et~al\mbox{.}}{2014}]%
        {zhu2014impact}
\bibfield{author}{\bibinfo{person}{Haiyi Zhu}, \bibinfo{person}{Robert~E
  Kraut}, {and} \bibinfo{person}{Aniket Kittur}.}
  \bibinfo{year}{2014}\natexlab{}.
\newblock \showarticletitle{The impact of membership overlap on the survival of
  online communities}. In \bibinfo{booktitle}{\emph{Proceedings of the SIGCHI
  Conference on Human Factors in Computing Systems}}.
  \bibinfo{pages}{281--290}.
\newblock


\bibitem[\protect\citeauthoryear{Zhu, Yu, Halfaker, and Terveen}{Zhu
  et~al\mbox{.}}{2018}]%
        {zhu2018value}
\bibfield{author}{\bibinfo{person}{Haiyi Zhu}, \bibinfo{person}{Bowen Yu},
  \bibinfo{person}{Aaron Halfaker}, {and} \bibinfo{person}{Loren Terveen}.}
  \bibinfo{year}{2018}\natexlab{}.
\newblock \showarticletitle{Value-sensitive algorithm design: Method, case
  study, and lessons}.
\newblock \bibinfo{journal}{\emph{Proceedings of the ACM on Human-Computer
  Interaction}} \bibinfo{volume}{2}, \bibinfo{number}{CSCW}
  (\bibinfo{year}{2018}), \bibinfo{pages}{1--23}.
\newblock


\bibitem[\protect\citeauthoryear{Zhu, Zhang, He, Kraut, and Kittur}{Zhu
  et~al\mbox{.}}{2013}]%
        {zhu2013effects}
\bibfield{author}{\bibinfo{person}{Haiyi Zhu}, \bibinfo{person}{Amy Zhang},
  \bibinfo{person}{Jiping He}, \bibinfo{person}{Robert~E Kraut}, {and}
  \bibinfo{person}{Aniket Kittur}.} \bibinfo{year}{2013}\natexlab{}.
\newblock \showarticletitle{Effects of peer feedback on contribution: a field
  experiment in Wikipedia}. In \bibinfo{booktitle}{\emph{Proceedings of the
  SIGCHI Conference on Human Factors in Computing Systems}}.
  \bibinfo{pages}{2253--2262}.
\newblock


\end{thebibliography}

\received{December 2020}
\received[revised]{January 2021}
\received[accepted]{February 2021}

\end{document}